\documentclass[12pt,a4paper]{article}

\usepackage{amsmath}
\usepackage{amssymb}
\usepackage{graphicx}
\usepackage{cite}
\usepackage{hyperref}

\begin{document}
\title{Thin and thick bubble walls I: vacuum phase transitions}
\author{
	Ariel M\'{e}gevand\thanks{Member of CONICET, Argentina. E-mail address:
		megevand@mdp.edu.ar}~ 
	and Federico Agust\'{\i}n Membiela\thanks{Member of CONICET, Argentina. E-mail
		address: membiela@mdp.edu.ar} \\[0.5cm]
	\normalsize \it IFIMAR (CONICET-UNMdP)\\
	\normalsize \it Departamento de F\'{\i}sica, Facultad de Ciencias Exactas
	y Naturales, \\
	\normalsize \it UNMdP, De\'{a}n Funes 3350, (7600) Mar del Plata, Argentina }
\date{}
\maketitle

\begin{abstract}
This is the first in a series of papers where we study the dynamics
of a bubble wall beyond usual approximations, such as the assumptions
of spherical bubbles and infinitely thin walls. In this paper, we
consider a vacuum phase transition. Thus, we describe a bubble as
a configuration of a scalar field whose equation of motion depends only
on the effective potential. 
The thin-wall approximation allows obtaining both
an effective equation of motion for the wall position and
a simplified equation for the field profile inside the wall. 
Several different assumptions are involved in this approximation. 
We discuss the conditions for the validity of each of them.
In particular, the minima of the effective potential must 
have approximately the same energy, and we discuss the correct implementation of this approximation.
We consider different improvements to the basic thin-wall approximation, 
such as an iterative method for finding the wall profile 
and a perturbative calculation in powers of the wall width.
We calculate the leading-order corrections.
Besides, we derive an equation of motion for the wall without any assumptions about its shape.
We present a suitable method to describe arbitrarily deformed walls from the spherical shape. 
We consider concrete examples and compare our approximations with numerical
solutions. In subsequent papers, we shall consider higher-order finite-width corrections, 
and we shall take into account the presence of the fluid.
\end{abstract}

\section{Introduction}

\label{sec:Intro}

A cosmological first-order phase transition occurs by nucleation and
expansion of bubbles of the stable phase in the sea of the metastable
phase. The propagation of the phase transition fronts may give rise
to the production of cosmological relics such as gravitational waves
\cite{h86,tw90,ktw92,kkt94}, magnetic fields \cite{co94,bbm96,soj97}, and baryons \cite{krs85,ckn93}. 
The dynamics of these bubble walls involves their
interaction with the surrounding hot plasma, which includes a variety
of phenomena such as friction \cite{lmt92,dlhll92,k92,a93,mp95}, bulk fluid motions \cite{s82,gkkm84}, and hydrodynamic
instabilities \cite{l92}. 
In the simplest case, we have a scalar
field with an effective potential $V(\phi)$ with two minima $\phi_{\pm}$
separated by a barrier. 
The absolute minimum $\phi_{-}$ corresponds
to the stable phase, while the local minimum $\phi_{+}$ corresponds
to the metastable phase. 
A newly nucleated bubble essentially consists 
of a spherical region where the field takes the value $\phi=\phi_{-}$, surrounded by the value
$\phi=\phi_{+}$. 
As a bubble expands, it
may lose its spherical shape for different reasons, such as 
the interaction with other bubbles 
(in particular, bubble collisions may give rise to the amplification of
small fluctuations\cite{bbm1,bbm2,bbm3})
or
the growth of perturbations due to hydrodynamic instabilities \cite{l92,hkllm93,mm14,mm14b}.
Moreover, as the phase transition develops and bubbles overlap, coalesce, and percolate,  the
concept of an individual bubble eventually loses its meaning. 
Nevertheless, the concept of bubble walls as the interfaces between the domains occupied by the two phases persists.

A bubble wall is very similar to a domain wall. 
The latter is a ``kink'' solution of a scalar field whose potential
has two minima $ \phi_\pm $ with the same energy \cite{vs00}. 
As regards its equation, the degeneracy of the potential is the essential difference with the phase-transition scenario. 
In this sense, the domain wall can be considered a specific case of the bubble wall.
Another difference is the bubble's spherical initial condition.
Both solutions of the field equation
correspond to a configuration $\phi(x^{\mu})$ in
which the field varies between the values $\phi_{-}$ and $\phi_{+}$. 
If the interface is thin, it can be described by a curved, time-dependent surface, parametrized
as $x^{\mu}=X^{\mu}(\xi^{a})$, where $\xi^{a}=(\xi^{0},\xi^{1},\xi^{2})$
are the world-volume coordinates of the hypersurface ($ \xi^0 $
is a temporal variable).

An effective equation for the wall can be obtained by proposing a solution of the form
$\phi=\phi(n)$, where $n$ is the coordinate perpendicular to the
hypersurface. 
Inserting into the field equation and making a few approximations
which are justified for a thin wall, one obtains an equation for the
field profile $\phi(n)$, as well as an equation for the wall surface $X^{\mu}(\xi^{a})$
(this procedure is also used for cosmic strings and other topological defects, 
and is often implemented at the action level rather than in the field equation \cite{vilenkin85,vs00,b94,hk95}). 
The thin-wall approximations used
here require the wall width, $l$, to be much smaller than the curvature
radius of the hypersurface, $L$. This is usually the case, since
the scale of $l$ is given by the inverse of the scale $v$ of the
theory, while $L$ is naturally of cosmological order and hence given
by the Hubble length, $H^{-1}\sim M_{P}/v^{2}$, where $M_{P}$ is
the Planck mass. 
However, it is worth emphasizing that the relevant curvature scale
$L$ here is that of the four-dimensional hypersurface. Thus, for
instance, for a collapsing spherical domain wall, the condition can
be violated well before the spatial radius $R$ becomes of order $l$,
since the time components of the curvature are relevant \cite{w89thick,ghg90}.

For a bubble wall, the potential energy density difference $\Delta V$ between minima
gives the pressure difference between phases, which makes the bubble grow.
Thus, in the equation of motion (EOM) for the wall surface,
$ \Delta V $ plays the role of a force that accelerates the wall.
On the other hand, this quantity is not as relevant for the wall profile. 
Indeed, the standard thin-wall approximation to calculate $ \phi(n) $ includes
approximating the effective potential $V$ by a degenerate potential $V_{0}$
\cite{c77}, which is equivalent to matching the profile
of the bubble wall with that of a domain wall.

It is usual to assume for simplicity that bubbles keep their initial
spherical shape as they expand. For a bubble of radius $ R $, the requirement 
that the wall width is much smaller than the curvature radius 
implies the condition $l\ll R$. 
This condition will eventually be satisfied during the phase transition
as $R$ grows from microscopic scale values $R\sim l\sim v^{-1}$ to cosmological
scale values $ R\sim H^{-1} $. 
We remark, however, that this is a necessary but not sufficient condition
as sometimes assumed in the literature. 
Like in the domain wall case, the four-dimensional curvature is relevant. 
Indeed,
the thin-wall approximation requires the potential difference $\Delta V$
to be small \cite{c77}. 
This condition does not involve the bubble radius at all and 
arises because the acceleration of the wall implies a curvature of the world volume.

The general form of the EOM has been extensively discussed in the literature
for domain walls and other topological defects, 
even beyond the thin-wall approximation 
(see, e.g., \cite{mt88,g88,w89thick,w89collapse,ghg90,gg90,g91,sm93,l93,cg95,al94,a95,a95b,a98}).
In contrast, for a bubble wall, the EOM has been scarcely discussed
beyond the planar, cylindrical, or spherical cases.
Small perturbations from these symmetric solutions have been considered, e.g., 
in Refs.~\cite{afw90,gv91} for bubbles in a vacuum phase transition 
(i.e., in the absence of fluid) 
and Refs.~\cite{l92,hkllm93,mm14,mm14b} for bubbles in a high-temperature 
phase transition 
(in the latter case, to study the linear stability of the hydrodynamic solutions).
To our knowledge, the effective EOM for a thick bubble wall has not been discussed.

The calculation of the scalar field profile of the bubble wall is also hardly addressed.
For phenomena involving the interactions of the scalar field with particles of the plasma%
\footnote{It is worth mentioning that, in this context, the concept of a thin or thick wall is related to the comparison with the mean free paths of the relevant particles (see, e.g., \cite{dlhll92}) and should not be confused with the thin-wall approximation discussed here.}  
(such as the friction of the wall with the plasma or electroweak baryogenesis),
a $\tanh$ ansatz is often used.
Although this may generally be 
a good approximation, one would expect some sensitivity to the function $\phi(n)$. 
Even if one is only interested in the wall motion, 
the problem does not fully separate into equations for $X^{\mu}(\xi^{a})$
and $\phi(n)$ since the surface tension
$\sigma$ (which depends on the profile) is a parameter in the wall EOM.
Hence, improving the description of the wall evolution requires improving the computation
of the profile%
\footnote{Some methods have been introduced 
	\cite{mr00,b16,e18} for calculating the
	spherical bounce beyond the thin-wall approximation, 
	since the latter is not reliable for computing the nucleation rate
	\cite{s88,sh91}. 	
	For these instantons
	(either the 4D bounce \cite{c77} or the 3D thermal instanton \cite{l81,l83}), 
	the calculation of the profile is essentially the same as for the corresponding nucleated bubble, 
	for which these techniques could in principle be adapted. We shall not discuss these methods here.
}. 

According to the above discussion, there are at least two situations where the thin-wall
approximation for a bubble wall breaks down. 
One occurs when the difference $\Delta V$ is large and is the case in a strongly first-order phase transition. 
The other possibility is that the bubble wall develops strong spatial curvatures due to instabilities. 
Both scenarios are of relevance
for the generation of gravitational waves (see, e.g., \cite{lm16b,mr17,eln19,elnv19,mm14,mm21a}). 

We aim to obtain an analytical description of the bubble wall dynamics, valid for arbitrary deformations from the spherical shape and thick walls. 
In particular, we analyze the assumptions usually made when using the thin-wall approximation, discuss the conditions for their validity, and derive both a general EOM for the wall and an equation for the profile beyond this approximation. 
In this paper, we focus on the case of a vacuum phase transition and corrections to first order in the wall width. 
In two companion papers \cite{mm2,mm3}, 
we discuss the higher-order corrections
and the case of a thermal phase transition. Considering a vacuum phase
transition certainly simplifies the problem. However, this can be
a good approximation for the case of a highly supercooled phase transition,
where the presence of the fluid has little effect on the wall. In
passing, we describe the oscillations of the field that originate
behind the phase-transition front and decay toward the bubble center.

The paper is organized as follows. After quickly reviewing the main
features of domain walls in Sec.~\ref{preliminares}, we apply the
thin-wall approximations to the field equation in Sec.~\ref{sec:Thin-wall} and obtain the equations
for the kink profile and the wall surface.
We discuss the requirements for the different approximations and
the construction of a degenerate potential $V_{0}$ approximating
a given effective potential $V$.
In Sec.~\ref{Monge}, we consider the Monge  parametrization of the hypersurface, 
which is suitable to describe local deformations from a given wall shape. 
In Sec.~\ref{sec:masalla}, we obtain
corrections to the wall EOM and the profile equation beyond the thin wall
approximation. In Sec.~\ref{sec_uso}, we test
our approximations with specific examples. Finally, we conclude in
Sec.~\ref{Conclu}. 
We discuss alternative derivations and previous results in App.~\ref{derivagral}, and
the field oscillations inside the bubble in App.~\ref{interior}.

\section{Planar and curved walls}

\label{preliminares}

In this section we review the kink profile and introduce a suitable
coordinate system associated to a curved wall surface. Either for
a domain wall or for a bubble wall, we shall assume that the field
satisfies the equation of motion
\begin{equation}
\nabla_{\mu}\nabla^{\mu}\phi+\frac{dV}{d\phi}=0,\label{eccampo0}
\end{equation}
where $\nabla_{\mu}$ is the covariant derivative. A damping term
can be added to this equation to take into account the friction force
with a surrounding fluid (see, e.g., \cite{ikkl94,h95}), and in the complete analysis
the fluid equations must be considered as well.
We shall discuss this extension in a separate paper \cite{mm3}. 
We look for solutions where $ \phi $ interpolates between the two potential minima $ \phi_\pm $.
For a bubble in a phase transition, we have a potential difference 
\begin{equation}
\Delta V\equiv V_{+}-V_{-},\label{defDV}
\end{equation}
where $V_{\pm}=V(\phi_{\pm})$ and $V_{-}<V_{+}$, and we shall assume that $\phi_{+}<\phi_{-}$. 
For a domain wall, we have $\Delta V=0$, in which case we shall denote the potential $V_{0}$ and its minima
$a_{\pm}$.

Let us consider the degenerate case.
The main features of the kink profile can be obtained by assuming
a static configuration in flat space and with the field varying in
a given direction, say, along the $z$ axis. Thus, Eq.~(\ref{eccampo0})
becomes $\phi''(z)=V_{0}'(\phi)$ (throughout this paper, a prime will
denote a derivative with respect to the explicit variable of a function).
Multiplying by $\phi'(z)$ and integrating with respect to $z$, with
the assumption that outside the wall the field is homogeneous, $\phi'(z)=0$,
we obtain the first-order equation $\frac{1}{2}\phi^{\prime2}=V_{0}(\phi)-V_{0}(a_{+})$.
Without loss of generality, we shall assume in this section that $V_{0}(a_{+})=0$.
The last equation can be readily integrated, and we obtain 
\begin{equation}
z-z_*=\pm\int_{\phi_*}^{\phi_0} d\phi/\sqrt{2V_{0}(\phi)},\label{solpl}
\end{equation}
which must be inverted to obtain the solution $\phi_{0}(z)$. Either sign can be used,
which expresses the reflection symmetry. Furthermore, the integration
introduced a constant which is due to the translation symmetry. 

The most familiar example  is that of the potential
$ V_0=\frac{\lambda}{4}(\phi^2-v^2)^2 $, which gives a $ \tanh $ profile
\cite{vs00}. Replacing the minima $ \pm v $ with $ a_\pm $, this quartic polynomial becomes
\begin{equation}
V_{0}(\phi)=\frac{\lambda}{4}\left(\phi-a_{+}\right)^{2}\left(\phi-a_{-}\right)^{2},\label{Vdeg}
\end{equation}
for which Eq.~(\ref{solpl}) gives
\begin{equation}
\phi_{0}(z)=\frac{a_{+}+a_{-}}{2}-\frac{a_{-}-a_{+}}{2}\tanh\left[\sqrt{\lambda/8}\left(a_{-}-a_{+}\right)z\right]\label{perfil-pl}
\end{equation}
(we assume $a_{+}<a_{-}$ and we have chosen the sign $\phi_{0}^{\prime}<0$
and the profile centered at $z=0$).
The solution $\phi(\mathbf{x})=\phi_{0}(z)$ gives the configuration of a planar interface separating domains where the field takes the values $a_{\pm}$. 
Although these values are reached asymptotically at $z=\pm\infty$, respectively, from Eq.~(\ref{perfil-pl}) we see that the field only varies in a range of $z$ of width $ l $, where
\begin{equation}
l\sim\left[\sqrt{\lambda/8}\left(a_{-}-a_{+}\right)\right]^{-1}.
\end{equation}
The separation between the minima is naturally given by the scale
of the theory, $a_{-}-a_{+}\sim v$, so we have roughly $l\sim v^{-1}$,
which, as already mentioned, is much smaller than the cosmological
scale $H^{-1}\sim M_{P}/v^{2}$. 

The energy density of the field is given by 
\begin{equation}
\rho=\frac{1}{2}\left(\partial_{t}\phi\right)^{2}+\frac{1}{2}\left(\nabla\phi\right)^{2}+V_{0},
\end{equation}
which in the static case reduces to $\rho=\frac{1}{2}\phi_{0}^{\prime2}+V_{0}$,
and with the aid of the field equation yields simply $\rho=\phi_{0}^{\prime2}$.
Hence, $\rho$ is concentrated in the range where the field varies,
i.e., inside the wall. The surface energy density is thus given by
\begin{equation}
\sigma=\int_{-\infty}^{+\infty}\phi_{0}^{\prime2}(z)dz=\int_{a_{+}}^{a_{-}}\sqrt{2V_{0}(\phi_{0})}d\phi_{0}.
\end{equation}
For the quartic potential we
have 
\begin{equation}
\sigma=\frac{1}{3}\sqrt{\lambda/8}\left(a_{-}-a_{+}\right)^{3}.\label{sigmapl}
\end{equation}
In this case the top of the potential barrier between the minima is
at the point $\phi=a_{\max}$, $V_{0}=V_{\max}$, with
\[
a_{\max}=\frac{1}{2}\left(a_{+}+a_{-}\right),\quad V_{\mathrm{\mathbf{\textrm{max}}}}=\frac{\lambda}{64}\left(a_{-}-a_{+}\right)^{4},
\]
so we have the relation
\begin{equation}
l\sim\sigma/V_{\textrm{max}}.\label{ancho}
\end{equation}

\begin{figure}[tb]
\centering%
\begin{minipage}[c]{0.4\columnwidth}%
\includegraphics[height=2cm]{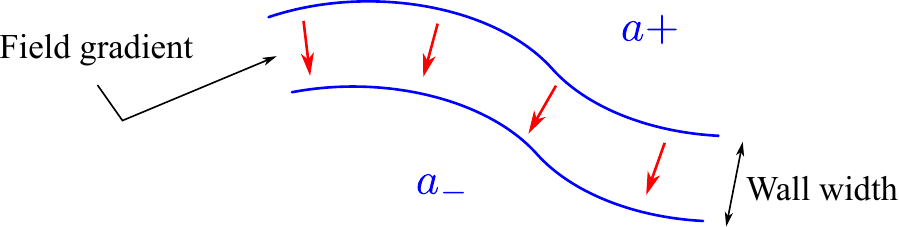}
\end{minipage}
\hfill%
\begin{minipage}[c]{0.4\columnwidth}%
\includegraphics[width=1\textwidth]{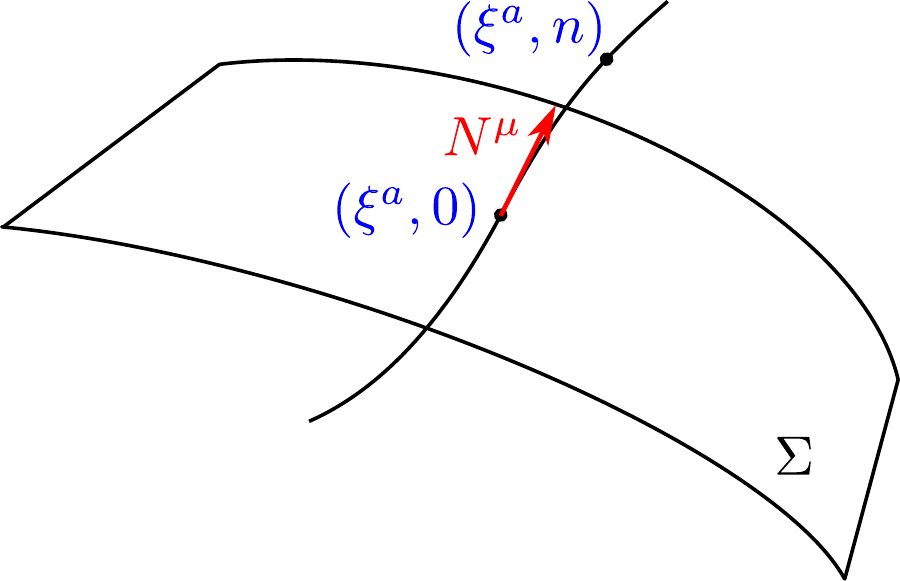}
\end{minipage}
\caption{Left: Schematically, hypersurfaces of constant $\phi$ and its gradient.
Right: The normal Gaussian coordinates on a hypersurface $\Sigma$.\label{figequipot}}
\end{figure}

Since the static and planar conditions greatly simplified the problem,
no approximation was necessary to obtain the solution (\ref{solpl}).
The general planar-symmetry solution is obtained by applying a boost, $\phi=\phi_{0}(n)$,
with $n=\gamma_{w}(z-v_{w}t)$ and $\gamma_{w}=1/\sqrt{1-v_{w}^{2}}$. 
More generally, we have a field configuration $\phi(x^{\mu})$ and
the wall is the zone where the scalar field varies from $a_{+}$ to
$a_{-}$ or, equivalently, the region where $\partial_{\mu}\phi\neq0$,
as sketched in Fig.~\ref{figequipot} (left). Since the values $a_{\pm}$
are reached asymptotically, we may consider two surfaces where $\phi$
takes constant values $b_{\pm}$ close to $a_{\pm}$ to define the
wall region. Although we will not need to use these values explicitly,
it is useful to have a mental image in which the wall is formally
bounded by these surfaces. Without any other reference scale, the
wall can be considered thin when the separation between these surfaces
is small compared to the local radius of curvature (which is approximately
the same for both surfaces if they are close enough). As already mentioned,
in most cases we will have $l\ll L$. 

In order to achieve an analytical
treatment, the strategy is to separate the physics of the two scales
$l$ and $L$, that is, to derive an equation for the field profile
$\phi(n)$ through the wall, and another one for the wall surface,
whose world-volume hypersurface $\Sigma$ is explicitly parametrized
as $x^{\mu}=X^{\mu}(\xi^{a})$. To that aim we need, in the first
place, to define the wall surface (say, as the locus of points where $ \phi $ takes some definite value between $ a_+ $ and $ a_- $). 
In the second place,
we must define a coordinate system $\bar{x}^{\mu}=(\xi^{a},n)$, where
the $\xi^{a}$ correspond to displacements over the surface and $n$
corresponds to displacements in the perpendicular direction (see Fig.~\ref{figequipot},
right).

It is convenient to use normal Gaussian coordinates, which are constructed
as follows \cite{synge}. From the parametrization $x^{\mu}=X^{\mu}(\xi^{a})$,
the variables $\xi^{a}$ determine a point on $\Sigma$. To these
points we assign the value $n=0$. To reach points outside $\Sigma$,
we draw geodesics normal to the hypersurface, i.e., tangent to the
normal vector $N_{\mu}$ which satisfies $N_{\mu}\partial_{a}X^{\mu}=0$
and $N_{\mu}N^{\mu}=-1$. The coordinate $n$ gives the proper distance
on the geodesic. Near the surface, the change of coordinates is given
by\footnote{The next term in the expansion is $-\frac{1}{2}\left(\Gamma_{\nu\rho}^{\mu}N^{\nu}N^{\rho}\right)_{\xi^{a}}n^{2}$
\cite{synge}, where the Christoffel symbols are given by 
$\Gamma_{\alpha\beta}^{\mu} =
\frac{1}{2}g^{\mu\nu}(\frac{\partial g_{\nu\alpha}}{\partial x^{\beta}}+\frac{\partial g_{\nu\beta}}{\partial x^{\alpha}}-\frac{\partial g_{\alpha\beta}}{\partial x^{\nu}})$.}
\begin{equation}
x^{\mu}=X^{\mu}(\xi^{a})+N^{\mu}(\xi^{a})n+\cdot\cdot\cdot.\label{gaussianas}
\end{equation}
In these coordinates the metric takes the form
\begin{equation}
ds^{2}=\overline{g}_{\mu\nu}d\overline{x}^{\mu}d\overline{x}^{\nu}=-dn^{2}+\overline{g}_{ab}d\xi^{a}d\xi^{b}.
\label{dsgauss}
\end{equation}
At $n=0$ we have $\overline{g}_{ab}=\gamma_{ab}$, where
\begin{equation}
\gamma_{ab}=g_{\mu\nu}\frac{\partial X^{\mu}}{\partial\xi^{a}}\frac{\partial X^{\nu}}{\partial\xi^{b}}
\end{equation}
is the induced metric. A similar coordinate system is often
used, defined by the simpler relation $x^{\mu}=X^{\mu}(\xi^{a})+N^{\mu}(\xi^{a})n$.
Both systems coincide close to the surface. The convenience of the
Gaussian coordinates lies in the fact that, along the geodesics, we have
$\partial_{n}g_{nn}=0$ and $\partial_{n}g_{an}=0$, which yields the
metric form (\ref{dsgauss}) not only at $ n=0 $. 

The extrinsic curvature of $\Sigma$ is defined through the change in the normal vector 
under an infinitesimal displacement on the hypersurface \cite{misner}. 
Since the vector is normalized, its variation is orthogonal to it, i.e., tangent to $\Sigma$. 
Thus, one can directly define the components $ K_{ab} $ in a coordinate basis tangent to the hypersurface.
For a covariant definition of the tensor $ K_{\mu\nu} $, 
it is more convenient to use the unit vector field $\tilde{N}^{\mu}$ tangent
to the geodesics (which gives the normal $N^{\mu}$ at $n=0$). Thus, the tensor \cite{wald}
\begin{equation}
	K_{\mu\nu}=-\nabla_{\mu}\tilde{N}_{\nu}
\end{equation}
gives the extrinsic curvature of all the surfaces $ \Sigma_n $ of constant $n$, not only of $ \Sigma\equiv\Sigma_0 $. 
In normal Gaussian coordinates, $ \tilde{N}_\mu $ takes the form
$\bar{N}_{\mu}=(0,0,0,-1)$, and Eq.~(\ref{dsgauss}) implies
$\bar{g}^{nn}=-1,$ $\bar{g}^{na}=0$, and
$\bar{\Gamma}_{nn}^{\mu}=\bar{\Gamma}_{\mu n}^{n}=0$, $\bar{\Gamma}_{ab}^{n}=\frac{1}{2}\partial_{n}\bar{g}_{ab}$.
Thus, the non-vanishing components of the extrinsic curvature tensor
are $\bar{K}_{ab}=-\bar{N}_{a;b}=-\bar{\Gamma}_{ab}^{n}$. The mean curvature
is often defined as the trace 
\begin{equation}
K\equiv\gamma^{ab}K_{ab}={K^{a}}_{a}.\label{defK}
\end{equation}
According to the above relations, we can write $ K $ in the various forms
\begin{equation}
K=-\bar{g}^{ab}\bar{N}_{a;b}=-\bar{g}^{ab}\bar{\Gamma}_{ab}^{n}=-\frac{1}{2}\bar{g}^{ab}\partial_{n}\bar{g}_{ab}=-g^{\mu\nu}\tilde{N}_{\mu;\nu}.\label{expresionesK}
\end{equation}
The last equality gives the covariant expression for the mean curvature
of $ \Sigma_n $. 

Since we are only interested in the extrinsic curvature of $\Sigma$, 
we do not actually need to consider the vector $ \tilde{N}^\mu $.
For any vector field $N^{\mu}$ which, at $ n=0 $,  gives the normal to $ \Sigma $,
we note that
\begin{equation}
\bar{N}_{a;b}|_{n=0}=\left.\frac{\partial x^{\mu}}{\partial\bar{x}^{a}}\frac{\partial x^{\nu}}{\partial\bar{x}^{b}}N_{\mu;\nu}\right|_{n=0}=\left.\frac{\partial X^{\mu}}{\partial\xi^{a}}\frac{\partial X^{\nu}}{\partial\xi^{b}}N_{\mu;\nu}\right|_{n=0}.
\end{equation}
Using also
\begin{equation}
g^{\mu\nu}|_{n=0}=\left.\bar{g}^{\rho\sigma}\frac{\partial x^{\mu}}{\partial\bar{x}^{\rho}}\frac{\partial x^{\nu}}{\partial\bar{x}^{\sigma}}\right|_{n=0}=\gamma^{ab}\frac{\partial X^{\mu}}{\partial\xi^{a}}\frac{\partial X^{\nu}}{\partial\xi^{b}}-N^{\mu}N^{\nu}|_{n=0},\label{gmnunugab}
\end{equation}
we obtain
\begin{equation}
K|_{n=0}=-\left(g^{\mu\nu}+N^{\mu}N^{\nu}\right)N_{\mu;\nu}|_{n=0}.\label{Ken0}
\end{equation}
Given an implicit equation for the hypersurface, $F(x^{\mu})=0$,
the vector $\partial_{\mu}F$ is perpendicular to $\Sigma$. Therefore,
we obtain the normal vector $N_{\mu}$ by dividing by $\sqrt{|F_{,\mu}F^{,\mu}|}$.
In addition, we must choose a sign for it. Thus, we define
\begin{equation}
N_{\mu} = -\partial_{\mu}F/s\,,\quad\textrm{where}\quad s=\sqrt{|F_{,\mu}F^{,\mu}|}.\label{Nmu}
\end{equation}
This vector satisfies the normalization condition $N_{\mu}N^{\mu}=-1$
everywhere, which implies $N^{\mu}N_{\mu;\nu}=0$. As a
consequence, we have
\begin{equation}
K|_{n=0}=-g^{\mu\nu}N_{\mu;\nu}|_{n=0}.\label{KN}
\end{equation}

The coordinate transformation (\ref{gaussianas}) can be inverted
order by order. We only need to consider the expression for $\overline{x}^{3}=n$
as a function of $x^{\mu}$ to lowest order in $n$. Multiplying (\ref{gaussianas})
by $N_{\mu}$, we obtain
\begin{equation}
n=-N_{\mu}(\xi^{a})\left(x^{\mu}-X^{\mu}(\xi^{a})\right)+\mathcal{O}(n^{2}).\label{gaussianasinvert}
\end{equation}
Using Eq.~(\ref{Nmu}), we obtain
\begin{equation}
n=s^{-1}\partial_{\mu}F(X^{\mu}(\xi^{a}))\left(x^{\mu}-X^{\mu}(\xi^{a})\right)+\mathcal{O}(n^{2})=F(x^{\mu})/s(x^{\mu})+\mathcal{O}(n^{2}).\label{ngral}
\end{equation}
In the last step we have used the expansion of $F(x^{\mu})$ around
$x^\mu=X^{\mu}(\xi^{a})$ and the fact that $F$ vanishes at $ \Sigma $.
In App.~\ref{derivagral} we calculate the term of order $ n^2 $.

\section{The thin-wall approximation}

\label{sec:Thin-wall}

At any time during a phase transition, we have domains of the stable
phase growing at the expense of the metastable phase, where the interfaces
form a system of moving walls. 
Assuming that the scalar field $\phi$ in the domains 
takes one of the fixed values $\phi_{\pm}$,
we can describe the evolution of the whole system by the motion of
these walls.  
Although this assumption does not refer explicitly to the wall width, 
we could regard it as the first of a series of assumptions usually made in the framework of the thin-wall approximation. 
To be rigorous, inside a recently nucleated bubble the field is not,
in general, at the minimum of its potential. As we will see in particular
examples, $\phi$ can undergo temporal and spatial oscillations before
reaching the value $\phi_{-}$. Therefore, if the initial value of
$\phi$ is not very close to $\phi_{-}$, it is to be expected that the
standard description will fail at the early evolution of the bubble. 
In particular,
the definition of the bubble wall may not be entirely clear. Nevertheless,
it is generally possible to identify the wall as a region where the
variation from $\phi_{\text{+}}$ to a value close to $\phi_{-}$
occurs. In our analysis, we shall make the standard assumption 
$\phi\equiv\phi_{-}$ inside the wall and check that it
leads to a reasonable approximation for the wall motion and profile
even when the assumption does not hold. 

\subsection{The wall equation of motion}

As usual in the treatment of a cosmological phase transition, we shall
neglect the backreaction of the bubbles on the background metric $g_{\mu\nu}$.
The treatment can be generalized to account for that effect, for example
by applying the Israel junction conditions on the wall (see, e.g., \cite{bkt87}). 
We could also assume the Friedmann-Robertson-Walker metric, or even the 
Minkowski metric as an approximation. However, it
is convenient to maintain a more general approach since, even for
flat space, it is of interest to consider a general coordinate system,
such as spherical coordinates. Therefore, we start from the field
equation (\ref{eccampo0}),
\begin{equation}
g^{\mu\nu}\left(\partial_{\mu}\partial_{\nu}\phi-\Gamma_{\mu\nu}^{\rho}\partial_{\rho}\phi\right)+V'(\phi)=0,\label{eccampo}
\end{equation}
and we shall derive an effective equation for the wall. 
Usually, the first assumption for a thin wall is that the field configuration is only a function of the normal
variable $n$. 
More precisely, the assumption is
that $\phi$ varies much more with $n$ than with $\xi^{a}$, so we can neglect $\partial_{a}\phi$ in front of $\partial_{n}\phi$.
Before using this approximation we shall discuss its validity and implications.

\begin{figure}[tb]
\centering
\includegraphics[width=0.4\textwidth]{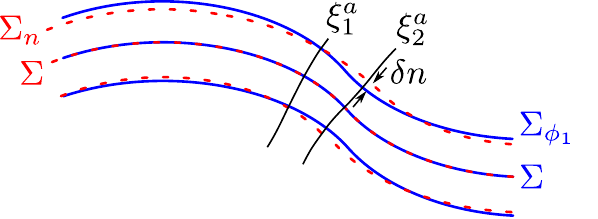} 
\hfill
\includegraphics[width=0.4\textwidth]{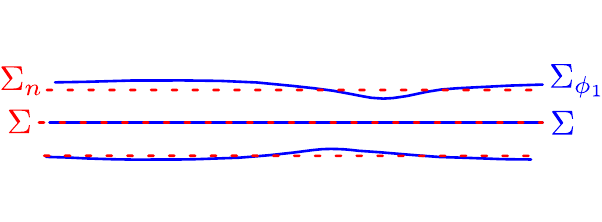}
\caption{Left: Surfaces of $n=$ constant (dotted lines) and surfaces of $\phi=$
constant (solid lines). Right: A visualization of the surfaces in the coordinates $(\xi^{a},n)$.%
\label{fignfi}}
\end{figure}

Intuitively, the condition $\partial_{a}\phi\ll\partial_{n}\phi$
will hold whenever the wall width $l$ is much smaller than the size
scale $L$ of surface undulations. Indeed, if we assume that $\partial_{a}\phi\sim\phi/L$
and $\partial_{n}\phi\sim\phi/l$, we have $\partial_{a}\phi/\partial_{n}\phi\sim l/L$.
However, the derivative $\partial_{a}\phi$ can be  smaller
than this rough estimate. 
This idea is implicit in the sketch on the left of Fig.~\ref{figequipot},
where the wall region is delimited by surfaces of $\phi=$ constant
and the gradient of $ \phi $ is perpendicular to these surfaces. 
To be more precise, let us define our reference surface $\Sigma$ by the condition $ \phi(x^\mu) =\phi_\Sigma$, where
$\phi_{\Sigma}$ is a conveniently chosen value between $\phi_{-}$ and $\phi_{+}$. 
By definition, $\Sigma$ is a surface of constant $\phi$, and in normal Gaussian coordinates we have $\partial_{a}\phi=0$ at $n=0$. 
The variable $n$ gives the physical distance in the direction perpendicular to $\Sigma$,
so a surface $\Sigma_{n}$ defined by $n(x^\mu)=$ constant lies at a fixed distance from $ \Sigma $.  
In contrast, a surface $\Sigma_{\phi}$ defined by $\phi(x^\mu)=$ constant will not be, in general, at a fixed distance from $ \Sigma $
(see the left panel of Fig.~\ref{fignfi}).
The change to normal Gaussian coordinates $(\xi^{a},n)$ can be seen as a change
of variables where we eliminate the undulations on the surfaces $ \Sigma_n $,
obtaining planes of $n=$ constant, while the undulations of $ \Sigma_\phi $ will only be partially eliminated (the idea is depicted in Fig.~\ref{fignfi})%
\footnote{We could build a coordinate system similar to the normal Gaussian
system but using $n=\phi$. By construction, in these coordinates
we will have $\partial_{a}\phi=0$ for all $n$. 
The problem with this setup is that now the variable $ n $ does not represent a physical distance and quickly becomes ill-defined as 
$\phi$ approaches a constant value near the borders of the wall.}.

If we move a distance $n$ from $\Sigma$, then at some given point
$\xi_{1}^{a}$ on $\Sigma_{n}$ the field will have a value $\phi_{1}$,
while at another point $\xi_{2}^{a}$ on this surface the field will
have a different value $\phi_{2}$. Let us call $\delta n(\xi_{2}^{a})$
the distance from the point $(\xi_{2}^{a},n)\in\Sigma_{n}$ to the
point where the normal geodesic through $\xi_{2}^{a}$ meets the surface
$\Sigma_{\phi_{1}}$ (see Fig.~\ref{fignfi}). The assumption that
$\phi$ only depends of $n$ implies that $ \phi_2=\phi_1 $, i.e., $\delta n(\xi_{2}^{a})=0$ and
$\Sigma_{\phi_{1}}=\Sigma_{n}$.
The variation of $\delta n(\xi^{a})$ gives the departures of the surface $\Sigma_{\phi_1}$
from $ \Sigma_n $. 
If the value $ \phi_1 $ is such that $ \Sigma_{\phi_1} $ formally defines the border of the wall,
the distance $ n $ in this construction is of the order of the wall width. 
Let us denote in this case $ n=l $. The separation of $\Sigma_{\phi_{1}}$ from $\Sigma_{l}$, namely, $\delta n\equiv\delta l$, is directly related to variations in the wall width. 
We shall thus refer to this
approximation as the incompressibility assumption.
\begin{description}
\item [{Assumption 1:}] The wall width is incompressible.
\end{description}

It is worth noting that the assumption $\delta n(\xi^{a})=0$ is not
necessarily an approximation, as there are exact solutions which satisfy
this condition. For example, this is the case for the planar domain
wall considered above and for a bubble with $O\left(3,1\right)$ symmetry
that we shall discuss later. For these cases we can assume, by symmetry,
that $\phi$ depends on a single variable. By doing that, we are restricting
ourselves to a particular type of solution, rather than making an
approximation. We cannot expect that there will always be solutions
which satisfy this condition. In general, with this assumption we
are disregarding the dynamics of the degrees of freedom associated
to the variation of the wall width, which makes sense if we assume
a large separation of the scales $l$ and $L$. 

To quantify these ideas, if the maximum variation $\delta\phi$ on the surface $\Sigma_{l}$
occurs in a distance scale $L$, then we roughly have $\partial_{a}\phi\sim\delta\phi/L$.
On the other hand, we have $\delta\phi\sim\partial_{n}\phi\delta l$,
which yields
\begin{equation}
\frac{\partial_{a}\phi}{\partial_{n}\phi}\sim\frac{\delta l}{L}\sim\frac{l}{L}\frac{\delta l}{l}.\label{errorordensuperior}
\end{equation}
Therefore, the precision of the approximation $\partial_{a}\phi\ll\partial_{n}\phi$
depends on the independent variable $\delta l$, and can be much
better than just $l/L$. In the worst case scenario we will
have $\delta l/l\sim1$ and $\partial_{a}\phi/\partial_{n}\phi\sim l/L$.
Let us assume that this is the case.
We shall use the incompressibility assumption by neglecting
terms containing derivatives $\partial_{a}\phi$ in the field equation
and in the energy-momentum tensor. 
In normal Gaussian coordinates, the leading-order term in these equations
contains two derivatives with respect to $n$, so it is of order $l^{-2}$.
In contrast, the terms containing derivatives $\partial_{a}\phi$
do not contain derivatives $\partial_{n}\phi$ (this is a consequence
of the fact that $\bar{g}^{na}=0$), and are of order $L^{-2}$. Therefore,
the terms that we will drop are at least of order $(l/L)^{2}$
with respect to the leading order.

In normal Gaussian coordinates the field equation (\ref{eccampo})
takes the form
\begin{equation}
-\partial_{n}^{2}\phi+\bar{g}^{ab}\left(\partial_{a}\partial_{b}\phi-\bar{\Gamma}_{ab}^{n}\partial_{n}\phi-\bar{\Gamma}_{ab}^{c}\partial_{c}\phi\right)+V'(\phi)=0.\label{ecfigauss}
\end{equation}
Using the assumption of incompressibility, this equation is reduced
to
\begin{equation}
\phi''(n)-K(n) \phi'(n)=V'(\phi),\label{ecfi0}
\end{equation}
where $K$ is the mean curvature of the surface $\Sigma_{n}$, given
by Eq.~(\ref{expresionesK}). Notice that, for consistency with the
assumption $\phi=\phi(n)$, the quantity $K$ in Eq.~(\ref{ecfi0})
must also depend on $n$ alone. Therefore, the mean curvature is a
constant on each hypersurface $\Sigma_{n}$. In particular, for the
wall hypersurface $\Sigma_{0}\equiv\Sigma$, the condition $K(0)=$
constant establishes a dynamic equation for the wall through Eq.~(\ref{Ken0}).
We only need to determine the value of this constant. If we apply
the usual trick of multiplying by $\partial_{n}\phi$ and then integrating
and using the boundary conditions $\phi(\pm\infty)=\phi_{\pm}$, we
obtain
\begin{equation}
-\int_{-\infty}^{+\infty}K(n)\phi^{\prime2}(n)dn=\Delta V,\label{1ra_EOM}
\end{equation}
where $\Delta V$ is defined in Eq.~(\ref{defDV}). The integral
on the left-hand side provides the average value of the curvature
of the surfaces $\Sigma_{n}$, weighted by the function $\phi'^{2}(n)$,
whose effective support essentially defines the wall region. The normalization
constant for such a weight is 
\begin{equation}
\sigma\equiv\int_{-\infty}^{+\infty}\phi^{\prime2}(n)dn.\label{def_sigma}
\end{equation}

The next usual assumption is that the variation of $K(n)$
through the thin wall can be neglected. Thus, we obtain the equation
for $\Sigma$,
\begin{equation}
-K(0)\sigma=\Delta V.\label{2daEOM}
\end{equation}
The last approximation is justified if the field
$\phi$ varies in a length scale much shorter than that associated
with $K$. More generally, we can state this assumption as follows.
\begin{description}
\item [{Assumption 2:}] Inside the wall, the field varies much
more than any other quantity.
\end{description}
The validity of this assumption depends of course on the quantity
we are considering. For the case of $K$ it fully makes sense from
the expression $K=-\frac{1}{2}\bar{g}^{ab}\partial_{n}\bar{g}_{ab}$,
since the metric in the normal Gaussian system depends on the physical
curvature of space and on the function $X^{\mu}(\xi^{a})$ used to
parametrize $\Sigma$, and in general will not vary as rapidly as
$\phi$. To quantitatively appreciate the accuracy of this approximation,
let us consider the expansion
\begin{equation}
K=K(0)+\partial_{n}K(0)n+\frac{1}{2}\partial_{n}^{2}K(0)n^{2}+\cdots.\label{expanK}
\end{equation}
If $K$ varies appreciably on a length scale $L$, when integrating
$K\phi'^{2}$ in Eq.~(\ref{1ra_EOM}), this expansion gives  a correction
of order $l/L$ to Eq.~(\ref{2daEOM}),
\begin{equation}
\int_{-\infty}^{+\infty}K(n)\phi^{\prime2}(n)dn=K(0)\sigma+\partial_{n}K(0)\int_{-\infty}^{+\infty}\phi^{\prime2}(n)ndn+\cdots\label{expangprom-1}
\end{equation}
We have regarded $\Sigma$ as a constant-field surface $\phi(x^{\mu})=\phi_{\Sigma}$,
but so far, we have not specified the value $\phi_{\Sigma}$. At this
point, it is convenient to associate $\Sigma$ with the ``center
of mass'' of the wall by demanding that
\begin{equation}
\int_{-\infty}^{+\infty}\phi^{\prime2}(n)ndn=0.\label{cero_n}
\end{equation}
Thus, the last term in Eq.~(\ref{expangprom-1}) vanishes, and the
error of this approximation is of order $(l/L)^{2}$.

Using the definition of $K$, Eq.~(\ref{defK}), we can write Eq.~(\ref{2daEOM})
in the more familiar form $-\sigma K_{ab}\gamma^{ab}=\Delta V$ (see
for example \cite{bkt87,gv91}), where the parameter $\sigma$ is the surface tension%
\footnote{The meaning of the wall equation $-\sigma K_{ab}\gamma^{ab}=\Delta V$
	becomes more intuitive if we consider the static case in which there
	are no time components and we have a two-dimensional surface in three
	dimensions. Since we have $p=-V$, we can write $p_{-}-p_{+}=-\sigma\mathrm{tr}({K^{a}}_{b})$,
	which yields the equation of static equilibrium where the surface
	tension balances the pressure difference. This equation is referred
	to as \emph{Laplace equation} in the physics of surfaces (see, e.g.,
	\cite{safran,millerneogi}).}
and to calculate its value, we need to find the profile $\phi(n)$.
For the particular case $V_{+}=V_{-}$
we obtain the well-known equation for a domain wall, $\gamma^{ab}K_{ab}=0$.
Using Eq.~(\ref{KN}), we can write the wall equation in terms
of the normal vector as 
\begin{equation}
g^{\mu\nu}N_{\mu;\nu}=\Delta V/\sigma.\label{EOMNmu}
\end{equation}
 Replacing Eq.~(\ref{Nmu}), we obtain the equation for the function
$F$ defining the surface,
\begin{equation}
\left.-g^{\mu\nu}\nabla_{\nu}\left(\partial_{\mu}F/s\right)\right|_{n=0}=\Delta V/\sigma.\label{EOM0}
\end{equation}
We could have derived this equation directly from the field equation
(\ref{eccampo}) in an arbitrary coordinate system by writting $ \phi=\phi(n(x^\mu)) $. 
The only information
we need about the normal Gaussian system is the function $n(x^{\mu})$.
However, there are subtleties with such a derivation, which
we discuss in App.~\ref{derivagral}. 

\subsection{The kink profile}

To obtain $\phi(n)$, we need to solve Eq.~(\ref{ecfi0}), where
the quantity $K$ is given by Eq.~(\ref{2daEOM}) and thus depends
on $\sigma$. In turn, $\sigma$ depends on $\phi(n)$. Therefore,
Eqs.~(\ref{ecfi0}), (\ref{def_sigma}), and (\ref{2daEOM}) give
a system of coupled equations for $\phi(n)$ and $\sigma$. In Sec.~\ref{sec:masalla}
we discuss an iterative method to solve this system. The usual approach,
though, is to just neglect the second term in Eq.~(\ref{ecfi0}).
The justification is that this term has only one derivative $\partial_{n}$,
and therefore is of order $l^{-1}$, while the first one has two derivatives
and is of order $l^{-2}$. If $L$ is the scale related to the 
curvature, we have $K\sim L^{-1}$, and the second term in (\ref{ecfi0})
is of order $l/L$ with respect to the first one. Then the condition
for this approximation to hold is $l/L\ll1$, which gives us the final
assumption regarding the thin-wall approximation.
\begin{description}
\item [{Assumption 3:}] The radius of curvature of the wall is much larger than its width.
\end{description}
It is worth remarking on the differences with the previous assumptions.
Here we assume that $K$ is small, while assumption 2 was that $K$
(as well as other quantities) changes little through the wall. Furthermore,
as long as $K$ is integrated with the weight $\phi^{\prime2}$, we
have seen that the terms neglected by assumption 2 are of order $(l/L)^{2}$.
In contrast, the third assumption discards terms of order $l/L$.
On the other hand, the first assumption ignores the variation of $\phi$
(and, as a consequence, of $K$) along the wall (i.e., with $ \xi^a $), and we argued that
the neglected terms are {at least} of order $(l/L)^{2}$. In
this sense, assumption 3 is the strongest one.

Since the scale of the wall width is set by the energy scale of the
theory, $l\sim v^{-1}$, the condition $l/L\ll1$ actually imposes
a constraint on the curvature scale $L$. In particular, according
to Eq.~(\ref{2daEOM}) we have $K=-\Delta V/\sigma$, and, hence,
\begin{equation}
L\sim\sigma/\Delta V.\label{curvatura}
\end{equation}
Therefore, we would naturally have $L\sim v^{-1}\sim l$ unless the
difference $V_{+}-V_{-}$ is much smaller than the scale $v^{4}$.
We can express this condition in terms of the potential shape if we
use the results for the planar domain wall obtained in Sec.~\ref{preliminares}.
Using Eq.~(\ref{ancho}) to introduce the height of the barrier in
Eq.~(\ref{curvatura}), we obtain
\begin{equation}
l/L\sim\Delta V/V_{\max}.\label{DVVmax}
\end{equation}
Consequently, the thin-wall approximation requires a nearly-degenerate
potential, $\Delta V\ll V_{\max}$. As already discussed in Sec.~\ref{sec:Intro},
the reason is that the potential difference causes an acceleration
and, hence, a curvature on the hypersurface in 3+1 dimensions. 

As an example of this fact, we may consider a planar wall in Minkowski
space. If we parametrize the
hypersurface as $z=z_{w}(t)$ (see Sec.~\ref{Monge}), we have $N_{\mu}=\gamma_{w}(\dot{z}_{w},0,0,-1)$,
with $\gamma_{w}=1/\sqrt{1-\dot{z}_{w}^{2}}$ and $\dot{z}_{w}=dz_{w}/dt$.
In this case, Eq.~(\ref{KN}) gives $K=-\partial_{t}(\gamma_{w}\dot{z}_{w})$,
showing that the curvature is directly related to the acceleration.
Thus, Eq.~(\ref{2daEOM}) links the acceleration to $\Delta V$,
as expected. It is also important to remark that the length $l$ to
consider here is a proper distance in a system moving with the wall,
while in a system at rest with the bubble center, this length can
be much smaller due to Lorentz contraction.

The requirement of a small $\Delta V$ becomes apparent as soon as
we use this third assumption to drop the second term in Eq.~(\ref{ecfi0}),
which gives 
\begin{equation}
\phi''(n)=V'(\phi).\label{ecfi0sup3}
\end{equation}
Multiplying by $\phi'$ and integrating from $n=-\infty$ to $n=+\infty$,
we obtain $V_{-}-V_{+}=0$. Thus, for consistency, this approximation
requires a degenerate potential. It is convenient, then, to approximate
the potential $V(\phi)$ by some degenerate potential $V_{0}(\phi)$
\cite{c77}. We are thus left with the equation for a static planar
wall described in Sec.~\ref{preliminares}. We shall denote $\phi_{0}(n)$
the field profile obtained with this approximation,
\begin{equation}
n=-\int_{\phi_{*}}^{\phi_{0}}\frac{d\phi}{\sqrt{2\left[V_{0}(\phi)-V_{0}(a_{\text{+}})\right]}}+n_{*},\label{formulaperfil}
\end{equation}
The minima $a_{\pm}$ of the approximating potential $V_{0}$ may differ from those
of $V$. Below we discuss the construction of the potential $V_{0}$
from a given $V$. For consistency, the solution $\phi_{0}$ satisfies
the boundary conditions $\phi_{0}(\pm\infty)=a_{\pm}$.
For convenience, we express the result (\ref{formulaperfil}) with
two arbitrary constants instead of just one. If we set $n_{*}=0$,
the value $\phi_{*}$ should be determined with the condition (\ref{cero_n}).
However, it will be more practical to use a conveniently chosen value
$\phi_{*}$ and then determine $n_{*}$ such that (\ref{cero_n})
is satisfied,
\begin{equation}
n_{*}=\sigma_{0}^{-1}\int_{a_{+}}^{a_{-}}d\phi_{0}\left[\sqrt{2\left[V_{0}(\phi_{0})-V_{0}(a_{+})\right]}\int_{\phi_{*}}^{\phi_{0}}\frac{d\phi}{\sqrt{2\left[V_{0}(\phi)-V_{0}(a_{+})\right]}}\right].\label{n0}
\end{equation}
With this approximation, the surface tension is given by
\begin{equation}
\sigma_{0}=\int_{-\infty}^{+\infty}\phi_{0}^{\prime2}(n)dn=\int_{a_{+}}^{a_{-}}\sqrt{2\left[V_{0}(\phi_{0})-V_{0}(a_{+})\right]}d\phi_{0}.\label{formulasigma}
\end{equation}
Although we calculated the profile with the approximation $\Delta V=0$,
we can now use the result (\ref{formulasigma}) in Eqs.~(\ref{2daEOM})
and (\ref{EOM0}) to obtain the curvature and the wall EOM to the
lowest non-trivial order in $\Delta V$. 

\subsection{Construction of the degenerate potential}

In Ref.~\cite{c77}, Coleman introduced the idea of considering
an approximately degenerate potential along with the thin-wall approximation.
He started from a potential $V_{0}$ with minima at $\phi=a_{\pm}$
such that $V_{0}(a_{+})=V_{0}(a_{-})$, and then added a term $V_{1}$
that produces a small energy difference between the minima,
\begin{equation}
V(\phi)=V_{0}(\phi)+V_{1}(\phi).\label{potpert}
\end{equation}
The  minima $\phi_{\pm}$ of $V$ will not coincide in general
with $a_{\pm}$, but if $V_{1}$ is small, the offset will remain
small too. As an example, Coleman considered a potential $V_{0}$
similar to that of Eq.~(\ref{Vdeg}) and a linear term $V_{1}$.
Thus, the approximation of dropping the second term in Eq.~(\ref{ecfi0})
is consistent with taking the zeroth-order potential $V_{0}$, which
leads to the results (\ref{formulaperfil})-(\ref{formulasigma}). 

It seems evident that, to use these results for any given potential
$V$, one should decompose $V$ as in Eq.~(\ref{potpert}). Thus, the potential
$V_{0}$ can be constructed by going the reverse way to what Coleman
did, that is, adding a linear term to $V$ to obtain a degenerate
$V_{0}$. Interestingly, less motivated approaches are often used
in the literature, directly using the potential $V$ instead of $V_{0}$.
Notice that we cannot just replace $\sqrt{V_{0}(\phi_{0})-V_{0}(a_{+})}$
with $\sqrt{V(\phi)-V(\phi_{+})}$ in Eqs.~(\ref{formulaperfil})-(\ref{formulasigma}),
since the latter becomes imaginary as $\phi$ approaches $\phi_{-}$.
To avoid this problem, some options include integrating only the
range of $\phi$ where $V>V_{+}$, using $V_{-}$ instead of $V_{+}$
inside the root, or just using $\sqrt{|V-V_{+}|}$ (see, e.g., \cite{bkt87,b18,sh91}).
These approximations may be reasonable as long as we have $V_{-}\simeq V_{+}$,
but they cannot be used if one wants to continue the calculation beyond
the zeroth order in $V_{1}$.

Here we shall approximate $V$ by a  potential $V_{0}=V-V_{1}$. The correction
$V_{1}$ is not necessarily linear. Its form can be chosen at will
and its parameters adjusted to obtain a degenerate $V_{0}$. The condition
$V_{0}'(a_{\pm})=0$ gives the equation for the minima,
\begin{equation}
V_{1}'(a_{\pm})=V'(a_{\pm}),\label{condminV0}
\end{equation}
while the condition $V_{0}(a_{+})=V_{0}(a_{-})$ gives the equation
\begin{equation}
V_{1}(a_{+})-V_{1}(a_{-})=V(a_{+})-V(a_{-}).\label{condV0deg}
\end{equation}
For instance, for a linear term $V_{1}=c\phi$, Eq.~(\ref{condminV0})
gives the minima $a_{\pm}$ as functions of $c$, and then Eq.~(\ref{condV0deg})
gives an equation for $c$. It should be noted that Coleman used the
specific value $c=\Delta V/(a_{+}-a_{-})$ to construct a potential
$V$ from a given $V_{0}$. This produces a difference of order $\Delta V$,
but not exactly $\Delta V$, between the minima of the full potential
$V$. For the inverse process of constructing $V_{0}$ from $V$,
our method gives an exactly degenerate $V_{0}$, which is suitable
to use in Eqs.~(\ref{formulaperfil})-(\ref{formulasigma}).

\section{Monge gauge}

\label{Monge}

We have rewritten the equation for the mean curvature, Eq.~(\ref{2daEOM}),
as an equation for the normal vector $N^{\mu}$, Eq.~(\ref{EOMNmu}),
and as an equation for the implicit function $F(x^{\mu})$ that defines
the surface, Eq.~(\ref{EOM0}). The explicit equation for the wall
position $x^{\mu}=X^{\mu}(\xi^{a})$ is often considered in the domain
wall case%
\footnote{The equations of motion for a domain wall are given by
	$\frac{1}{\sqrt{\gamma}}\partial_{a}\left(\sqrt{\gamma}\gamma^{ab}\partial_{b}x^{\mu}\right)+
	\Gamma_{\nu\sigma}^{\mu}\gamma^{ab}\partial_{a}x^{\nu}\partial_{b}x^{\sigma}=0$
	\cite{vs00}. However, only the motion transverse to the surface is observable. 
	Projecting this vector equation on $N_{\mu}$ and using properties such as $N_{\mu}\partial_{a}x^{\mu}=0$, we obtain $-\gamma^{ab}\partial_{b}x^{\rho}\partial_{a}x^{\mu}\nabla_{\rho}N_{\mu}=0$.
	Taking into account the relations (\ref{expresionesK})-(\ref{KN}), this last equation is equivalent to our equation (\ref{EOMNmu}) for $ \Delta V=0 $.\label{foot_wallEOM}}. 
We do not actually need to consider the four variables $x^{\mu}$,
since we can exploit the invariance under reparametrizations of the
surface to eliminate some degrees of freedom. We shall adopt a \emph{Monge
representation} in which the implicit definition of the surface is
of the form 
\begin{equation}
	x^{3}-x^3_w(x^{0},x^{1},x^{2})=0.
\end{equation}	
Such a parametrization is always possible locally, and is widely used
in the form $z=z_w(t,x,y)$ when dealing with small deformations of
planar surfaces (see for example \cite{npw89}). An advantage of
the Monge gauge is that we can deal with (small or large) deformations
of a given surface without specifying it explicitly, by suitably choosing
the coordinate system. 
For example, for dealing with deformations of a spherically-symmetric surface it is clearly convenient to
describe the surface by $r=r_w(t,\theta,\phi)$,
although the description $z=z_w(t,x,y)$ is also valid.

Thus, the explicit
representation is given by $X^{\mu}(\xi^{a})=(\xi^{0},\xi^{1},\xi^{2},x_w^3(\xi^{a}))$,
the implicit function is  given by
$F(x^\mu)=x^3-x^3_{w}(x^a)$, with $ a=0,1,2 $,
and we have 
\begin{equation}
	N_{\mu} = - s^{-1} \partial_{\mu}(x^3-x^3_w) 
		    = s^{-1} (\partial_{0}x^3_{w},\partial_{1}x^3_{w},\partial_{2}x^3_{w},-1),
\end{equation} 
with
$s^{2}=-g^{33}+2g^{3a}\partial_{a}x^3_{w}-g^{ab}\partial_{a}x^3_{w}\partial_{b}x^3_{w}$.
Inserting these expressions in Eq.~($\ref{EOM0}$), we obtain 
\begin{multline}
\left[
  g^{ab}\left(\partial_{b}\frac{\partial_{a}x^3_{w}}{s} - \Gamma_{ab}^{c}\frac{\partial_{c}x^3_{w}}{s}\right)
  - g^{3a} \left( \partial_{a}\frac{1}{s} - \frac{2\Gamma_{3a}^{3}}{s} - \partial_{3}\frac{1}{s}\partial_{a}x^3_{w} + 2\Gamma_{3a}^{b}\frac{\partial_{b}x^3_{w}}{s} \right)
\right.  
\\
\left.
 +  \frac{g^{ab}\Gamma_{ab}^{3}}{s}
 -g^{33}\Gamma_{33}^{a}\frac{\partial_{a}x^3_{w}}{s} - g^{33} \left( \partial_{3}\frac{1}{s} - \Gamma_{33}^{3} \frac{1}{s} \right) 
 \right]_{x^3=x^3_{w}}
= \frac{\Delta V}{\sigma}.
\label{EOMtot}
\end{multline}
This equation for the wall position $ x^3_w $ is the complete equation of motion for a thin wall in
curved space-time without making any assumptions about the wall shape.
We note that covariance is lost since the Monge parameterization is
tied to a particular coordinate system, although the system is not
explicitly specified in Eq.~(\ref{EOMtot}). 

To avoid cumbersome expressions, from now on we will restrict ourselves
to coordinate systems where the metric is diagonal in $x^3$ 
\begin{equation}
	g^{3a}=0, \label{metricadiagbloques}
\end{equation}
Thus, we have 
\begin{equation}
	\left[g^{ab}\partial_{b}\frac{\partial_{a}x_{w}^{3}}{s}-g^{\mu\nu}\Gamma_{\mu\nu}^{c}\frac{\partial_{c}x_{w}^{3}}{s}-g^{33}\partial_{3}\frac{1}{s}+g^{\mu\nu}\Gamma_{\mu\nu}^{3}\frac{1}{s}\right]_{x^{3}=x_{w}^{3}}=\frac{\Delta V}{\sigma}.
\end{equation}
with 
$ s^{2}=-g^{33}-g^{ab}\partial_{a}x_{w}^{3}\partial_{b}x_{w}^{3} $.
The quantity $ s $ depends on $ x^3 $  only through the metric components.
We have 
$ \partial_{3}s^{-1}= 
\frac{1}{2}s^{-3}(\partial_{3}g^{33}+\partial_{3}g^{ab}\partial_{a}x_{w}^{3}\partial_{b}x_{w}^{3}) $. 
The derivatives of $ g^{\mu\nu} $ can be expressed in terms of those of $ g_{\mu\nu} $ by differentiating the relations
$ g^{33}=1/g_{33} $, $g^{ab}g_{bc}=\delta_{c}^{a}$ (since we are considering a block-diagonal metric). 
We have
$ \partial_{3}g^{33}=-(g^{33})^2\partial_{3}g_{33} $,
$ \partial_{3}g^{ab}=-g^{ad}\partial_{3}g_{dc}g^{cb} $. 
Besides, for this metric the Christoffel symbols have simple expressions in terms of these derivatives,
$ \Gamma_{33}^{3}=\frac{1}{2}g^{33}\partial_{3}g_{33} $, $ \Gamma_{ab}^{3}=-\frac{1}{2}g^{33}\partial_{3}g_{ab} $.
We can then write 
\begin{equation}
	\partial_{3}s^{-1}=
	{s^{-3}}\left(-g^{33}\Gamma_{33}^{3} + g_{33} g^{ad} \Gamma_{dc}^{3}g^{cb}\partial_{a}x_{w}^{3}\partial_{b}x_{w}^{3}\right),
\end{equation}
and we obtain
\begin{equation} \label{EOMdiagonal}
g^{ab}\partial_{b}\frac{\partial_{a}x_{w}^{3}}{s}-g^{\mu\nu}\Gamma_{\mu\nu}^{c}\frac{\partial_{c}x_{w}^{3}}{s}-\frac{g^{ad}\Gamma_{dc}^{3}g^{cb}\partial_{a}x_{w}^{3}\partial_{b}x_{w}^{3}}{s^{3}}+\frac{(g^{33})^{2}\Gamma_{33}^{3}}{s^{3}}+\frac{g^{\mu\nu}\Gamma_{\mu\nu}^{3}}{s}=\frac{\Delta V}{\sigma},
\end{equation}
where $s$, the metric, and the Christoffel symbols are evaluated at $x^3=x^3_{w}$. 

\subsection{The Friedmann-Robertson-Walker metric}

Let us consider the specific case of the Robertson-Walker metric
\begin{equation} \label{frw}
	ds^{2}=dt^{2}-a^{2}(t)\tilde{g}_{ij}dx^{i}dx^{j},
\end{equation}
where $ \tilde{g}_{ij} $ is the purely spatial metric 
(note that we use letters $ a,b,c,\ldots $ for the indices 0,1,2 and 
letters $ i,j,k,\ldots $ for the indices 1,2,3).
In quasi Cartesian and spherical coordinates we have
\begin{equation} \label{frwspace}
	\tilde{g}_{ij}dx^{i}dx^{j}=
	\left(\delta_{ij}+k\frac{x_{i}x_{j}}{1-k\mathbf{x}^{2}}\right)dx^{i}dx^{j}=
	\frac{dr^{2}}{1-kr^{2}}+r^{2}(d\theta^{2}+\sin^{2}\theta d\varphi^{2}),
\end{equation}
where $ k=0,\pm 1 $ is the sign of the spatial curvature.
The non-zero components of the affine connection are given by \cite{weinbergCosmology}
\begin{equation}
	\Gamma_{ij}^{0}=
	a\dot{a}\tilde{g}_{ij},\quad\Gamma_{0j}^{i}=(\dot{a}/a)\delta_{ij},\quad\Gamma_{jl}^{i}=\tilde{\Gamma}_{jl}^{i},
\end{equation}
where $ \tilde{\Gamma}_{jl}^{i} $ are the affine connections obtained from the metric $ \tilde{g}_{ij} $.

\subsubsection{Deformations of a planar wall}

In  Cartesian coordinates, the metric (\ref{frw})-(\ref{frwspace}) is not of the form (\ref{metricadiagbloques}), except when $ k=0 $, 
so let us consider first the spatially flat case. We have $ \tilde{g}_{ij}=\delta_{ij} $ 
and $ \tilde{\Gamma}_{jl}^{i}=0 $, and we write the Monge parametrization as  $z=z_{w}(t,x,y)$. 
Thus, Eq.~(\ref{EOMdiagonal}) becomes
\begin{equation} \label{EOMplFRW}
	\partial_{t}\frac{\partial_{t}z_{w}}{s}-a^{-2}\partial_{x}\frac{\partial_{x}z_{w}}{s}-a^{-2}\partial_{y}\frac{\partial_{y}z_{w}}{s}+3H\frac{\partial_{t}z_{w}}{s}=\frac{\Delta V}{\sigma},
\end{equation}
with $ H=\dot{a}/a $ and 
$ s^{2}=a^{-2}-(\partial_{t}z_{w})^{2}+a^{-2}(\partial_{x}z_{w})^{2}+a^{-2}(\partial_{y}z_{w})^{2} $.
In particular, in flat space, the wall EOM becomes 
\begin{equation} 
\partial_{t}\frac{\partial_{t}z_{w}}{s}-\partial_{x}\frac{\partial_{x}z_{w}}{s}-\partial_{y}\frac{\partial_{y}z_{w}}{s}=\frac{\Delta V}{\sigma},
\label{EOMrect}
\end{equation}
with 
$ s^{2}=1-(\partial_{t}z_{w})^{2}+(\partial_{x}z_{w})^{2}+(\partial_{y}z_{w})^{2} $.
The equations of motion (\ref{EOMplFRW}) or (\ref{EOMrect}) describe an arbitrary wall surface at least locally, i.e., as long as the parametrization $z=z_{w}(t,x,y)$ is valid. 
However, this parametrization will be most useful in the case of deformations from a planar wall.

Let us consider some particular cases.
For  $\Delta V=0$ we obtain a form of the domain-wall equation which can be found, e.g., in \cite{vilenkin85}.
The case $ \Delta V=0 $ also applies to phases in equilibrium.
For a static interface we obtain the equation
\begin{equation}
	\partial_{x}\frac{\partial_{x}z_{w}}{s}+\partial_{y}\frac{\partial_{y}z_{w}}{s}=0,
\end{equation} 
which is often used in the physics of surfaces \cite{safran}. 
On the other hand,
in the limit of small deformations and small velocities $(\partial_{a}z_{w})^{2}\ll1$,
Eq.~(\ref{EOMrect}) becomes
\begin{equation}
\sigma\left(\partial_{t}^{2}-\partial_{x}^{2}-\partial_{y}^{2}\right)z_{w}=\Delta V.
\end{equation}
This equation was discussed in Ref.~\cite{gv91}. 
For a planar wall without deformations but arbitrary velocity, 
Eq.~(\ref{EOMrect}) gives
\begin{equation}
\sigma\frac{d}{dt}\left(\gamma_{w}\dot{z}_{w}\right)=\Delta V,\label{EOMpl}
\end{equation}
with $s^{2}=1-\dot{z}_{w}^{2}\equiv\gamma_{w}^{-2}$. 
This equation is similar to the equation for a relativistic particle in one
dimension, where $\Delta V$ plays the role of the force and $\sigma$
that of the mass. 
In this case, the more general equation (\ref{EOMplFRW}) gives
\begin{equation} 
	\partial_{t}(s^{-1}\dot{z}_{w})+3H(s^{-1}\dot{z}_{w})={\Delta V}/{\sigma},
\end{equation}
with $ s^{2}=a^{-2}-(\dot{z}_{w})^{2} $. 
In terms of the physical distance from the bubble center to the wall, $ z_{w}'=\int_0^t adz_{w} $,
we have $ \dot{z}_{w}=a^{-1}\dot{z}_{w}' $, $ s^{2}=a^{-2}(1-\dot{z}_{w}^{\prime2})\equiv a^{-2}\gamma_{w}^{\prime-2} $,
and we obtain for $ z_w' $ the same equation (\ref{EOMpl}) but
with the typical damping term proportional to $ 3H $.

\subsubsection{Deformations of a spherical wall.}

For a phase transition, a bubble wall which is deformed
from a spherical surface is more realistic. 
In this case, the Monge
parametrization $r=r_{w}(t,\theta,\varphi)$ in spherical coordinates
is the most reasonable choice. 
Let us consider the general Robertson-Walker metric (\ref{frw})-(\ref{frwspace}).
We have
\begin{equation}
	\tilde{g}_{rr}=({1-kr^{2}})^{-1},
	\;\tilde{g}_{\theta\theta}=r^{2},
	\;\tilde{g}_{\varphi\varphi}=r^{2}\sin^{2}\theta;
	\;\tilde{g}^{rr}=1-kr^{2},\;\tilde{g}^{\theta\theta}=r^{-2},
	\;\tilde{g}^{\varphi\varphi}=r^{-2}\sin^{-2}\theta,
\end{equation}
and the non-zero components of the spatial connections are given by 
\begin{equation}
	\begin{split}
		& \tilde{\Gamma}_{rr}^{r}	=\frac{kr}{1-kr^{2}},
		\quad \tilde{\Gamma}_{\theta\theta}^{r}=-r(1-kr^{2}),
		\quad\tilde{\Gamma}_{\varphi\varphi}^{r}=-r(1-kr^{2})\sin^{2}\theta,
		\\
		& \tilde{\Gamma}_{r\theta}^{\theta}	=r^{-1},
		\quad \tilde{\Gamma}_{\varphi\varphi}^{\theta}=-\sin\theta\cos\theta,
		\quad\tilde{\Gamma}_{r\varphi}^{\varphi}=r^{-1},
		\quad \tilde{\Gamma}_{\theta\varphi}^{\varphi}=\cot\theta.
	\end{split}
\end{equation}
The wall equation (\ref{EOMdiagonal}) becomes
\begin{multline}
	\partial_{t}\frac{\partial_{t}r_{w}}{s} - a^{-2}r_{w}^{-2}\partial_{\theta}\frac{\partial_{\theta}r_{w}}{s} - a^{-2}r_{w}^{-2}\sin^{-2}\theta\partial_{\varphi}\frac{\partial_{\varphi}r_{w}}{s} + 3H\frac{\partial_{t}r_{w}}{s} 
	- \frac{\cot\theta}{a^{2}r_{w}^{2}}\frac{\partial_{\theta}r_{w}}{s} 
	\\
	+ (1-kr^{2})\frac{(\partial_{\theta}r_{w})^{2} + \sin^{-2}\theta(\partial_{\varphi}r_{w})^{2}}{a^{4}r_{w}^{3}s^{3}} + \frac{kr_{w}(1-kr_{w}^{2})}{a^{4}s^{3}}+\frac{2-3kr_{w}^{2}}{a^{2}r_{w}s}=\frac{\Delta V}{\sigma},
\end{multline}
with 
$ s^{2} = a^{-2}(1-kr_{w}^{2}) 
- (\partial_{t}r_{w})^{2} + a^{-2}r_{w}^{-2}(\partial_{\theta}r_{w})^{2} + a^{-2}r^{-2}\sin^{-2}\theta(\partial_{\varphi}r_{w})^{2} $.
In particular, in flat space we have
\begin{multline}
  \partial_{t}\frac{\partial_{t}r_{w}}{s}-r_{w}^{-2}\partial_{\theta}\frac{\partial_{\theta}r_{w}}{s}-r_{w}^{-2}\sin^{-2}\theta\partial_{\varphi}\frac{\partial_{\varphi}r_{w}}{s} 
  \\
-\frac{\cot\theta}{r_{w}^{2}}\frac{\partial_{\theta}r_{w}}{s}
+\frac{\left(\partial_{\theta}r_{w}\right)^{2}+\sin^{-2}\theta\left(\partial_{\varphi}r_{w}\right)^{2}}{s^{3}r_{w}^{3}}
+\frac{2}{r_{w}s}  
=\frac{\Delta V}{\sigma},
\label{EOMesf}
\end{multline}
with 
$
s^{2} = 1 - (\partial_{t}r_{w})^{2} + r_{w}^{-2}(\partial_{\theta}r_{w})^{2} + r_{w}^{-2}\sin^{-2}\theta(\partial_{\varphi}r_{w})^{2}
$.
In the limit $r_{w}\to\infty$, the operators $r_{w}^{-1}\partial_{\theta}$
and $r_{w}^{-1}\partial_{\varphi}$ behave as the derivatives $\partial_{x}$
and $\partial_{y}$ on a planar background surface. Therefore, the
first three terms in Eq.~(\ref{EOMesf}) in this limit give the result
(\ref{EOMrect}), while the other terms have an extra factor of $r_{w}^{-1}$
and vanish. The last term on the left-hand side is due exclusively
to the surface tension of the spherical background. In the absence
of deformations we have $s^{2}=1-\dot{r}_{w}^{2}=\gamma_{w}^{-2}$,
and the equation of motion becomes
\begin{equation}
\frac{d}{dt}\left(\sigma\gamma_{w}\dot{r}_{w}\right)+\frac{2\sigma\gamma_{w}}{r_{w}}=\Delta V.\label{EOMbbesf}
\end{equation}
The first term is like in the planar case, while the second term gives
a force that opposes the acceleration and increases for a smaller
bubble radius. 

\section{Beyond the thin-wall approximation}

\label{sec:masalla}

We shall now discuss two different ways of improving the thin-wall
approximation.

\subsection{Iterative method}

The assumption of incompressibility already simplifies the field equation
(\ref{ecfigauss}) considerably and leads to an ordinary differential
equation in the variable $n$, Eq.~(\ref{ecfi0}). This equation
involves the additional function $K(n)$. Nevertheless, this function
is partially specified by the construction of the normal Gaussian
system (as can be seen from the expression $K=\frac{1}{2}\bar{g}^{ab}\partial_{n}\bar{g}_{ab}$),
and we can determine it from the condition (\ref{1ra_EOM}). Notice
that the latter is derived from Eq.~(\ref{ecfi0}), so the solutions
for $\phi$ and $K$ come from the same equation. 
To better understand how this
is possible, let us simplify the problem by neglecting the variation
of $K$ through the wall (second assumption). Then, we need to solve
the equation 
\begin{equation}
	\phi''(n) + K\phi'(n) = V'(\phi) \label{ecfi_Kcte}
\end{equation}
with the boundary conditions $\phi(\pm\infty)=\phi_{\pm}$.
The latter are only attainable for a specific value of $K$. 
Indeed, Eq.~(\ref{ecfi_Kcte}) is very similar to the exact equation with spherical symmetry discussed below.
It is well known that this kind of problem has a mechanical analogy in which $ \phi $ is the position of a particle in a potential $ -V(\phi) $ and $ n $ is the time. 
In the case of Eq.~(\ref{ecfi_Kcte}), we also have a friction force with a constant friction coefficient $ K $. 
According to our boundary conditions, at time $ -\infty $ this particle is released at rest on top of the hill at the position $ \phi_- $ and must come at rest at time $ +\infty $ on top of the lower hill at $ \phi_+ $
(see the central panel of Fig.~\ref{figpot}).
However, if the friction coefficient $ K $ is too low, the particle will overshoot and pass $ \phi_+ $ at a finite time, while if $ K $ is too high, it will undershoot and never reach $ \phi_+ $.

To implement this overshoot-undershoot method, 
the first thing to notice is that the initial condition $ \phi(-\infty) =\phi_-$ cannot be achieved in practice.
Due to the time translation symmetry of this equivalent problem, we can move this condition to time $ 0 $. Then the kink will occur at infinity, which, in the equivalent problem, means that the particle will not roll down the hill if we release it exactly at the top of the hill.
Therefore, we release the particle at $ \phi =(1-\varepsilon)\phi_- $. 
The smaller the value of $ \varepsilon $, the farther away the kink will occur, but its shape will not change appreciably when $ \varepsilon $ is small enough.
Thus, we solve the equation for some value of $K$ (e.g., the value $K=-\Delta V/\sigma_{0}$ obtained with the
thin-wall approximation) and check whether $\phi$ approaches the value $\phi_{+}$ for large $n$. 
If the solution undershoots, we decrease the effective friction $ K $. If it overshoots, we increase $ K $. Then we solve the equation again and repeat the procedure until $ \phi $ stays close to $ \phi_- $ for a long time.
After a few steps the value of $K$ changes very little with each iteration.

The solution $ \phi_r(n) $ obtained with this recursive method gives an approximation for the wall profile which interpolates between the values $ \phi_\pm $ with arbitrary precision
(in contrast, the solution $ \phi_0 $ given by the thin-wall approximation interpolates between the values $ a_\pm $). 
The kink will be at an arbitrary position $ n_r $ that depends on the chosen value of $ \varepsilon $ and can be evaluated as $ n_r =\int_{-\infty}^{+\infty}\phi_r^{\prime 2}(n)ndn$.
On the other hand, we can obtain the correct wall position from the wall EOM.
Since we have used both assumptions 1 and 2, Eq.~(\ref{2daEOM}) holds, and 
the wall EOM has the same form as in the thin-wall approximation, namely
$\gamma^{ab}K_{ab}=-\Delta V/\sigma_r$, 
where the surface tension is now given by  $\sigma_r=\int_{-\infty}^{+\infty}\phi_r^{\prime2}(n)dn$.
Moreover, for consistency, this integral should coincide with the value $\sigma_r=-\Delta V/K_r$, where $ K_r $ is the value of $ K $ obtained in the last iteration.
In Sec.~\ref{sec_uso} we compare this approximation with the exact result for a few cases.

This method can be extended to the case where the field
does not have a fixed value inside the bubble. To that aim, let us consider
an interior point $n_{i}$ (possibly time-dependent) where $\partial_{n}\phi=0$.
The value of $n_{i}$ is irrelevant due to translation symmetry, while
the value of $\phi_{i}=\phi(n_{i})$ can be suitably defined to represent
the boundary between the wall and the bubble interior. Thus, we can replace Eq.~(\ref{1ra_EOM})
with
\begin{equation}
-\int_{n_{i}}^{\infty}K(n)\phi^{\prime2}(n)dn=V_{+}-V(\phi_{i}).\label{1ra_EOM_fi}
\end{equation}
We could further use the approximation of neglecting the variation
of $K$ to obtain an equation analogous to Eq.~(\ref{2daEOM}). If we do not make this simplification (either here or in
the case $\phi_{i}=\phi_{-}$), we must solve a set of integro-differential
equations. Yet, the fact that we are dealing with a single variable
is in principle a great simplification over the problem of solving
$\phi(x^{\mu})$ using lattice calculations. 
We shall explore these possibilities elsewhere.
On the other hand, our
procedure involves the shooting method, which becomes unusable when
there is more than one scalar field \cite{j99}. 

\subsection{The next order in the wall width}

In the method described above, we only make the first and second assumptions
and solve exactly the resulting equation for the profile $\phi(n)$
and the value of $K$, which is assumed to be constant. 
Alternatively, we can consider perturbative
corrections to the three assumptions. This approach has been used
for the case of domain walls, where the equation $\gamma^{ab}K_{ab}=0$
is regarded as the zero-thickness approximation for the wall EOM.
This equation is a generalization of the EOM for strings obtained from the Nambu-Goto action \cite{vs00}
(see also footnote \ref{foot_wallEOM}).
For
domain walls and strings, the leading-order finite-width corrections
have been discussed in a number of papers 
\cite{g88,mt88,w89thick,gg90,ghg90,g91,sm93,l93,cg95,al94,a95,a95b,a98}. 
There is some discrepancy in the results, possibly due
to ambiguities in the definition of the problem \cite{hk95}. 
Hence
the importance of a clear definition of the wall surface and the clear
statement of the assumptions made. In any case, it turns out that
the finite-width corrections are of order $(l/L)^{2}$. In the case
of bubble walls, we have seen that the non-vanishing potential difference
sets a lower bound on the magnitude of this expansion parameter, $ l/L\sim\Delta V/V_{\max} $. 
Moreover, we shall see that in this case we have corrections of order $ l/L $. 

To obtain an expansion for both the profile and surface equations
in powers of $l/L$, we begin by expressing the field as
\begin{equation}
\phi=\phi_{0}+\phi_{1}+\phi_{2}+\cdots,\label{expanfi}
\end{equation}
where $\phi_{0}$ is the zeroth-order solution given by Eq.~(\ref{formulaperfil}),
and each term is of order $l/L$ with respect to the previous one.
To introduce this expansion in the field equation, the potential $V$
needs to be expanded in powers of $\phi$. This would be enough for
the domain-wall case, but we will also write the potential in the
form (\ref{potpert}), $V=V_{0}+V_{1}$, where $V_{0}$ is degenerate
and $V_{1}$ causes the gap between $V_{-}$ and $V_{+}$. Here we
shall consider the leading-order corrections, for which we continue
using the assumption $\phi=\phi(n)$. We shall verify that this assumption
is consistent. 

Our starting point are thus Eqs.~(\ref{ecfi0}) and (\ref{1ra_EOM}).
Let us consider first the latter. To improve the slowly-varying approximation
for $K$, we use the expansion (\ref{expanK}), where we must note
that $K\sim L^{-1}$, $\partial_{n}K\sim L^{-2}$, and so on. Each
factor of $n$ in this expansion generates a factor of order $l$
upon integration in (\ref{1ra_EOM}). On the other hand, taking the
derivative of the field expansion (\ref{expanfi}) and squaring, we
obtain
\begin{equation}
\phi^{\prime2}(n)=\phi_{0}^{\prime 2}(n)+2\phi_{0}'(n) \phi_{1}'(n)+\mathcal{O}(L^{-2}l^{-2}).
\end{equation}
Here, each derivative generates a factor of order $l^{-1}$, and we
must also take into account that $\phi_{0}\sim v\sim l^{-1}$ and
$\phi_{1}\sim\phi_{0}l/L\sim L^{-1}$. Inserting these expansions
in (\ref{1ra_EOM}) and taking into account that $\phi_{0}$ fulfills
the condition (\ref{cero_n}), we obtain
\begin{equation}
K(0)\left(\sigma_{0}+\delta\sigma\right)+\mathcal{O}(L^{-3}l^{-1})=-\Delta V,\label{segundaap}
\end{equation}
with $\sigma_{0}$ given by Eq.~(\ref{formulasigma}), and
\begin{equation}
\delta\sigma=\int_{-\infty}^{+\infty} 2\phi_{0}'(n)\phi_{1}'(n)dn.\label{deltasigmadef}
\end{equation}
Notice that $\delta\sigma$ is of order $l^{-2}L^{-1}\sim(l/L)\sigma_{0}$,
while the omitted terms in Eq.~(\ref{segundaap}) are of order $L^{-3}l^{-1}\sim K\sigma_{0}(l/L)^{2}$.
We see that, to this order, the equation of motion has the same form
as at zeroth order,
\begin{equation}
-\gamma^{ab}K_{ab}=\Delta V/\sigma_{1},\label{EOM_NLO}
\end{equation}
where $\sigma_{1}=\sigma_{0}+\delta\sigma$. At higher orders, there
will be further corrections to the parameter $\sigma$, and we will
also have terms containing derivatives of $K$ from the expansion
(\ref{expanK}), which will modify the form of the wall equation.
We shall address these corrections in a subsequent paper \cite{mm2}. 

To find the correction to the kink profile, we need to expand Eq.~(\ref{ecfi0})
to order $l/L$. On the right-hand side, we have
\begin{equation}
	\begin{split}
		V'(\phi_{0}+\phi_{1})
		&= V'(\phi_{0})+V''(\phi_{0})\phi_{1}+\cdots
		\\
		&= V_{0}'(\phi_{0})+V_{1}'(\phi_{0})+V_{0}''(\phi_{0})\phi_{1}+V_{1}''(\phi_{0})\phi_{1}+\cdots.
	\end{split}
\end{equation}
Dimensionally, we have $V_{0}(\phi_{0})\sim l^{-4}$, $V_{0}'(\phi_{0})\sim l^{-3}$,
$V_{0}''(\phi_{0})\sim l^{-2}$, and we consider that $V_{1}$ is
of order $l/L$ higher than $V_{0}$, so we have $V_{1}(\phi_{0})\sim l^{-4}(l/L)$,
$V_{1}'(\phi_{0})\sim l^{-3}(l/L)$, and so on. To expand the left-hand
side of Eq.~(\ref{ecfi0}) to the same order, we use the expansions for $ \phi $ and $ K $, and we 
take into account the result (\ref{segundaap})
in the form
\begin{equation}
-K(0)\sigma_{0}=\left[V_{1}(a_{+})-V_{1}(a_{-})+\mathcal{O}(a_{\pm}-\phi_{\pm})^{2}\right]\left(1-\delta\sigma/\sigma_{0}\right),\label{Krecurs}
\end{equation}
where we used the condition (\ref{condV0deg}) and the expansion of
$V$ around its minima. Taking also into account that the lowest-order
solution $\phi_{0}$ satisfies Eq.~(\ref{ecfi0sup3}),
$\phi_{0}''(n)=V_{0}'(\phi_{0})$,
the next order in Eq.~(\ref{ecfi0}) yields
\begin{equation}
\phi_{1}''(n)+\frac{V_{1}(a_{+})-V_{1}(a_{-})}{\sigma_{0}}\phi_{0}'(n)
= V_{0}''(\phi_{0})\phi_{1}(n) + V_{1}'(\phi_{0}).\label{ecfiuno}
\end{equation}
This is a second-order differential equation for $\phi_{1}$, where
$\phi_{0}$ is given by Eq.~(\ref{formulaperfil}). Since its coefficients
are either constants or functions of $n$, it is consistent to assume
that $\phi_{1}$ is a function of $n$ alone. 

Multiplying Eq.~(\ref{ecfiuno}) by $\phi_{0}'$ and integrating, we obtain
\begin{equation}
\int_{\infty}^{n}\phi_{0}'(\tilde{n})\phi_{1}''(\tilde{n})d\tilde{n}
-\left[V_{1}(a_{+})-V_{1}(a_{-})\right]G(n)
= \int_{\infty}^{n}\frac{dV_{0}'(\phi_{0})}{dn}\phi_{1}(\tilde{n})d\tilde{n}
+ \int_{\infty}^{n}\frac{dV_{1}(\phi_{0})}{dn} d\tilde{n},
\label{intecfiuno}
\end{equation}
where we introduced the function
\begin{equation}
G(n) = \sigma_{0}^{-1}\int_{n}^{\infty} \phi_{0}^{\prime 2}(\tilde{n})d\tilde{n}
=\sigma_{0}^{-1}\int_{a_{\text{+}}}^{\phi_{0}}\sqrt{2\left(V_{0}(\phi)-V_{0}(a_{+})\right)}d\phi,
\end{equation}
which ranges between $0$ and $1$. Integrating by parts the first
and third terms in Eq.~(\ref{intecfiuno}) and using again the equality $\phi_{0}''(n)=V_{0}'(\phi_{0})$,
we obtain
\begin{equation}
\phi_{0}'(n)\phi_{1}'(n)-\phi_{0}''(n)\phi_{1}(n)=\left[V_{1}(a_{+})-V_{1}(a_{-})\right]G(\phi_{0})+V_{1}(\phi_{0})-V_{1}(a_{+}).\label{ecdif_fi1}
\end{equation}
This is now a first-order differential equation for $\phi_{1}$. Before
attempting to solve it, we note that if we integrate the left-hand
side, after an integration by parts we obtain the integral (\ref{deltasigmadef}). 
Therefore, we have
\begin{equation}
\delta\sigma=\int_{a_{+}}^{a_{-}}\frac{V_{1}(\phi_{0})-V_{1}(a_{+})-\left[V_{1}(a_{-})-V_{1}(a_{+})\right]G(\phi_{0})}{\sqrt{2\left(V_{0}(\phi_{0})-V_{0}(a_{+})\right)}}d\phi_{0},\label{deltasigma}
\end{equation}
where we used Eq.~(\ref{formulaperfil}) to change the integration variable
from $n$ to $\phi_{0}$. 

Returning to Eq.~(\ref{ecdif_fi1}), its general solution is given by 
\begin{equation}
\phi_{1}=\phi_{p}+\phi_{h},\label{fi1tot}
\end{equation}
where
\begin{equation}
\phi_{h}=C\phi_{0}'(n)=C\sqrt{2\left(V_{0}(\phi_{0})-V_{0}(a_{+})\right)},\label{filh}
\end{equation}
with $C$ an arbitrary constant, and
\begin{equation}
\begin{split}
	\phi_{p} 
	& = \phi_{0}'(n) \int_{n_{*}}^{n} \frac{\left[V_{1}(a_{-}) - V_{1}(a_{+})\right] G(\tilde{n})+V_{1}(\phi_{0})-V_{1}(a_{+})}{\phi_{0}^{\prime 2}(\tilde{n})} d\tilde{n}
	\\
	&= \sqrt{2\left(V_{0}(\phi_{0})-V_{0}(a_{+})\right)} \int_{\phi_{*}}^{\phi_{0}}\frac{V_{1}(\phi)-V_{1}(a_{+}) - \left[V_{1}(a_{-})-V_{1}(a_{+})\right] G(\phi)}{\left[2\left(V_{0}(\phi)-V_{0}(a_{+})\right)\right]^{3/2}} d\phi.
\end{split}
\label{filp}
\end{equation}
The relation between the constants $\phi_{*}$ and $n_{*}$ is given
by Eq.~(\ref{n0}), so that Eq.~(\ref{cero_n}) is satisfied at
zeroth order. The function $\phi_{h}$, which is the solution of the
homogeneous equation, satisfies boundary conditions $\phi_{h}(\pm\infty)=0$,
so the particular solution $\phi_{p}$ should fulfill the boundary
conditions for $\phi_{1}$ at $n=\pm\infty$, namely, $\phi_{1}=\phi_{\pm}-a_{\pm}+\mathcal{O}(l/L)$.
While the first factor in (\ref{filp}) vanishes at $\phi_{0}=a_{\pm}$,
the second factor diverges. Applying L'Hospital's rule twice, we obtain
\begin{equation}
\lim_{\phi_{0}\to a_{\pm}}\phi_{1}=-\frac{V_{1}'(a_{\pm})}{V_{0}''(a_{\pm})}.\label{limfi1}
\end{equation}
Recalling that a small $V_{1}$ implies a slight displacement of the
potential minima, we can use the expansion of $V$ around its minima
in Eq.~(\ref{condminV0}), and we obtain $V_{1}'(a_{\pm})=V''(\phi_{\pm})(a_{\pm}-\phi_{\pm})$.
Furthermore, since $V''=V_{0}''+V_{1}'',$ to this order we can replace
$V''$ for $V_{0}''$, and, inserting in Eq.~(\ref{limfi1}) we see
that $ \phi_1 $ gives the correct limit. The constant
of integration $C$ is fixed by condition (\ref{cero_n}), which,
at this order, gives
\begin{equation}
\int_{-\infty}^{+\infty}\phi_{0}'(n)\phi_{1}'(n)ndn=0.\label{cero_n_1}
\end{equation}
Integrating by parts and writing $\phi_{1}=\phi_{p}+C\phi_{0}'$,
we obtain 
\begin{align}
C & = -2\sigma_{0}^{-1} \int_{-\infty}^{+\infty} \left[\phi_{0}''(n)n + \phi_{0}'(n)\right]\phi_{p}(n)dn
\label{formulaC_0}
\\
 & =2\sigma_{0}^{-1} \int_{a_{+}}^{a_{-}} \phi_{p}(\phi_{0}) \left[-\frac{V_{0}'(\phi_{0})}{\sqrt{2\left(V_{0}(\phi_{0})-V_{0}(a_{+})\right)}}n(\phi_{0})+1\right] d\phi_{0},
 \label{formulaC}
\end{align}
with $n(\phi_{0})$ given by Eq.~(\ref{formulaperfil}). 

It is interesting to notice that the term $\phi_{h}$ in Eq.~(\ref{fi1tot})
does not contribute to the surface tension, as can be readily seen
by replacing $\phi_{1}=\phi_{h}$ in Eq.~(\ref{deltasigmadef}).
The solution $\partial_{n}\phi_{0}$ is due to the translation symmetry
of the theory, which implies that the function $\phi_{0}(n+\delta n)$
should be a solution too \cite{gj75}.

When $V_{+}=V_{-}$ (i.e., for a domain wall), we have $V_{1}=0$
and $K=0$, so we obtain $\phi_{p}=0$, which also implies $C=0$.
This confirms that the correction for that case is of higher order.
For $V_{+}\neq V_{-}$, we have linear-order corrections to the profile
and the parameter $\sigma$, but the EOM retains the same form.

\section{Surface stress-energy tensor}

\label{Tmunu}

The approximations we used to obtain the wall EOM can be used to derive
the energy-momentum tensor of the wall. For the scalar field we have
\begin{equation}
T_{\mu\nu}=
\partial_{\mu}\phi\partial_{\nu}\phi-\frac{1}{2}g_{\mu\nu}\partial_{\alpha}\phi\partial^{\alpha}\phi+g_{\mu\nu}V(\phi).
\label{Tmunucpo}
\end{equation}
Thus, outside the wall we have $T_{\mu\nu}=V_{\pm}\,g_{\mu\nu}$.
In normal Gaussian coordinates we have $V=V_{+}$ for $n>0$ and $V=V_{-}$
for $n<0$. In covariant form and in the limit of an infinitely thin
wall, we can write this part of $T_{\mu\nu}$ which omits the wall
as 
\begin{equation}
T_{\mu\nu}^{\mathrm{bulk}}=g_{\mu\nu}\left[V_{+}\Theta\left(F(x^{\mu})\right)+V_{-}\Theta\left(-F(x^{\mu})\right)\right].\label{Tmunubulk}
\end{equation}
On the other hand, if we use the first assumption, $\partial_{a}\phi=0$,
we have, in normal Gaussian coordinates
\begin{equation}
\bar{T}_{ab}=\bar{g}_{ab}\left(\frac{1}{2}\phi^{\prime2}+V\right)\,,\quad\bar{T}_{an}=0\,,\quad\bar{T}_{nn}=\frac{1}{2}\phi^{\prime2}-V.\label{Tab_0}
\end{equation}
The quantities $\phi'$ and $V$ are related by Eq.~(\ref{ecfi0}). Multiplying
that equation by $\phi'$ and integrating, we obtain its first integral,
which we write in the form 
\begin{equation}
\frac{1}{2}\phi^{\prime2}+V_{+}+\int_{n}^{\infty}K(\tilde{n})\phi^{\prime2}(\tilde{n})d\tilde{n}=V(\phi),\label{1ra_int}
\end{equation}
where we used the boundary condition $\phi\equiv\phi_{+}$ outside
the bubble. Using this result, we can write Eq.~(\ref{Tab_0}) as
\begin{equation}
\bar{T}_{ab}=\bar{g}_{ab}\left[\phi^{\prime2}+V_{+}+\int_{n}^{\infty}K(\tilde{n})\phi^{\prime2}(\tilde{n})d\tilde{n}\right],
\quad\bar{T}_{nn}=\bar{g}_{nn}\left[V_{+}+\int_{n}^{\infty}K(\tilde{n})\phi^{\prime2}(\tilde{n})d\tilde{n}\right].
\label{Tab}
\end{equation}
The potential $V$ on the right-hand side of Eq.~(\ref{1ra_int})
varies from the value $V_{+}$ outside the bubble to the value $V_{-}$
inside, but takes higher values at the wall, as $\phi$ crosses the
potential barrier. On the left-hand side, we may assume that the function
$K(n)$ does not change its sign over the short width of the wall.
Therefore, the integral in Eq.~(\ref{1ra_int}) is a monotonic function
of $n$, and the last two terms on the left-hand side interpolate
between the values $V_{+}$ and $V_{-}$. These terms appear in Eq.~(\ref{Tab}),
and, in the thin-wall limit, give the result (\ref{Tmunubulk}). Hence,
beyond the thin-wall approximation we define
\begin{equation}
T_{\mu\nu}^{\mathrm{bulk}}=
g_{\mu\nu}\left[V_{+}+\int_{n}^{\infty}K(\tilde{n})\phi^{\prime2}(\tilde{n})d\tilde{n}\right].
\end{equation}

In the equality (\ref{1ra_int}), the term $\frac{1}{2}\phi^{\prime2}$
reproduces the shape of the barrier without the potential jump. The
quantity $\phi^{\prime2}$ vanishes outside the wall, so in Eq.~(\ref{Tab}) we identify the energy-momentum tensor
of the wall, 
\begin{equation}
\bar{T}_{ab}^{\mathrm{wall}}=\bar{g}_{ab}\phi^{\prime2}\,,\quad\bar{T}_{an}^{\mathrm{wall}}=0\,,\quad\bar{T}_{nn}^{\mathrm{wall}}=0.\label{Tabwall}
\end{equation}
So far we have only used the first assumption. According to the second
assumption, $\phi$ varies more than any other quantity inside the
wall. In the limit of an infinitely thin wall, the function $\phi^{\prime2}(n)$
behaves like a delta function, and we can write Eq.~(\ref{Tabwall})
as
\begin{equation}
\bar{T}_{ab}^{\mathrm{wall}}=\gamma_{ab}\sigma\delta(n),\quad\bar{T}_{n\mu}^{\mathrm{wall}}=0.\label{Tabdelta}
\end{equation}
Using Eqs.~(\ref{gmnunugab}) and (\ref{ngral}) we can write these
expressions in covariant form,
\begin{equation}
T_{\mu\nu}^{\mathrm{wall}}=\sigma\left(g_{\mu\nu}+N_{\mu}N_{\nu}\right)\delta(F/s)=\sigma s\left(g_{\mu\nu}+N_{\mu}N_{\nu}\right)\delta(F).\label{Tmunuwall}
\end{equation}
Finally, if we use the third assumption (approximating $\phi$ by
$\phi_{0}$), we have $\sigma=\sigma_{0}$, which is given by
Eq.~(\ref{formulasigma}). The leading-order finite-width corrections
discussed in Sec.~\ref{sec:masalla} modify the value of $\sigma$
but not the expression for $T_{\mu\nu}^{\mathrm{wall}}$. In App.~\ref{derivagral}
we discuss a derivation of Eq.~(\ref{Tmunuwall}) avoiding the use
of normal Gaussian coordinates.

In the Monge gauge, we can write Eq.~(\ref{Tmunuwall}) in the form
\begin{equation}
	T_{\mu\nu}^{\mathrm{wall}}=\sigma\left[s g_{\mu\nu}+s^{-1} \partial_{\mu}(x^3-x^3_{w}) 
	\partial_{\nu}(x^3-x^3_{w})\right] 	\delta(x^3-x^3_w).
\end{equation}
In Minkowski space, we have the parametrization $ z=z_w(t,x,y) $ and we obtain
\begin{equation}
	T_{\mu\nu}^{\mathrm{wall}}=\sigma\left[s \eta_{\mu\nu}+s^{-1} 
	(\hat{z}_\mu-\partial_{\mu}z_{w}) (\hat{z}_\nu -\partial_{\nu}z_{w})\right] \delta(z-z_w),
	\label{TmunuMink}
\end{equation}
where $ \hat{z} =(0,0,0,1) $ and we have $ s^{2}=1-\dot{z}_{w}^{2} + (\nabla z_{w})^{2}$. 
For a static planar wall at $z_{w}=0$ we obtain
the well-known result $T_{\mu\nu}^{\mathrm{wall}}=\sigma\mathrm{diag}(1,-1,-1,0)\delta(z)$.
In the non-static case, we have non-diagonal terms and, besides, $ s^{-1} $ gives the gamma factor $\gamma_w=1/\sqrt{1-\dot{z}_{w}^{2}}$. 
In particular, we have $T_{00}^{\mathrm{wall}}=\sigma\gamma_{w}\delta(z-z_{w})$.
If we integrate $T_{00}^{\mathrm{wall}}$ with respect to $z$, we
obtain the surface energy density $E_{w}=\sigma\gamma_{w}$. As expected,
it differs in a gamma factor with respect to an inertial observer
in which the wall is instantaneously at rest. 

In spherical coordinates, the wall parametrization is $ r=r_w(t,\theta,\varphi) $ and we have
\begin{equation}
	T_{\mu\nu}^{\mathrm{wall}} = \sigma\left[s g_{\mu\nu} + 
	s^{-1} (\hat{r}_{\mu}-\partial_{\mu}r_{w}) (\hat{r}_{\nu}-\partial_{\nu}r_{w})\right] \delta(r-r_w),
	\label{Tmunuesf}
\end{equation}
with $g_{\mu\nu}=\mathrm{diag}(1,-1,-r^{2},-r^{2}\sin^{2}\theta)$ 
and 
\begin{equation}
s^{2} = 1 - (\partial_{t}r_{w})^{2} + r_{w}^{-2}(\partial_{\theta}r_{w})^{2} + r_{w}^{-2}\sin^{-2}\theta(\partial_{\varphi}r_{w})^{2}
=
1-\dot{r}_{w}^{2} + (\nabla r_{w})^{2}.
\end{equation}
In Eq.~(\ref{Tmunuesf}), the covariant vectors $ \hat{r}_\mu =(0,0,0,1)$ and
$\partial_{\mu}r_w=(\partial_{t}r_{w},\partial_{\theta}r_{w},\partial_{\phi}r_{w},0)$
are the components of the one-forms $ dr $ and $dr_{w}$, respectively, in the spherical coordinate basis $\omega^{0}=dt$, $\omega^{\theta}=d\theta$, $\omega^{\phi}=d\phi$, $\omega^{r}=dr$ \cite{misner}. 
Geometrically, $ \hat{r} $ is the unit vector in the radial direction.
Furthermore, in the orthonormal basis $\omega^{\hat{0}}=dt$, $\omega^{\hat{\theta}}=rd\theta$,
$\omega^{\hat{\phi}}=r\sin\theta d\phi$, $\omega^{\hat{r}}=dr$, 
the metric tensor is $ \eta_{\mu\nu} $ and
the space components of $ \partial_\mu r_w $ give
the usual form of the gradient $ \nabla r_w $ in spherical coordinates. 
Therefore, expressing the result (\ref{Tmunuesf}) in this orthonormal basis and rotating to the standard basis
(associated to Minkowski coordinates), we obtain
\begin{equation}
	T_{\mu\nu}^{\mathrm{wall}} = \sigma\left[s \eta_{\mu\nu} + 
	s^{-1} (\hat{r}_\mu-\partial_{\mu}r_{w}) (\hat{r}_{\nu}-\partial_{\nu}r_{w})\right] \delta(r-r_w).
	\label{Tmunu_hibrido}
\end{equation}
where now we have $ \partial_\mu = (\partial_t,\nabla) $ and $ \hat{r}_\mu = \mathbf{r}/r=(0,\sin\theta\cos\varphi,\sin\theta\sin\varphi,\cos\theta) $.
This expression is equivalent to (\ref{TmunuMink}) and (\ref{Tmunuesf}) 
but is more useful for the calculation of gravitational waves in a cosmological phase transition.
In App.~\ref{derivagral} we compare Eq.~(\ref{Tmunu_hibrido}) with previous results.

\section{Specific examples }

\label{sec_uso}

We will now test the accuracy of the various levels of approximation
we have discussed for a thin wall, including the approximation of
the potential by a degenerate one. To this end, we will compare the
approximations with exact numerical solutions for specific potentials.
To keep the numerical computations to a minimum, we will consider
the simple cases of planar and spherical walls. 

\subsection{Symmetric solutions}

\begin{figure}[tb]
\includegraphics[width=0.45\textwidth]{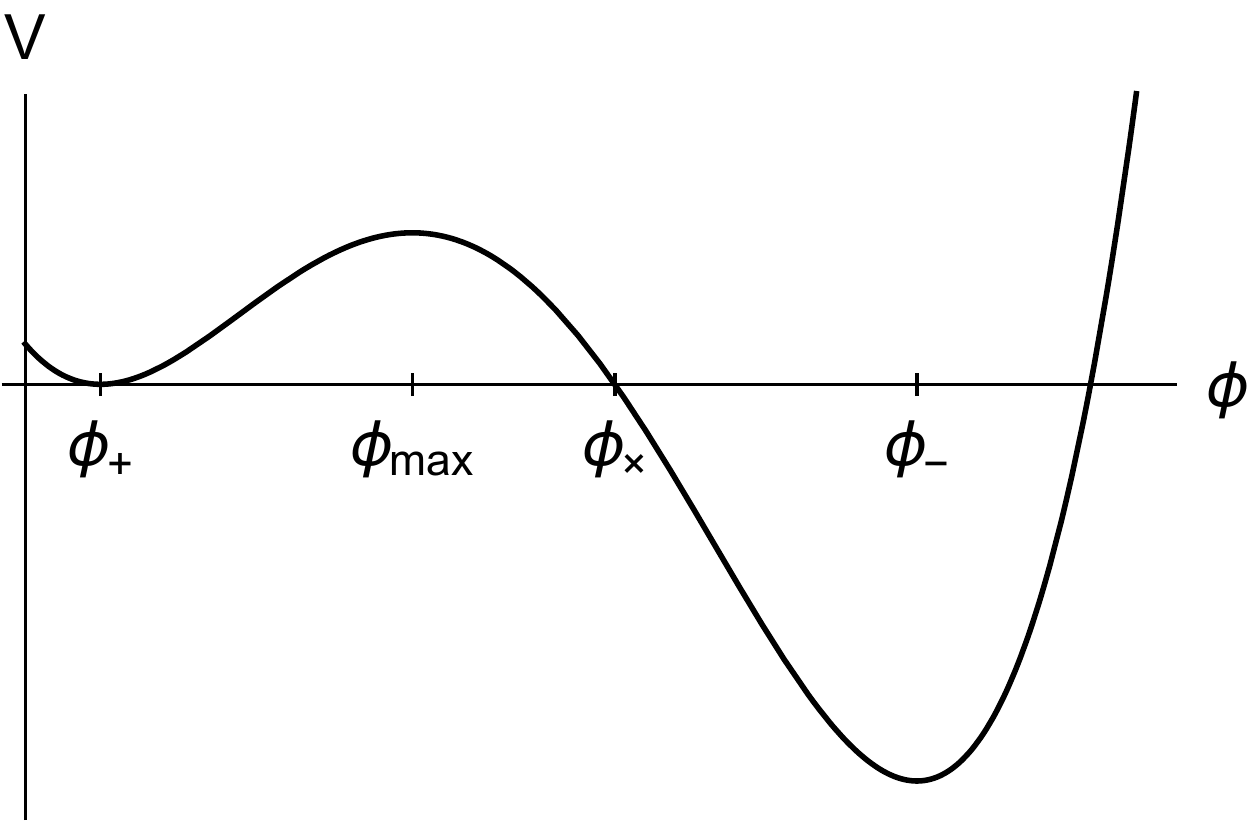}
\hfill
\includegraphics[width=0.45\textwidth]{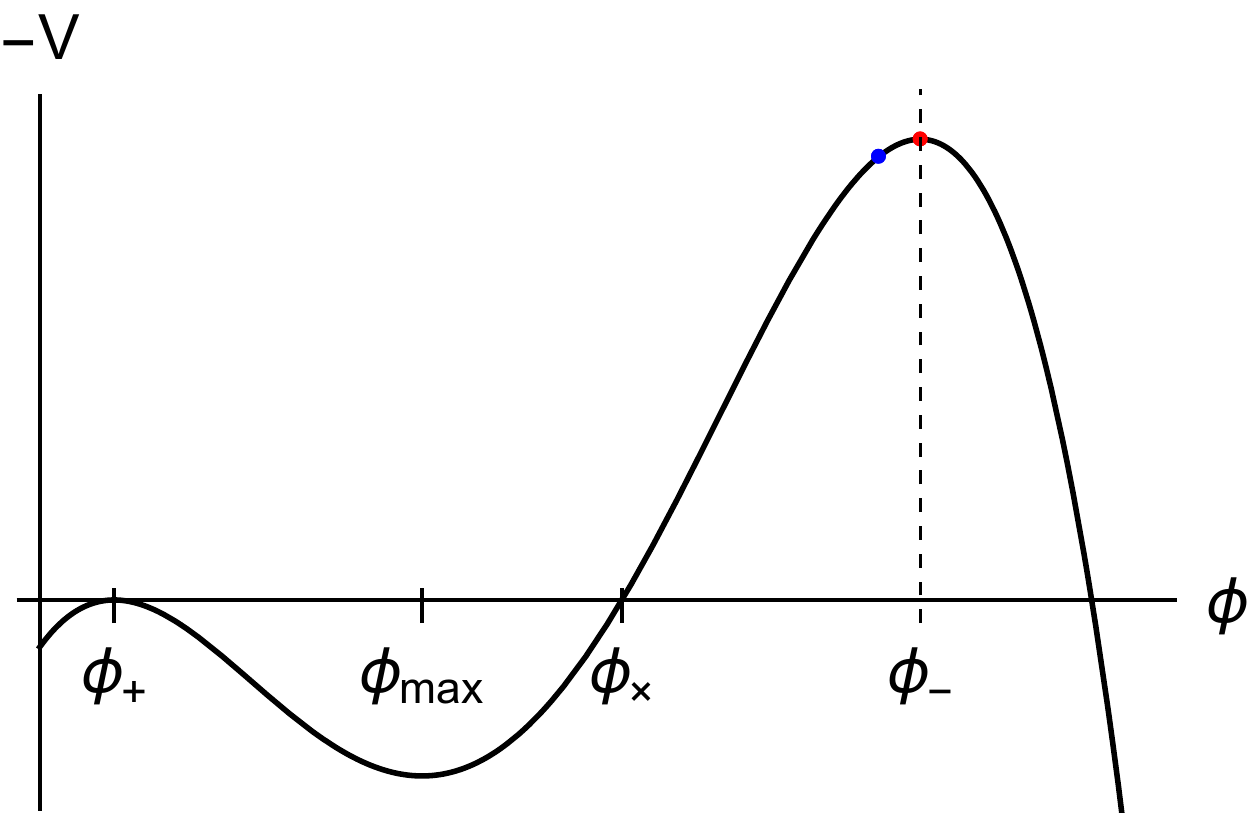}
\caption{A broken-symmetry potential $ V $
and the inverted potential $-V$.
The points on the curve of $-V$ indicate the value of $\phi_{i}$
for the planar and spherical symmetry cases.\label{figpot}}
\end{figure}

We will consider a solution with O(3,1) symmetry, so that we only need
to solve an ordinary differential equation. 
This is the case discussed by Coleman in Ref.~\cite{c77}, as it corresponds to the
bounce solution when changing from real to imaginary time. Let us
consider Eq.~(\ref{eccampo}) in Minkowski space,
\begin{equation}
\frac{\partial^{2}\phi}{\partial t^{2}}-\nabla^{2}\phi=-V'(\phi),
\label{ecfiMink}
\end{equation}
and assume that the solution has the form $\phi(t,\mathbf{x})=\bar\phi(\rho)$,
with $\rho=\sqrt{r^{2}-t^{2}}$ and $r=|\mathbf{x}|$. In the thin-wall
limit, such a solution will describe an accelerated spherical wall.
We are also interested in the 1+1-dimensional case, which is equivalent
to a planar symmetry solution in 3+1 dimensions. 
Therefore, we write Eq.~(\ref{ecfiMink}) in $j+1$ dimensions, with $ j=1$ or $3 $
(the following derivations are also valid for a cylindrical wall for $j=2$),
\begin{equation}
\frac{d^{2}\bar\phi}{d\rho^{2}}+\frac{j}{\rho}\frac{d\bar\phi}{d\rho}=V'(\bar\phi).
\label{ec4d}
\end{equation}
The boundary conditions are
\begin{equation}
\lim_{\rho\to\infty}\bar\phi(\rho)=\phi_{+}\,,\quad\frac{d\bar\phi}{d\rho}(0)=0.\label{cc}
\end{equation}
As already mentioned, this equation has a mechanical analogy in which
$\bar\phi$ is the position of a particle and $\rho$ is the time. This
particle moves in a potential $-V$ and is subject to a friction force
that decreases with time. The conditions (\ref{cc}) imply that the
particle is released at rest at time zero and reaches $\phi_{+}$
at time infinity. To achieve the latter condition, the initial position
$\phi_{i}$ should lie between $\phi_{-}$ and the point $\phi_{\times}$
where $V(\phi_\times)=V(\phi_{+})$ (see Fig.~\ref{figpot}). For the spherically
symmetric case the friction is higher, which makes it necessary to
release the particle from closer to $\phi_{-}$.

\begin{figure}[tb]
	\includegraphics[width=0.47\textwidth]{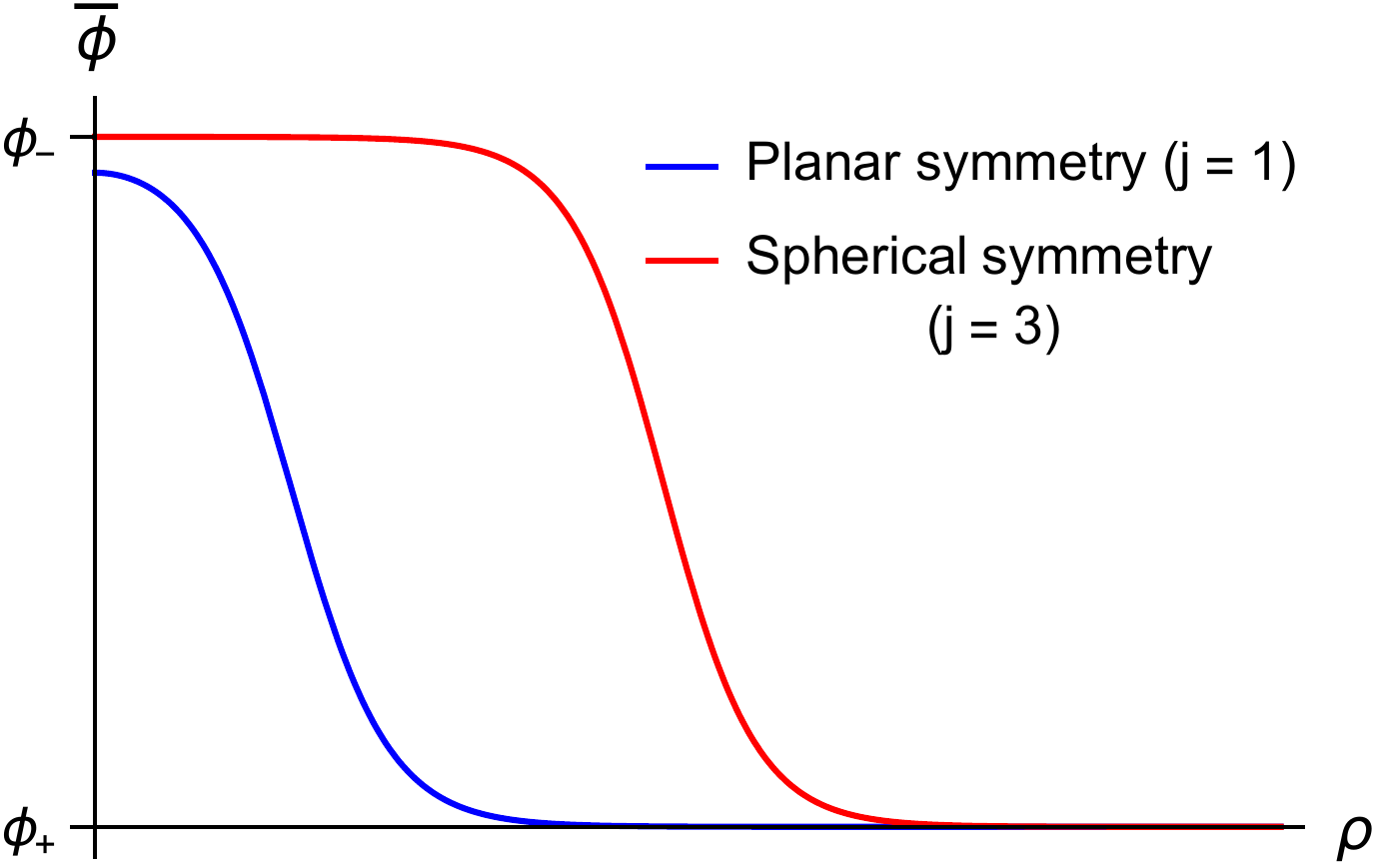}
	\hfill
	\includegraphics[width=0.47\textwidth]{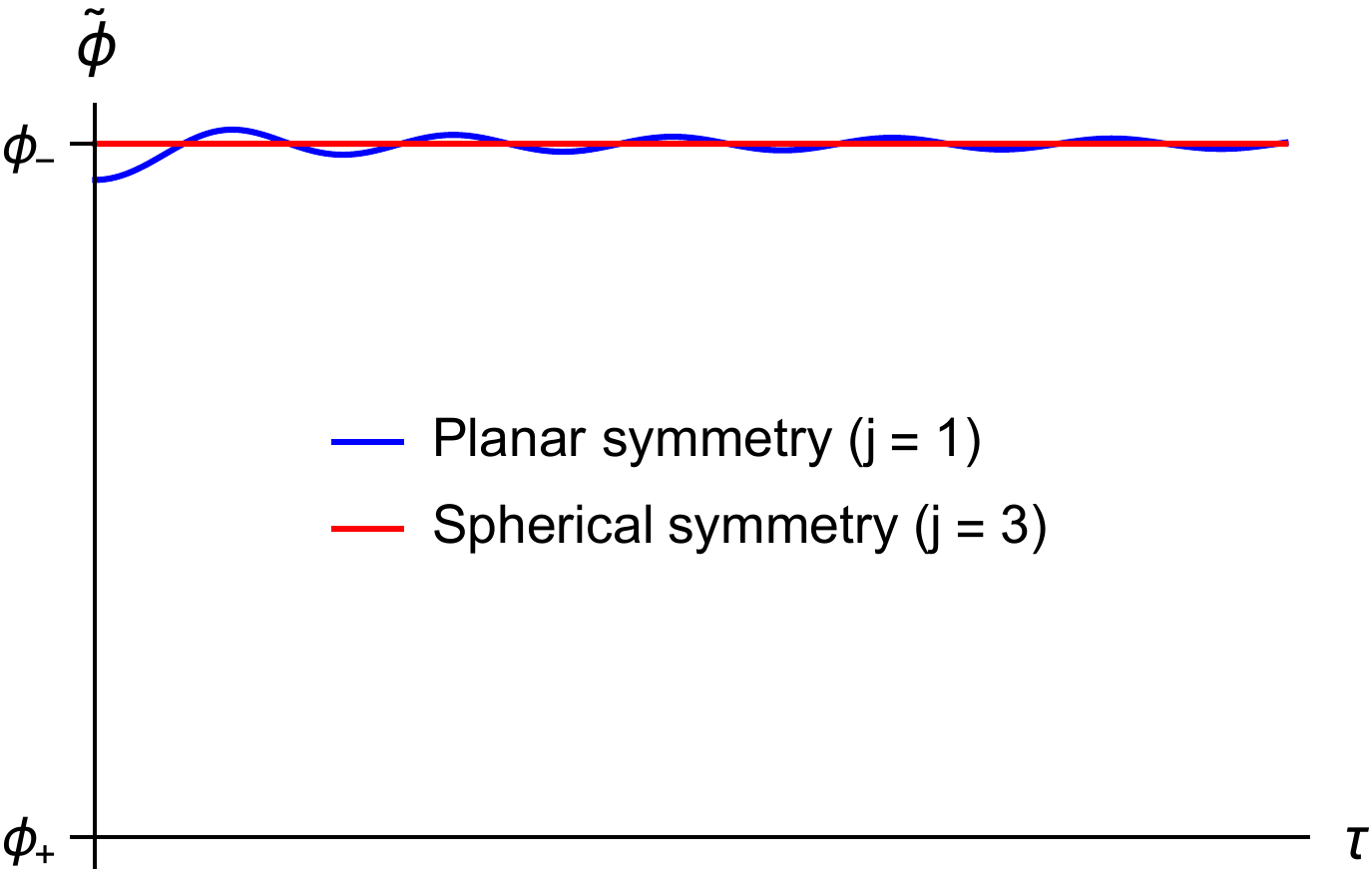}
	\caption{The solutions $\bar\phi(\rho)$ and $\tilde{\phi}(\tau)$.%
	\label{figperfil0}}
\end{figure}

The standard procedure to solve Eq.~(\ref{ec4d}) with the boundary
conditions (\ref{cc}) is the following shooting method. The particle
is released at rest from some position $\phi_{i}$ and the equation
is numerically solved. If $\phi_{i}$ is too close to $\phi_{-}$,
after a finite time the particle will reach positions beyond $\phi_{+}$.
On the other hand, if $\phi_{i}$ is too close to $\phi_{\times}$,
after a finite time the particle returns towards $\phi_{\max}$ before
reaching $\phi_{+}$. Thus, the initial value $\phi_{i}$ must be
adjusted so that the particle gets as close to $\phi_{+}$ as possible.
A few iterations are needed to find the correct $\phi_{i}$ with great
precision. The solution is shown in the left panel of Fig.~\ref{figperfil0}
for a potential of the form $V=\frac{1}{2}m^{2}\phi^{2}-\frac{e}{3}\phi^{3}+\frac{\lambda}{4}\phi^{4}$
(see subsection \ref{specificpots} for details).

\begin{figure}[tb]
	\includegraphics[width=0.47\textwidth]{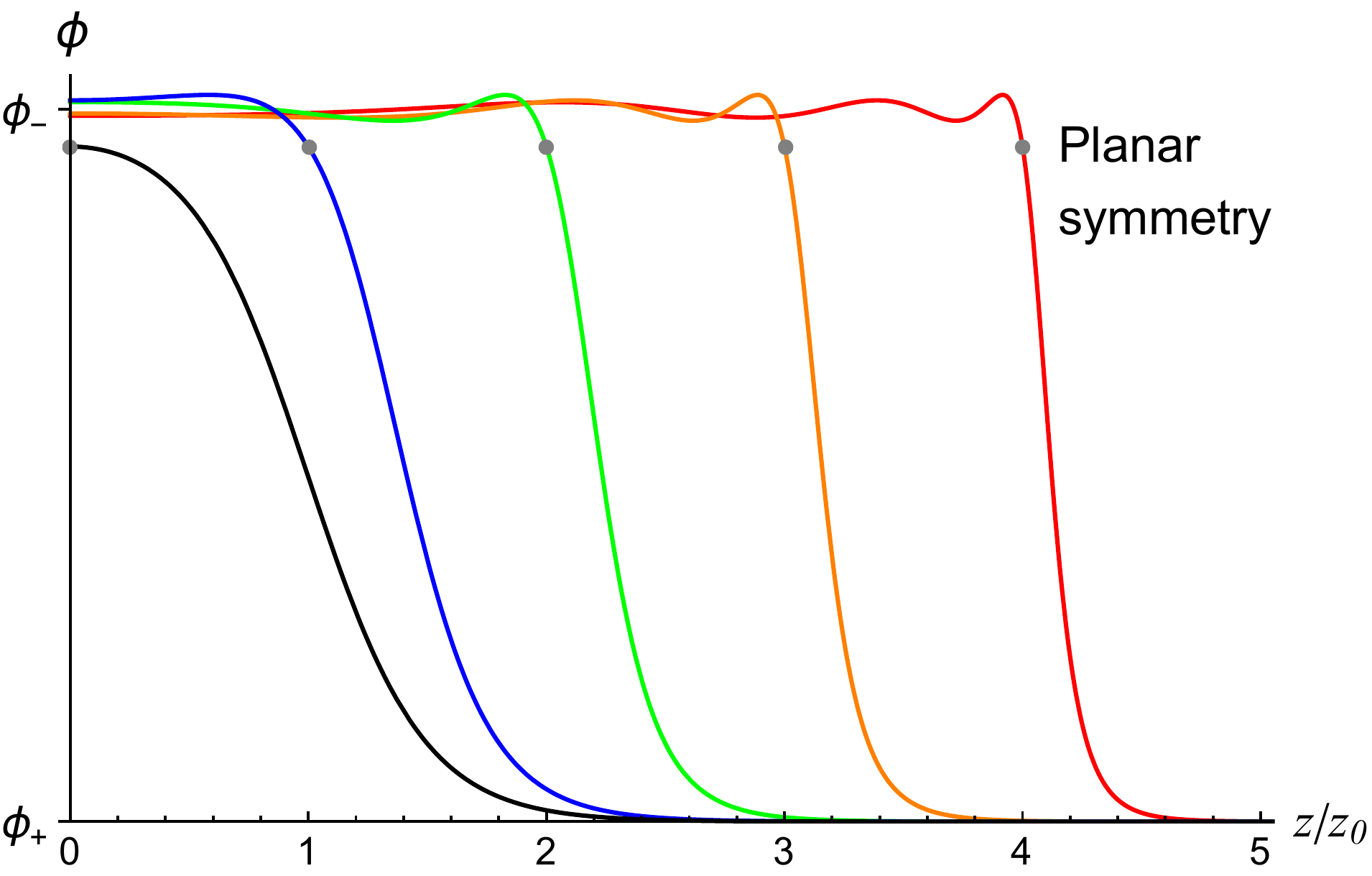}
	\hfill
	\includegraphics[width=0.47\textwidth]{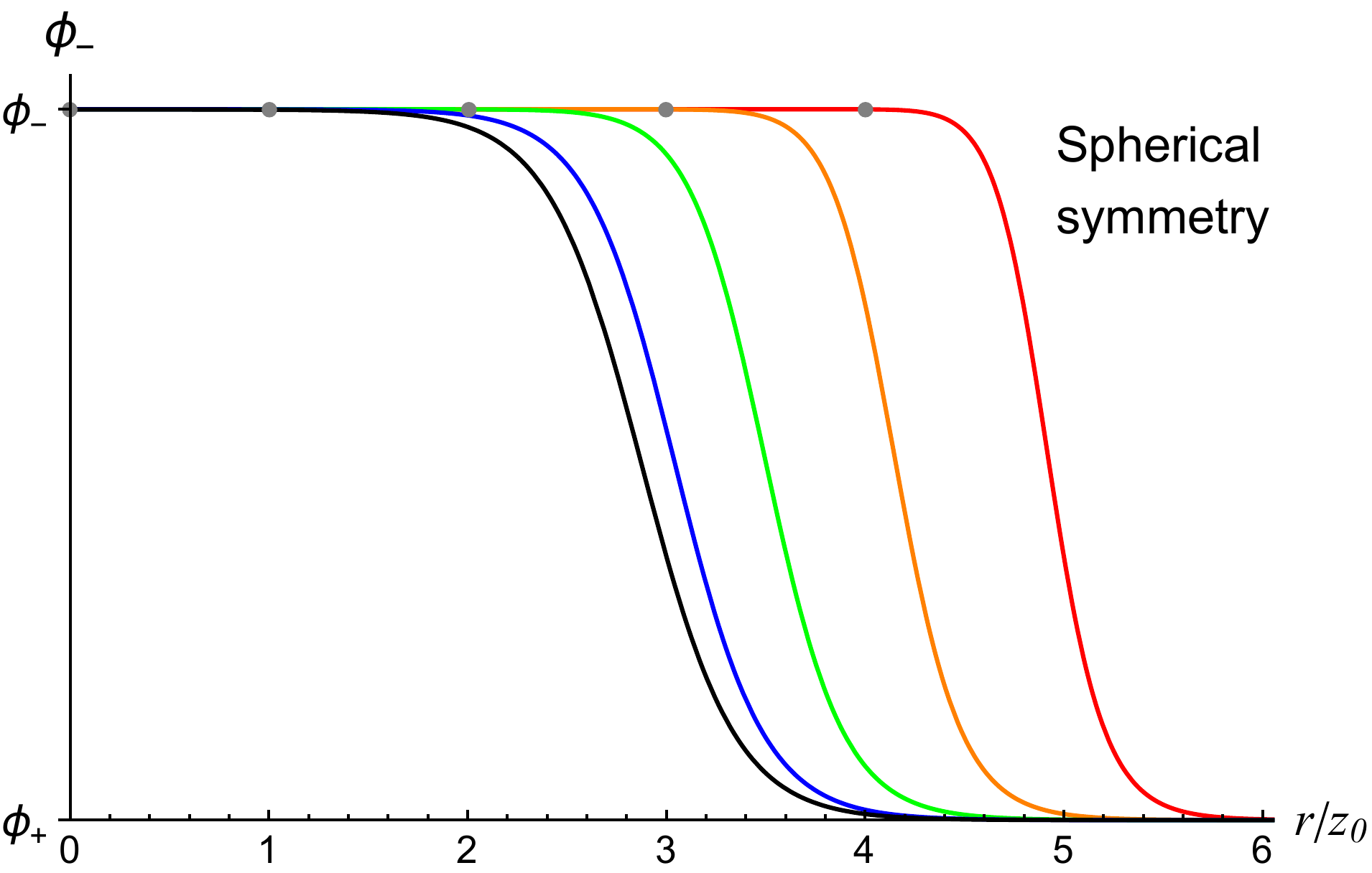}
	\caption{The complete
		profile $\phi(t,r)$ at times $t=0,z_0,2z_0,3z_0$ and $4z_0$ for the planar and spherical symmetry cases.
		The gray dots indicate the point $r=t$, $\phi=\phi_{i}$.\label{figperfilt}}
\end{figure}

At $t=0$ we have $\rho=r$, and the graph of $\bar\phi(\rho)$ gives
the complete profile of the newly nucleated bubble. 
In principle, once the function $\bar\phi(\rho)$ has been computed, 
to obtain the field profile at any subsequent time $t$ we only need to
evaluate $\bar\phi(\rho)$ at $\rho=\sqrt{r^{2}-t^{2}}$.
However, the numerical result obtained with the shooting method gives $\bar\phi(\rho)$
only for real values of $\rho$, so it cannot be used for $r<t$. 
Thus, at $t>0$ we only obtain a part of the bubble profile, namely, the part to the right of the gray dots in Fig.~\ref{figperfilt}.
We may define this part of the bubble profile as the wall profile (which includes the outside of the bubble as well).

One way to calculate the complete bubble profile $\phi(r,t)$ at time $t$ is to solve Eq.~(\ref{ecfiMink}) 
with the initial condition $\phi(r,0)=\bar\phi(r)$
using a lattice calculation.
Such a calculation has been done, e.g., in Ref.~\cite{cghw20}.
Due to spherical symmetry, this is a lattice field theory simulation in only 1 spatial dimension.
We remark that in this case the symmetry of the bubbles is not O(3) but O(3,1), so
the field profile must be a function of the single variable $\rho$, and
such a 1D simulation should not be necessary (a much simpler 0D calculation should be enough).
Theoretically, the solution for $r<t$ (which we may call the bubble interior) can be obtained  by analytic continuation of $\bar\phi(\rho)$ to imaginary $\rho$.
To accomplish this in practice, 
we write $\phi(t,\mathbf{x})=\bar\phi(i\tau)\equiv\tilde{\phi}(\tau)$, with $\tau=\sqrt{t^{2}-r^{2}}$,
and we solve the equation for the function $\tilde{\phi}(\tau)$,
\begin{equation}
\frac{d^{2}\tilde{\phi}}{d\tau^{2}}+\frac{j}{\tau}\frac{d\tilde{\phi}}{d\tau}=-V'(\tilde{\phi}),\label{ec4dint}
\end{equation}
with the matching conditions at $\tau=\rho=0$
\begin{equation}
\tilde{\phi}(0)=\phi_{i}\,,\quad\frac{d\tilde{\phi}}{d\tau}(0)=0.\label{condec4dint}
\end{equation}
We still have a mechanical analogy, but the potential is now $+V$ and the particle
will fall from $\phi_{i}$ toward the minimum $\phi_{-}$. The right
panel of Fig.~\ref{figperfil0} shows the solution.

At $r=0$ we have $\tau=t$, and the graph of $\tilde{\phi}(\tau)$
gives the evolution of $\phi$ at the bubble center. The field falls
from its initial value toward the minimum and undergoes damped oscillations.
For the spherical case, these oscillations are not appreciable 
since $\phi_{i}$ is very close to $\phi_{-}$. As we shall see
later, for a potential with a higher difference between minima relative
to the barrier height, the oscillations are appreciable also for this case. 
The complete solution $\phi(t,r)$ is shown at several equally-spaced times in
Fig.~\ref{figperfilt}. 
The initial acceleration of the wall can be appreciated.
We use as a length and time scale the value $z_0$ corresponding to the average wall position at $t=0$ for the planar case, defined as
\begin{equation}
	z_0= \sigma^{-1} \int_{0}^{\infty}\left({d\bar\phi}/{d\rho}\right)^{2}\rho \,d\rho ,
\end{equation}
where $\sigma=\int_{0}^{\infty}\left({d\bar\phi}/{d\rho}\right)^{2}d\rho$.
In the planar case, we observe oscillations that generate behind the wall and
decay toward the bubble center, in agreement with previous lattice calculations \cite{cghw20}. 
To understand the form of these spatial oscillations, notice that when the field profile passes through a given point $z$ in space, $\phi$ initially grows from $\phi_+$ towards the true minimum $\phi_-$. 
However, when it reaches this minimum, $\phi$ oscillates around it. 
The initial amplitude of these time oscillations at a given $z$ is $\phi_ - -\phi_i$, while at previous points $z'<z$, the amplitude has already decreased.
A gray dot indicates the value $\phi_i$ in Fig.~\ref{figperfilt}. 
This point divides
the solutions $\phi=\tilde{\phi}(\sqrt{t^{2}-r^{2}})$ and $\phi=\bar\phi(\sqrt{r^{2}-t^{2}})$
and, thus, provides a way for separating the bubble interior from the wall profile.
In App.~\ref{interior}, we discuss in more detail the bubble profile.

\subsection{The thin-wall approximation}

Since we are considering a specific, highly symmetric case, the solution
is a function of a single variable, $\phi(t,\mathbf{x})=\phi(\rho)$.
This means that the incompressibility assumption will be valid even
for a thick wall. The variable $\rho$ must then have a close relationship
with the variable $n$ of the normal Gaussian coordinates. Indeed,
the thin-wall approximation can be directly implemented on the function
$\phi(\rho)$ \cite{c77}. Defining the wall hypersurface as that
corresponding to a fixed value $\rho=r_{c}$, we immediately obtain
Coleman's solution for the wall position,
\begin{equation}
r_{w}^{2}=r_c^{2}+t^{2}.\label{rwcoleman}
\end{equation}
Due to the symmetry assumption, it was not necessary to obtain an
equation of motion. 
The parameter $r_c$ gives the initial bubble
radius, which is not arbitrary for these symmetrical solutions. 
Indeed, if we multiply Eq.~(\ref{ec4d})
by $d\phi/d\rho$ and integrate, assuming that $\rho\simeq r_c$
inside the wall (second assumption for a thin wall) we obtain the value
\begin{equation}
r_c=j\sigma/\left[V_{+}-V(\phi_{i})\right]\label{rho0}.
\end{equation}
If, in addition, we have $\phi_{i}=\phi_{-}$, we obtain
\begin{equation}
r_c =j\sigma/\Delta V .\label{rc}
\end{equation}
This result coincides with the radius of the newly nucleated bubble obtained from the bounce solution \cite{c77}. 
Finally, we may estimate $\sigma$ and the wall profile
using the third thin-wall assumption, which in this approach consists
of dropping the term with a single derivative in Eq.~(\ref{ec4d})
(its validity condition can be expressed as $l\ll r_c$).

Thus, for a solution with O(3,1) symmetry, the initial bubble radius is given by $r_0=r_c\simeq 3\sigma/\Delta V$, while in the 1+1 dimensional case we have $z_0\simeq\sigma/\Delta V$. 
The relation $r_0\simeq 3z_0$ can be appreciated in Fig.~\ref{figperfilt}.
The treatment of Sec.~\ref{sec:Thin-wall} is, of course, much broader.
Even the specific cases of planar and spherical walls
without deformations are more general than those considered here.
The corresponding equations of motion, Eqs.~(\ref{EOMpl}) and (\ref{EOMbbesf}),
can be treated together by writing
\begin{equation}
\frac{d}{dt}\left(\sigma\gamma_{w}\dot{r}_{w}\right)+\frac{(j-1)\sigma\gamma_{w}}{r_{w}}=\Delta V.
\end{equation}
Using the equalities $\gamma_{w}^{3}\ddot{r}_{w}=\dot{\gamma}_{w}/\dot{r}_{w}=d\gamma_{w}/dr_{w}$,
we can write the previous equation as
\begin{equation}
\frac{d\gamma_{w}}{dr_{w}}+\frac{(j-1)\gamma_{w}}{r_{w}}=\frac{\Delta V}{\sigma}.
\end{equation}
For an initial condition $\gamma_{w}(r_{0})=\gamma_{0}$, the solution
is given by
\begin{equation}
-\Delta Vr_{w}^{j}/j+\sigma\gamma_{w}r_{w}^{j-1}=-\Delta Vr_{0}^{j}/j+\sigma\gamma_{0}r_{0}^{j-1}.\label{consener}
\end{equation}
Recalling that the surface energy density is given by $E_{w}=\sigma\gamma_{w}$,
this expression is nothing other than the energy conservation equation
(see also Ref.~\cite{bkt83}). Solving for $\dot{r}_{w}$,
we have
\begin{equation}
\dot{r}_{w}=\sqrt{1-\left[{r_{w}}/{r_{c}}-\left({r_{0}}/{r_{c}}
	-\gamma_{0}\right)\left({r_{0}}/{r_{w}}\right)^{j-1}\right]^{-2}},\label{rwdot}
\end{equation}
with $r_{c}$ given by Eq.~(\ref{rc}). The solution is given by
the integral
\begin{equation}
t = \int_{r_{0}}^{r_{w}}
\frac{dr}{ \sqrt{1 - \left[{r}/{r_{c}} - \left({r_{0}}/{r_{c}}-\gamma_{0}\right) \left({r_{0}}/{r}\right)^{j-1}\right]^{-2}}}.
\label{soltrw}
\end{equation}
In particular, for the initial conditions $\gamma_{0}=1$ and $r_{0}=r_{c}$,
we obtain the result (\ref{rwcoleman}).

Regarding the wall profile, there is a subtle difference between the
two approaches. With assumption 3, Eq.~(\ref{ec4d}) takes the same
form as Eq.~(\ref{ecfi0sup3}) and leads to the result (\ref{formulaperfil}),
but as a function of $\rho$ instead of $n$. The resulting wall profile,
$ \phi_{0}(\rho-r_c) 
$,
with $\rho= \sqrt{r^{2}-t^{2}} $,
is not the same as
$\phi_{0}(n)=\phi_{0}\left(\gamma_{w}\left(r-r_{w}\right)\right)$. 
Nevertheless, using (\ref{rwcoleman}), we can write $\rho=\sqrt{r^{2}-r_{w}^{2}+r_{c}^{2}}$.
Expanding the square root for $r-r_{w}\ll r_{c}$ 
and using the relation
$\gamma_{w}=r_{w}/r_{c}$, we obtain $\rho \simeq \gamma_w(r-r_{w})+ r_{c} $. Hence, the two results agree to lowest order in $ n $.

\subsection{Specific potentials}
\label{specificpots}

We now apply our results to specific potentials.

\subsubsection{Polynomial potential}

The simplest example which shows the general characteristics
of an effective potential for a first-order phase transition is the quartic potential
\begin{equation}
V(\phi)=\frac{1}{2}m^{2}\phi^{2}-\frac{e}{3}\phi^{3}+\frac{\lambda}{4}\phi^{4},\label{pot}
\end{equation}
where all the parameters are positive. The minima are given by $\phi_{+}=0$
with $V_{+}=0$ and
\begin{align}
\phi_{-} & =(e/2\lambda)\left(1+\sqrt{1-4\lambda m^{2}/e^{2}}\right)
\end{align}
with $V_{-}<V_{+}$ for $\lambda m^{2}/e^{2}<2/9$. The peak of the
potential barrier is at $\phi_{\max}=e/\lambda-\phi_{-}$. Let us
consider a linear correction $V_{1}=c\phi$ to obtain a degenerate
potential $V_{0}=V-V_{1}$. We use Eqs.~(\ref{condminV0})-(\ref{condV0deg}),
which yield 
\begin{equation}
c=\frac{9e\lambda m^{2}-2e^{3}}{27\lambda^{2}},\quad a_{\pm}=\frac{e}{3\lambda}\left(1\mp\sqrt{3-9\lambda m^{2}/e^{2}}\right).\label{apm}
\end{equation}
The resulting potential $V_{0}$ is of the form (\ref{Vdeg}),
\begin{equation}
V_{0}(\phi) = \frac{\lambda}{4}\left(\phi-a_{+}\right)^{2}\left(\phi-a_{-}\right)^{2}-\frac{\lambda}{4}a_{+}^{2}a_{-}^{2}.
\label{potdeglin}
\end{equation}
Therefore, we obtain the same expressions as in Sec.~\ref{preliminares}
for the profile and the surface tension in the thin-wall approximation,
Eqs.~(\ref{perfil-pl}) and (\ref{sigmapl}). Replacing the above
values of $a_{\pm}$, we have
\begin{equation}
\sigma_{0}=\frac{1}{3}\sqrt{\frac{\lambda}{8}}\left(\frac{2e}{3\lambda}\sqrt{3-9\lambda m^{2}/e^{2}}\right)^{3},\label{sigma}
\end{equation}
\begin{equation}
\phi_{0}(n)=\frac{e}{3\lambda}\left[1-\sqrt{3-9\lambda m^{2}/e^{2}}\tanh\left(\sqrt{e^{2}/6\lambda-m^{2}/2}\; n\right)\right].
\label{perf_pd_n}
\end{equation}
We obtain the evolution of the kink by replacing  $n=\gamma_{w}(r-r_{w})$,
where $r_{w}$ is the solution of the EOM, which in this case is given
by $r_{w}=\sqrt{r_{c}^{2}+t^{2}}$, with $r_{c}=j\sigma/\Delta V$.

The next order corrections are given by Eqs.~(\ref{deltasigma})-(\ref{filp}).
We obtain $\delta\sigma=0$ and
\begin{equation}
\phi_{1}=-\left(2c/\lambda\right)\left(a_{-}-a_{+}\right)^{-2}.
\end{equation}
To understand these simple results, notice that the correction $ \phi_1 $ 
interpolates between the values  $\phi_{+}-a_{+}$ and $\phi_{-}-a_{-}$ 
[to lowest order; see discussion around Eq.~(\ref{limfi1})]
For the linear modification of the polynomial potential, the displacement of the two minima is
the same (to this order), $\phi_{+}-a_{+}=\phi_{-}-a_{-}$. 
As a consequence, we obtain a constant solution $\phi_{p}$,
which implies that both the constant $C$ and $\delta\sigma$
vanish, as can be easily seen from Eqs.~(\ref{formulaC_0}) and (\ref{deltasigmadef}). 

\begin{figure}[tb]
	\centering
	\includegraphics[width=0.5\textwidth]{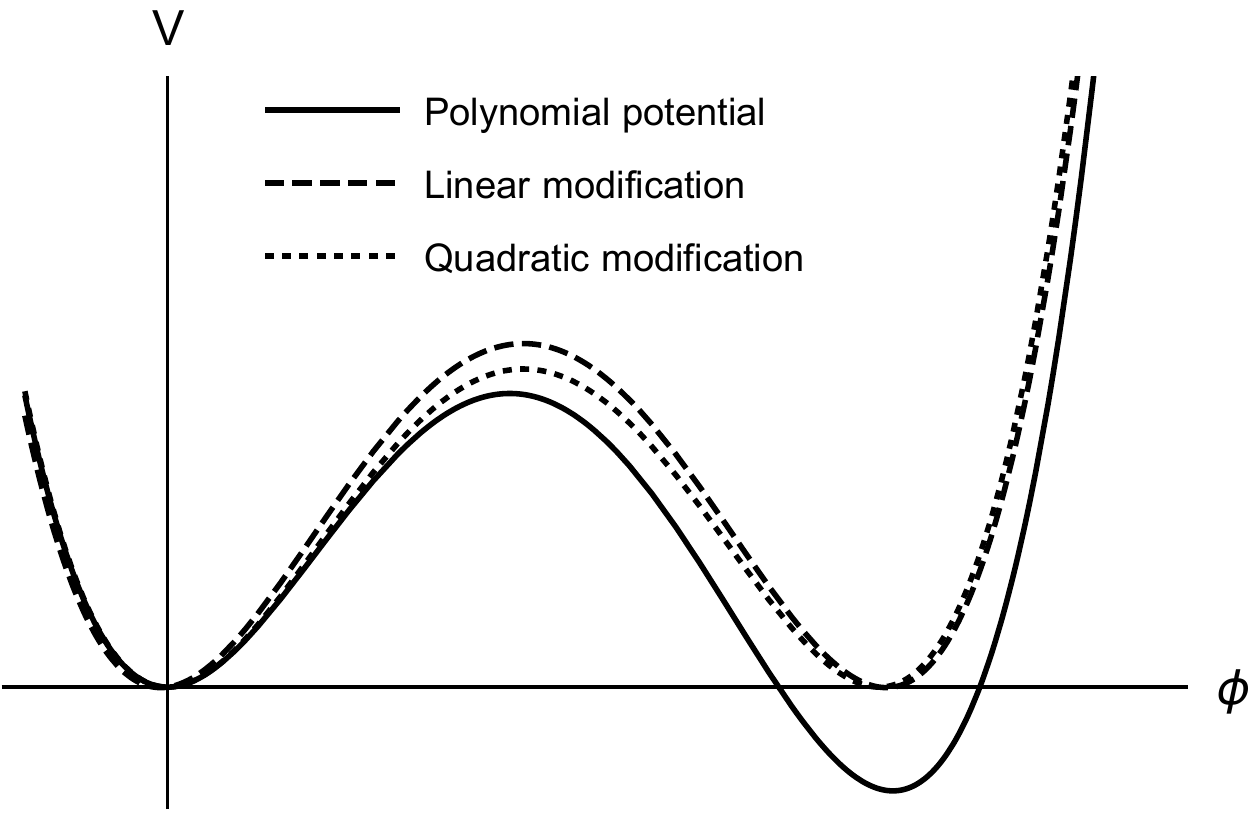}
	\caption{The polynomial potential (\ref{pot}) and the potential $V_{0}$
		obtained from two different modifications $V_{1}$. The parameters
		of the potential are such that $\Delta V/V_{\max}\simeq0.35$.\label{fig_potdelgada}}
\end{figure}

\begin{figure}[tb]
\centering
\includegraphics[width=0.49\textwidth]{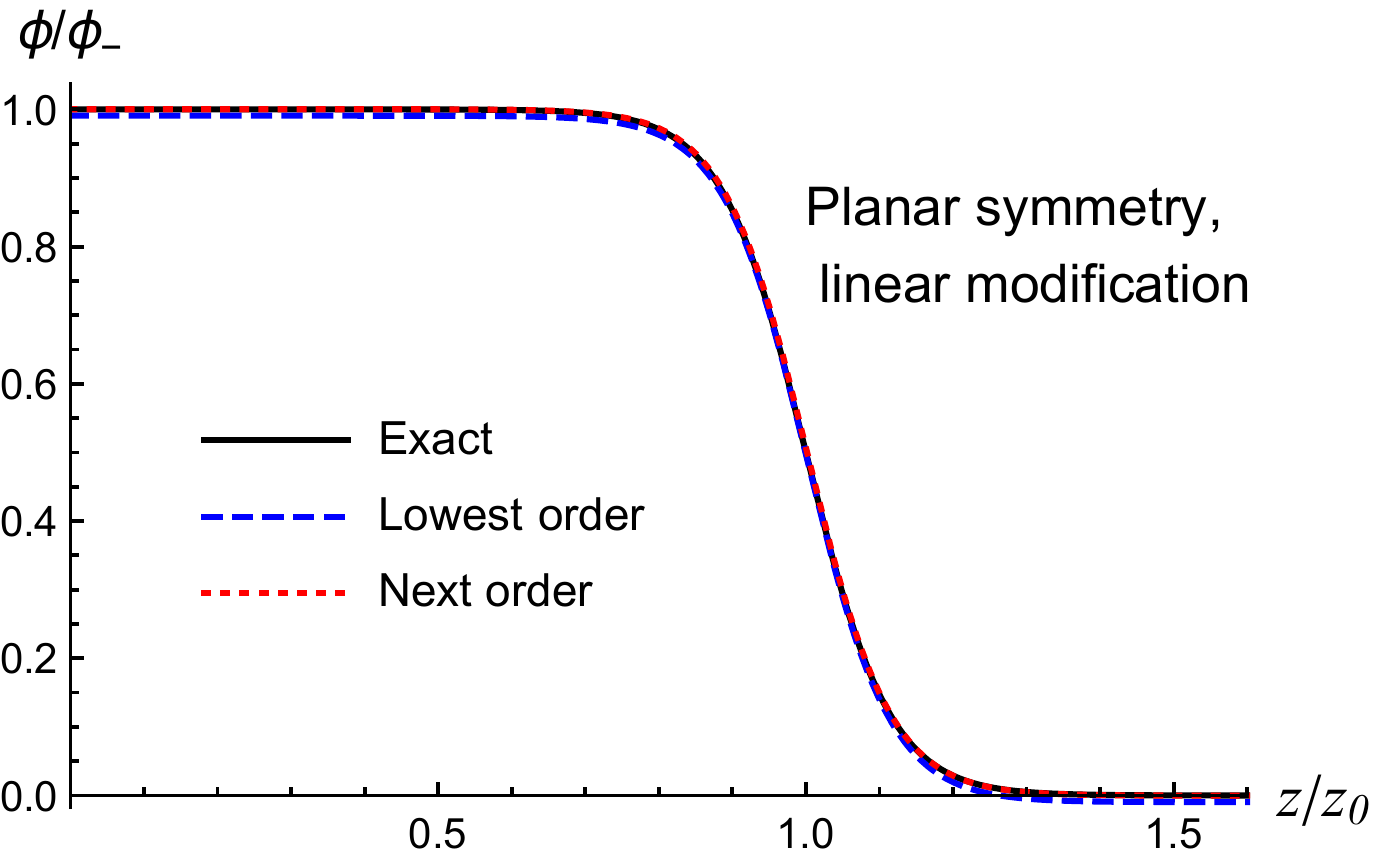}
\hfill
\includegraphics[width=0.49\textwidth]{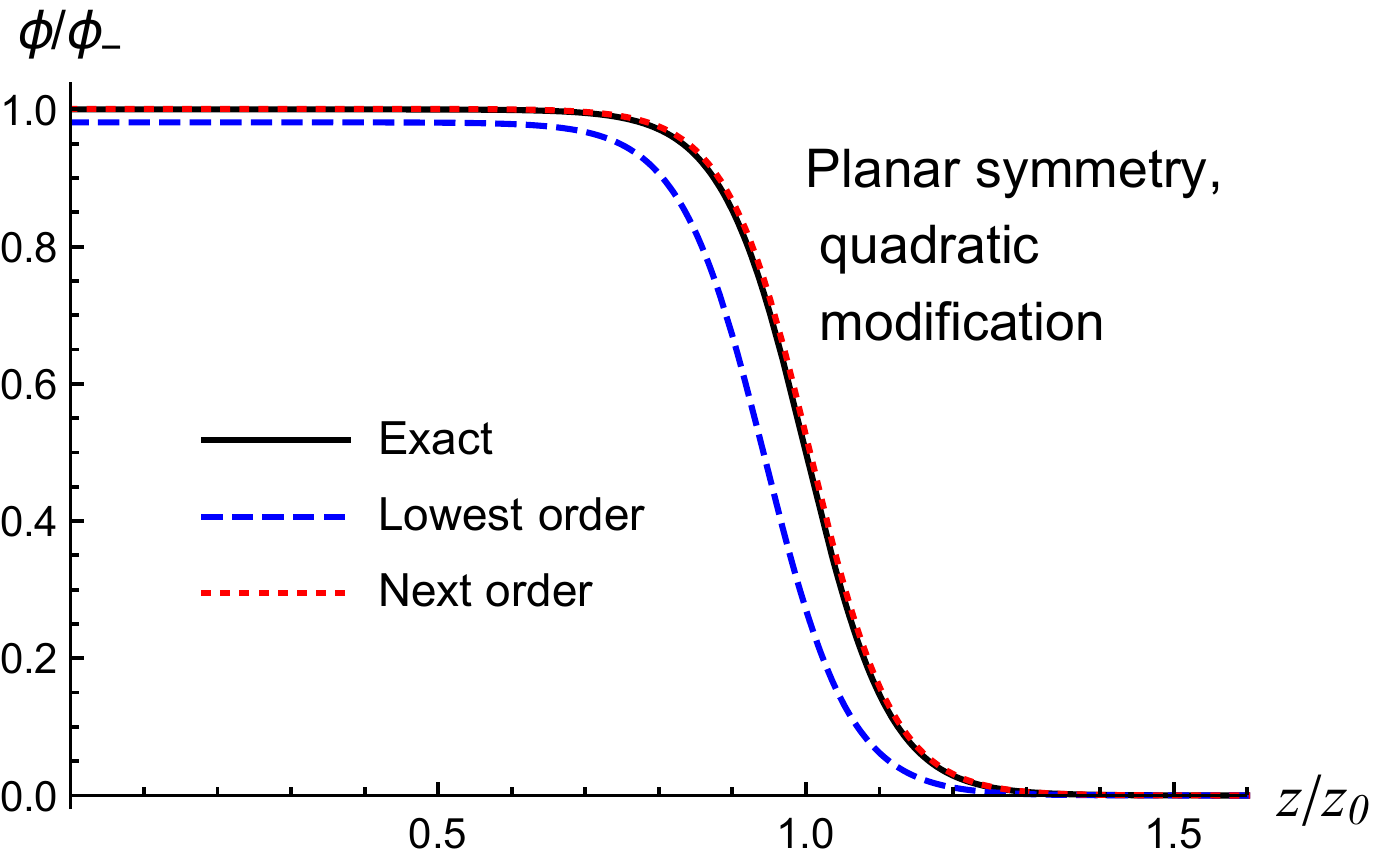}
\caption{The planar-symmetry solution at $t=0$ for the potential of Fig.~\ref{fig_potdelgada} 
and the approximations for the linear $V_{1}$ (left) and for the quadratic $V_{1}$ (right).\label{fig_casodelgada}}
\end{figure}

According to Eq.(\ref{DVVmax}), we expect the thin-wall approximation to
work well when $\Delta V/V_{\max}\ll1$. As a first example, let us
consider the potential plotted in Fig.~\ref{fig_potdelgada} (solid line), for
which we have $\Delta V/V_{\max}\simeq0.35$. The dashed line corresponds
to the potential $V_{0}$ of Eq.~(\ref{potdeglin}). 
In the left panel of Fig.~\ref{fig_casodelgada}, we show the exact
solution for $\phi$ a $t=0$, together with the approximations $\phi_{0}$
and $\phi_{0}+\phi_{1}$. Here we consider only the planar case since
the spherical case is very similar. We see that the approximation
is very good even to the lowest order. This is a particular feature
of the polynomial potential in combination with the linear modification.

If we use instead a quadratic modification $V_{1}=\frac{1}{2}\delta m^{2}\phi^{2}$,
most of the expressions are simpler. We have $\delta m^{2}=m^{2}-2e^{2}/9\lambda$,
$a_{+}=0$, $a_{-}=2e/3\lambda\equiv a$, and 
\[
V_{0}=\frac{\lambda}{4}\phi^{2}\left(\phi-a\right)^{2}.
\]
The lowest-order thin-wall approximation for the kink profile is given
by 
\begin{equation}
\phi_{0}=a\left[1+\exp\left(\sqrt{\lambda/2}\,a\,n\right)\right]^{-1},
\end{equation}
and the surface tension is given by $\sigma_{0}=\frac{1}{3}\sqrt{\lambda/8}a^{3}$.
In this case we have $a_{\text{+}}=\phi_{+}$, so $\phi_{0}$ matches
the exact value of $\phi$ outside the bubble. However, the difference
$a_{-}-\phi_{-}$ is higher than with the linear modification, as
can be seen in the right panel of Fig.~\ref{fig_casodelgada} (dashed
line). The wall position also has a larger error, which is related to the error
in the surface tension through the relation $r_{w}=j\sigma/\Delta V$
(at $t=0$). For the present case we have $(\sigma-\sigma_{0})/\sigma\simeq0.05$.
Nevertheless, to the next order we obtain a non-vanishing correction
$\delta\sigma=-\delta m^{2}a/\sqrt{2\lambda}$, which decreases the
error, $(\sigma-\sigma_{1})/\sigma\simeq-0.006$. The correction to
the profile is given by 
\begin{equation}
\phi_{1}=-\frac{2\delta m^{2}}{\lambda a^{3}}\phi_{0}\left[a+\left(a-\phi_{0}\right)\left(4-\ln\frac{a-\phi_{0}}{\phi_{0}}\right)\right].
\end{equation}
 As can be seen in Fig.~\ref{fig_casodelgada}, the approximation
is quite better to this order. 

\begin{figure}[tb]
	\centering
	\includegraphics[width=0.5\textwidth]{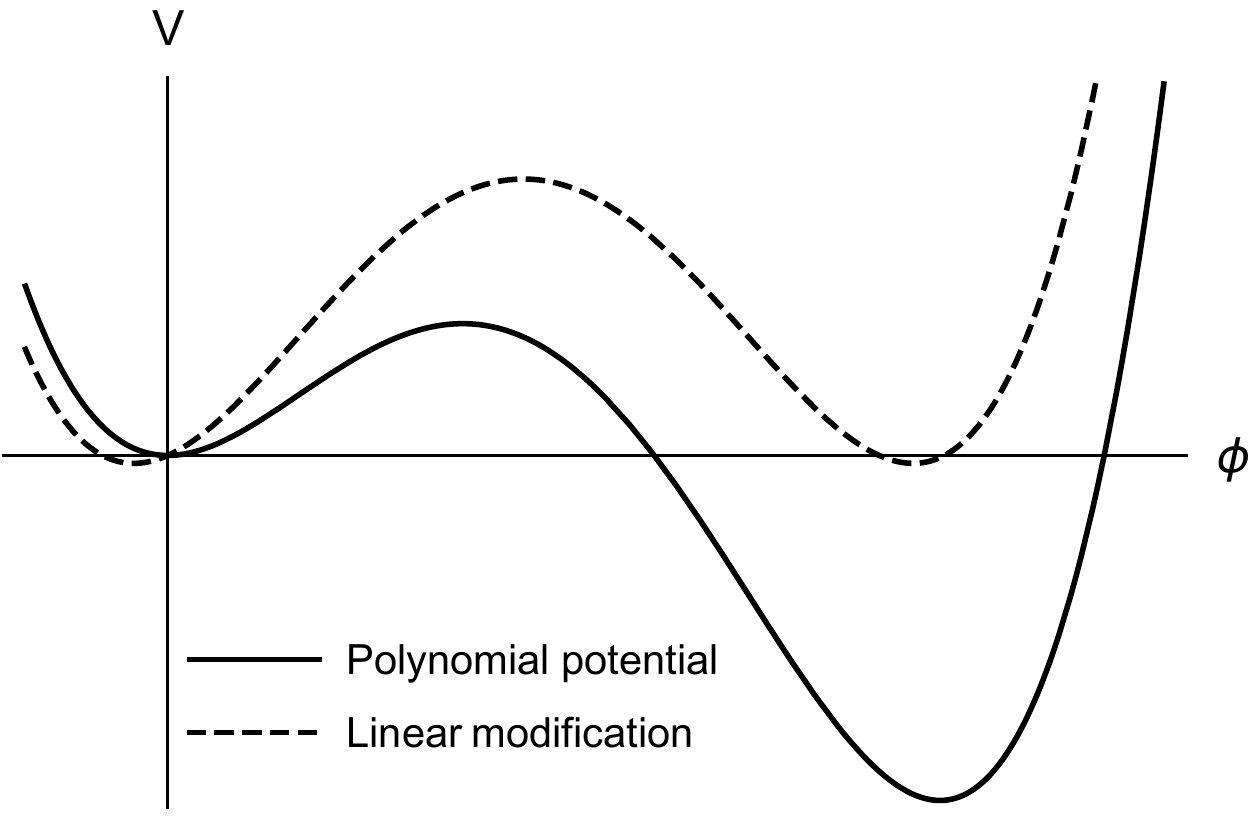}
	\caption{The polynomial potential (\ref{pot}) with parameters such that
		$\Delta V/V_{\max}\simeq2.6$ and the potential $V_{0}$ obtained
		with a linear $V_{1}$.\label{fig_potgruesa}}
\end{figure}

\begin{figure}[tb]
	\includegraphics[width=0.49\textwidth]{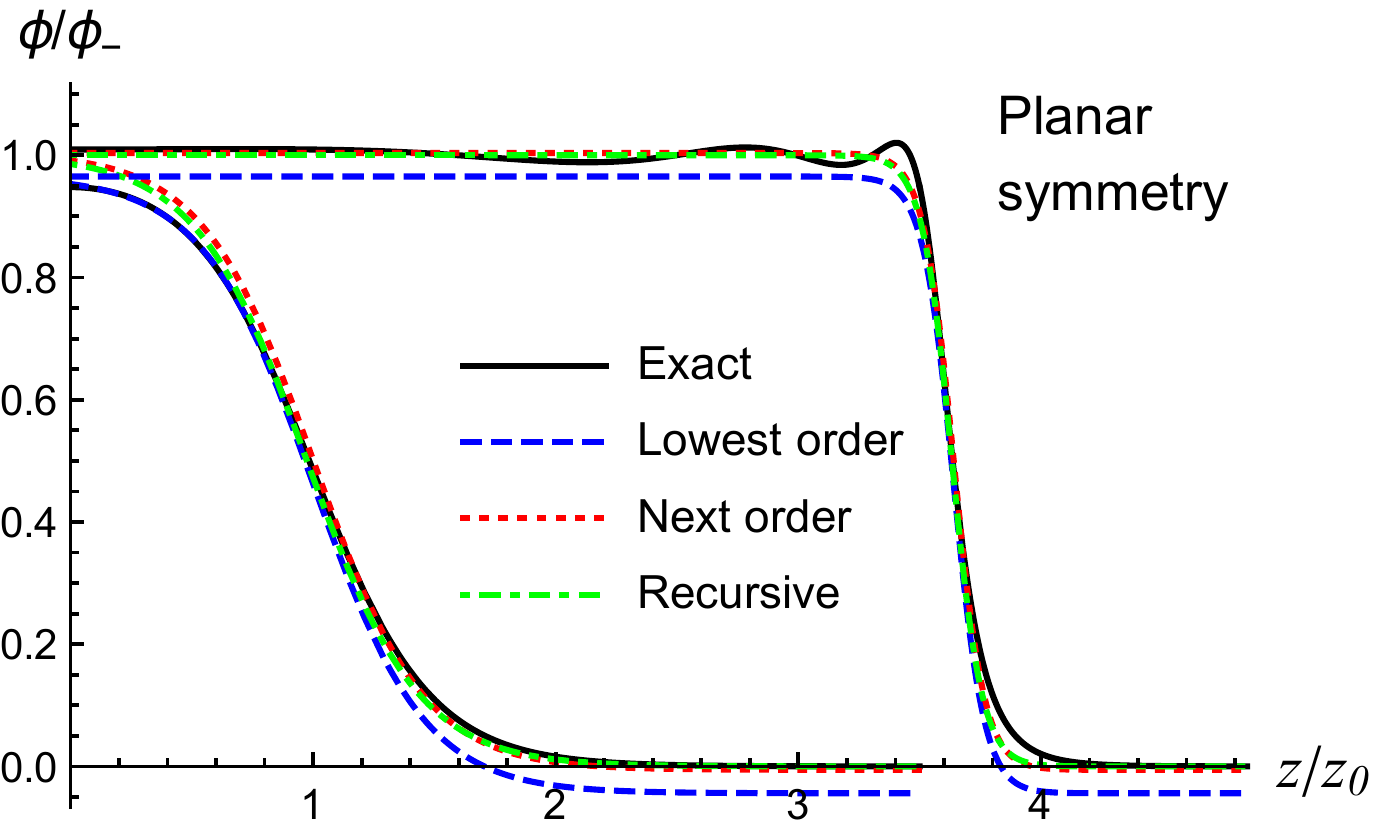}
	\hfill
	\includegraphics[width=0.49\textwidth]{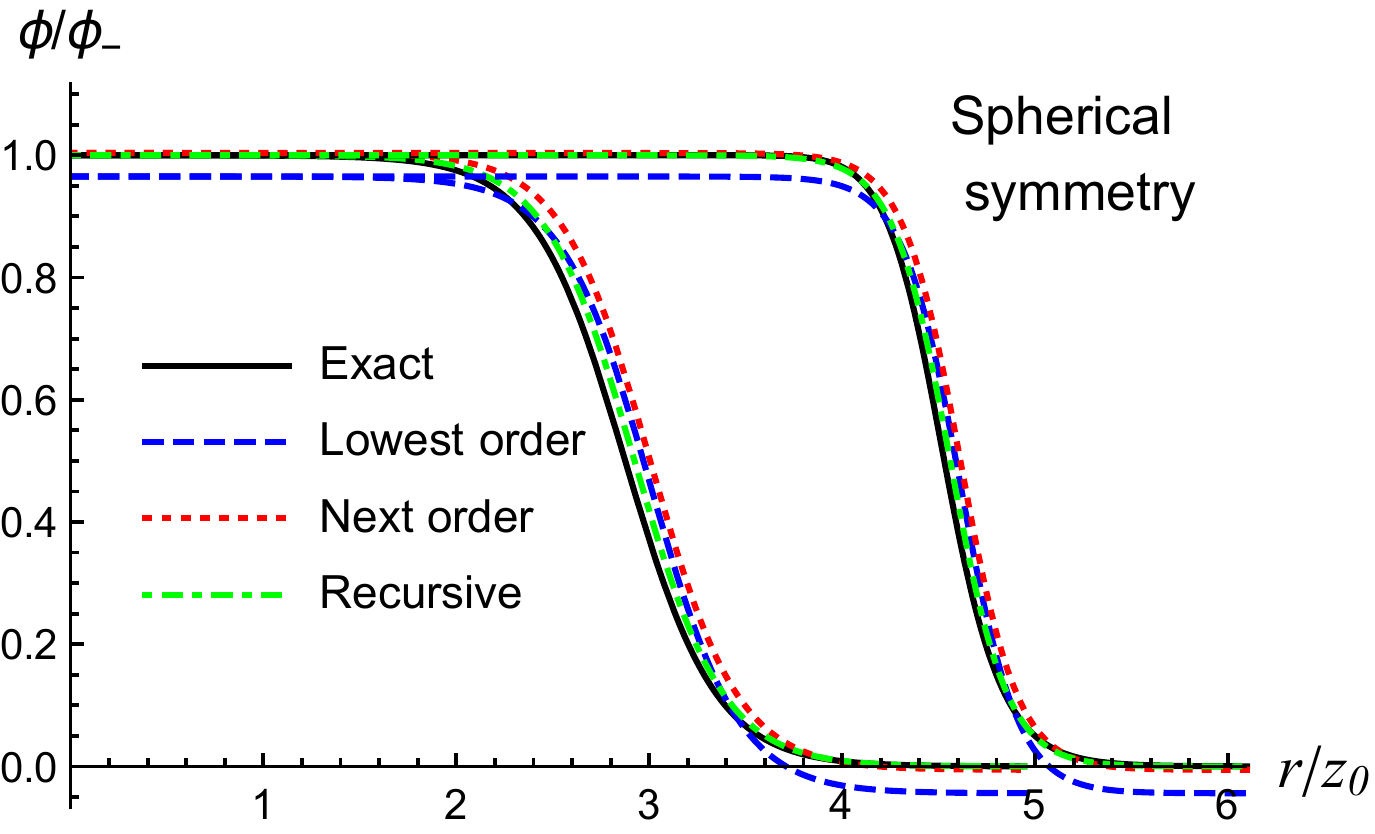}
	\caption{Numerical solutions and their approximations at $t=0$ and $t=3.5z_0$ for the potential of Fig.~\ref{fig_potgruesa}.%
		\label{fig_casogruesa}}
\end{figure}

Let us now test the approximation for a potential which departs significantly
from the degenerate case, namely, $\Delta V/V_{\max}\simeq2.6$,
shown in Fig.~\ref{fig_potgruesa}. 
We consider both the planar
and spherical cases, but only the linear modification. We have verified
that the quadratic modification gives similar results. 
In Fig.~\ref{fig_casogruesa}, we plot the
wall profiles at 
nucleation and at a subsequent time.
Despite the fact that the potential difference
between the minima is quite large, we see that the approximations still
work quite well, for both the planar and spherical cases. 
In particular, the
approximation $\phi_{0}$ fits quite well the shape and position of the wall, while
the correction $\phi_{1}$ improves significantly the values of
the boundary conditions $\phi=\phi_{\pm}$.
On the other hand, 
the solution obtained from the recursive method discussed
in Sec.~\ref{sec:masalla} matches these values with arbitrary precision.
For the planar case, the analytic approximation $\sigma_0$
has an error of 15\%, while for the spherical case the error is 5\%. 

\begin{figure}[tb]
	\includegraphics[width=0.49\textwidth]{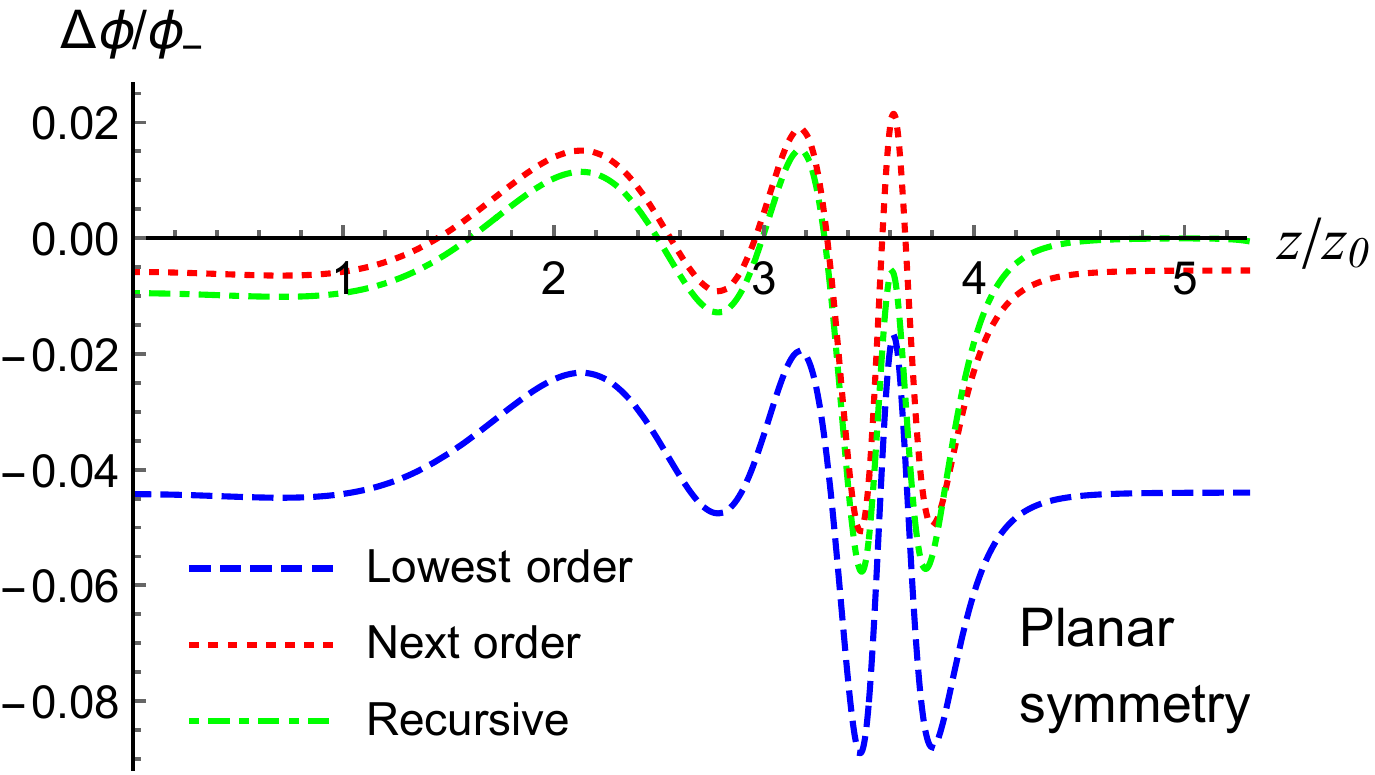}
	\hfill
	\includegraphics[width=0.49\textwidth]{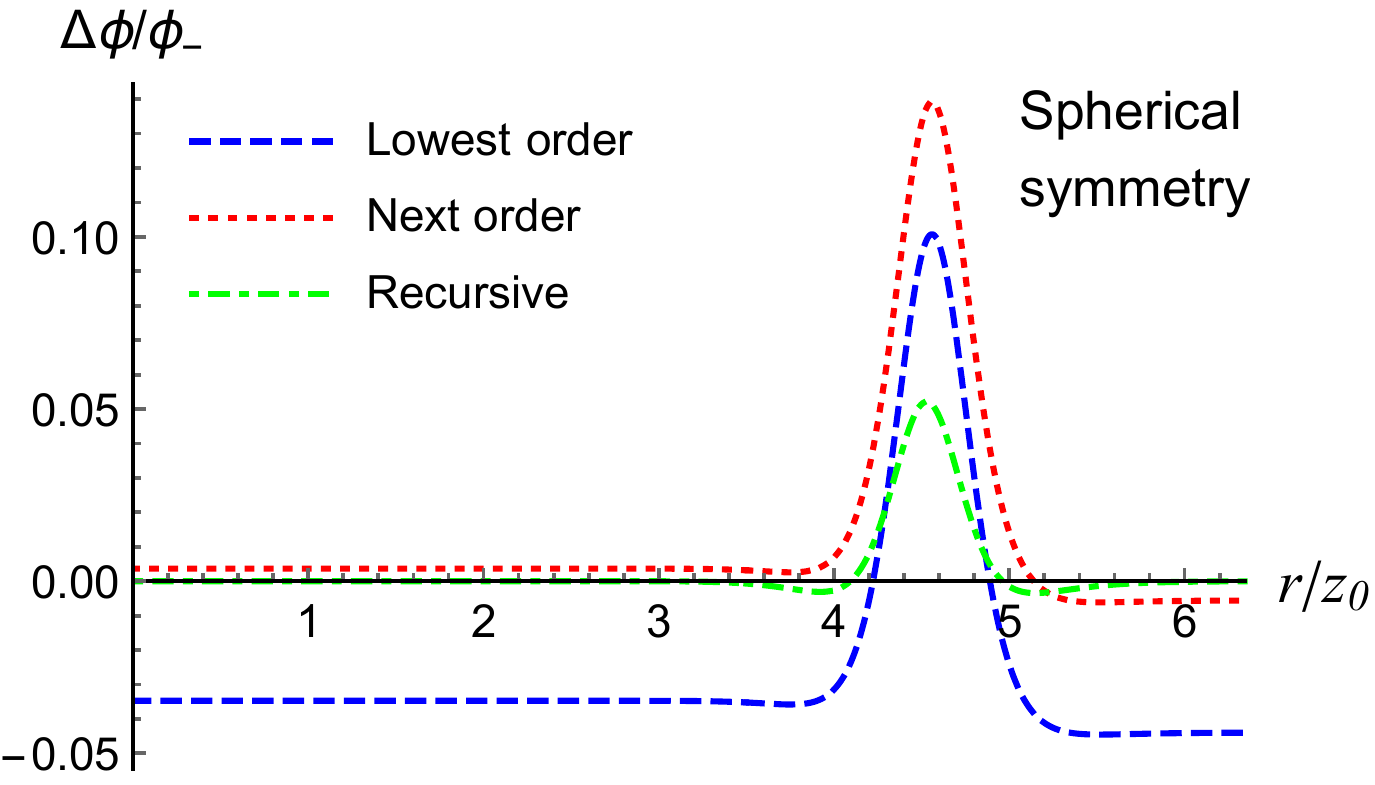}
	\caption{The error of the approximations in Fig.~\ref{fig_casogruesa} at $t=3.5z_0$, normalized to the value $\phi_-$.%
	\label{fig_casogruesadif}}
\end{figure}

In Fig.~\ref{fig_casogruesadif} we plot the differences $\Delta\phi$ between the approximations and the numerical solution, divided by the value $\phi_-$ (for the case $t>0$ of Fig.~\ref{fig_casogruesa}).
For the planar case, we see that the next order approximation is very close to the recursive approximation. 
Their errors are below a 5\% (relative to $\phi_-$). 
Of course, these approximations do not reproduce the oscillatory behavior of the field inside the bubble.
For the spherical case, the recursive approximation is clearly better than the perturbative ones.
On the other hand, the next-order correction improves the approximation for the boundary values $\phi_\pm$
with respect to the lowest order, 
but worsens the approximation for the field inside the wall. 
This is because in this case the correction $\phi_1$ is just a constant
(the shape of the profile is the same at the two orders).

\begin{figure}[tb]
	\centering
\includegraphics[width=0.5\textwidth]{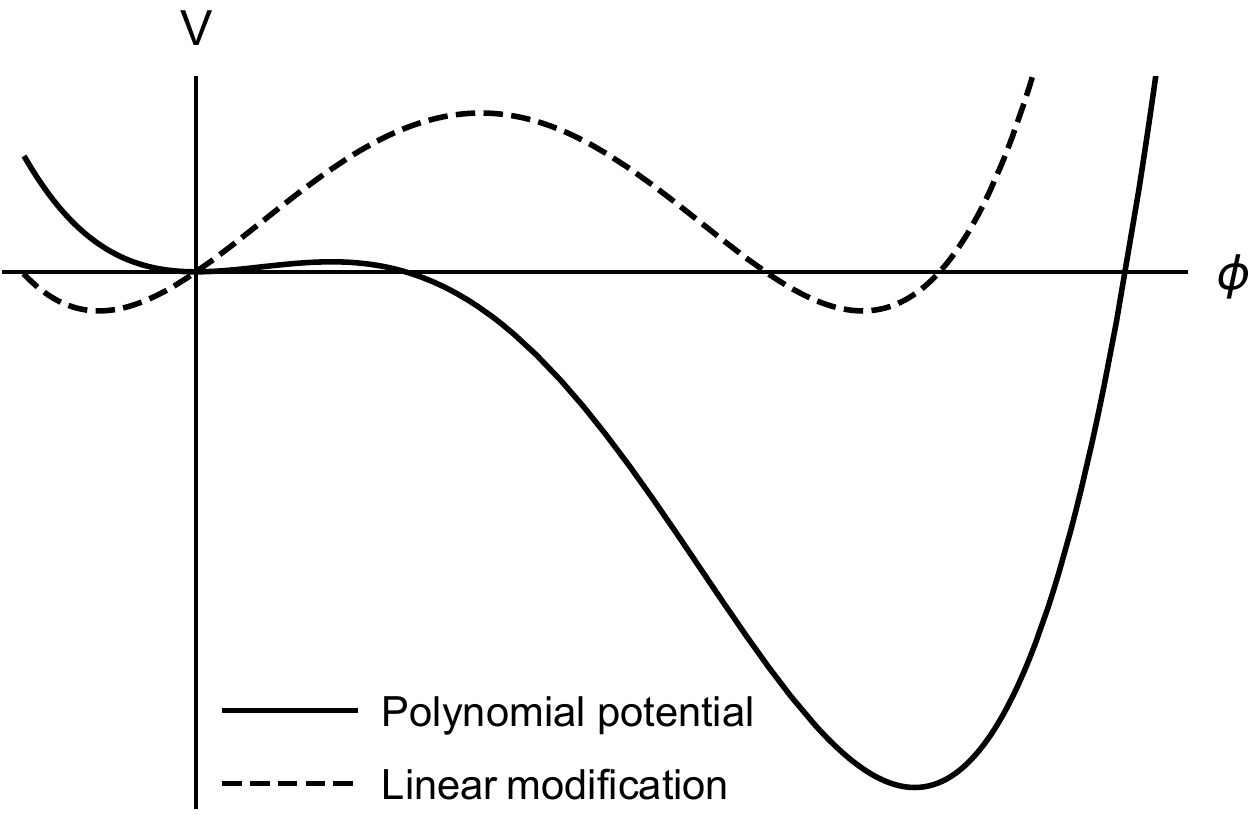}
\caption{The polynomial potential (\ref{pot}) with $\Delta V/V_{\max}\simeq52$
and the degenerate potential $V_{0}$ for a linear term $V_{1}$.\label{fig_potmuygruesa}}
\end{figure}

In order to test the limits of the approximations, we shall also consider
the extreme case where the potential difference between the minima
is much larger than the height of the barrier. This situation, which
is close to spinodal decomposition, is not very usual in a thermal
phase transition, where the potential is degenerate at the critical
temperature. As the temperature decreases, the height of the barrier
decreases and the energy difference between the minima increases,
but bubble nucleation usually takes place as soon as the barrier reaches
a height comparable to the potential difference. We consider the potential
shown in Fig.~\ref{fig_potmuygruesa}, corresponding to $\Delta V/V_{\max}\simeq52$,
and a linear modification term. 

\begin{figure}[tb]
	\includegraphics[width=0.49\textwidth]{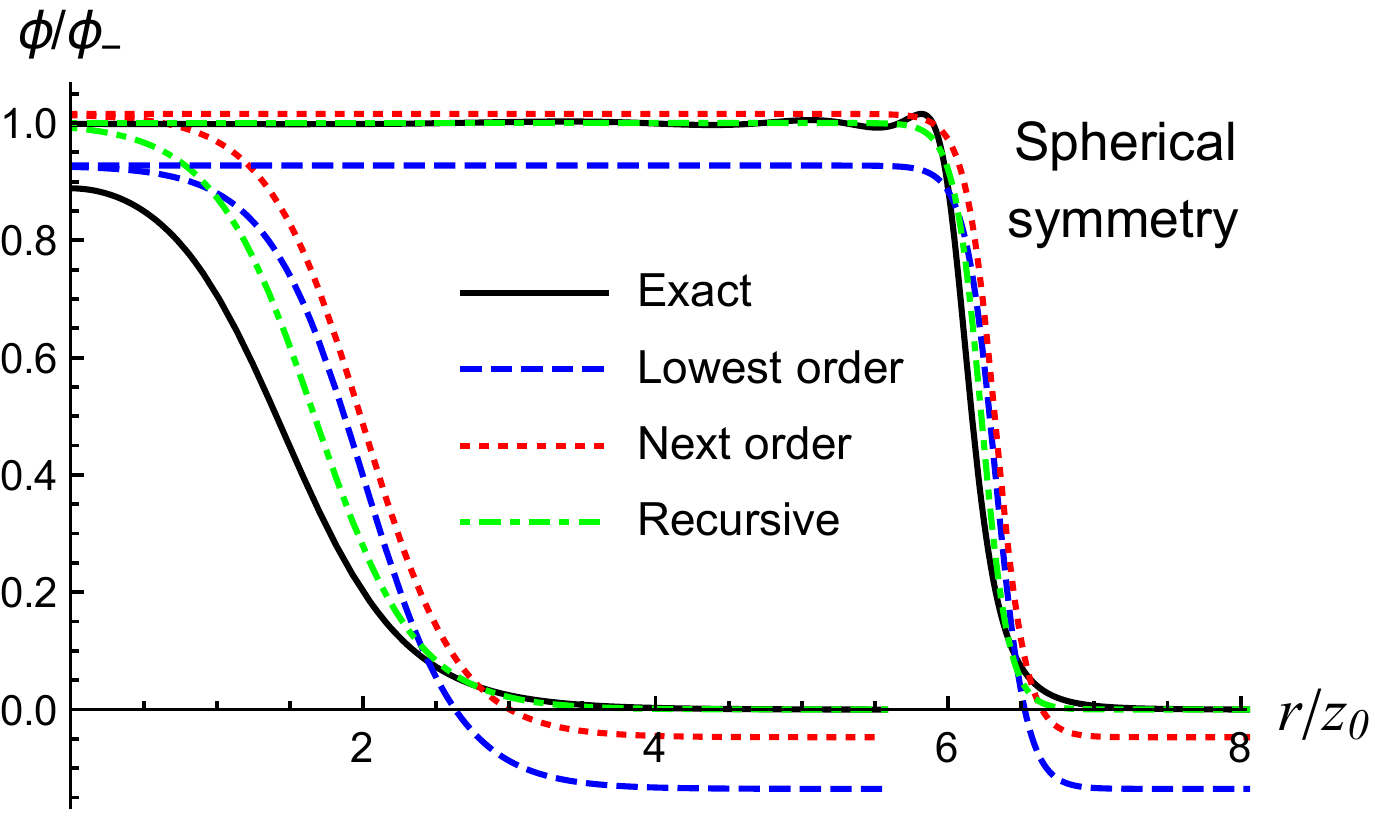}
	\hfill
	\includegraphics[width=0.49\textwidth]{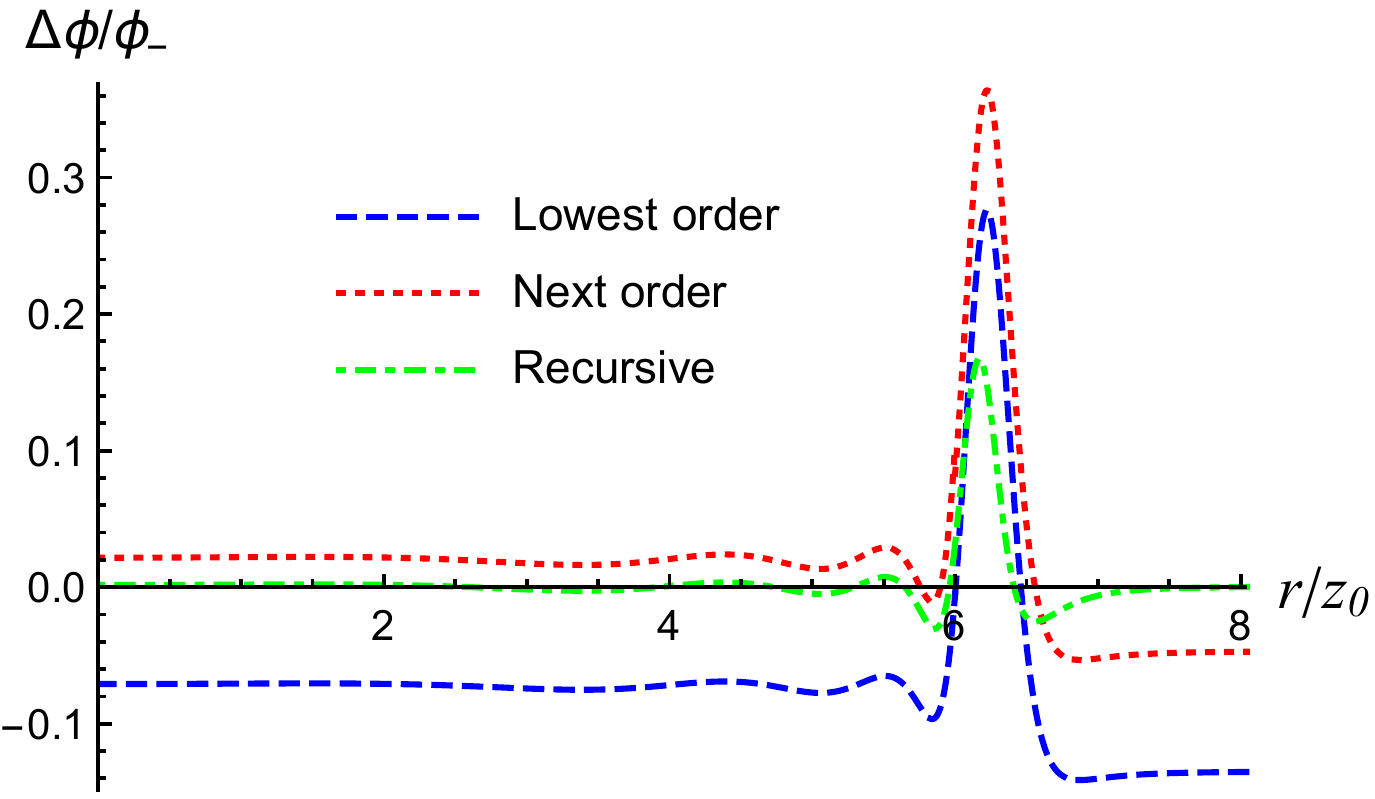}
	\caption{Left: The numerical solution for the spherical case and its analytic approximations
		at $t=0$ and at $t=6z_0$ for the potential of Fig.~\ref{fig_potmuygruesa}. Right: The error of the approximations at $t=6z_0$, normalized to the value $\phi_-$.\label{fig_muygruesa}}
\end{figure}

We plot the solution for the spherical case at two times in the left panel of
Fig.~\ref{fig_muygruesa}. 
We discuss the planar case in App.~\ref{interior}
(in that case, the exact solution has stronger oscillations of the field).
The right panel of Fig.~\ref{fig_muygruesa} shows the differences between the approximations and the exact solution
divided by $\phi_-$ for the case
when the field has already reached the value $ \phi_- $ at the bubble center. 
We see that the next order
approximation is quite good, and even the lowest order approximation
is not so bad, depending on the precision needed. 
Again, the recursive solution approximates better the wall
profile and matches the boundary values $\phi_{\pm}$ with arbitrary
precision. 

\subsubsection{Coleman-Weinberg potential}

\begin{figure}[tb]
	\centering
\includegraphics[width=0.45\textwidth]{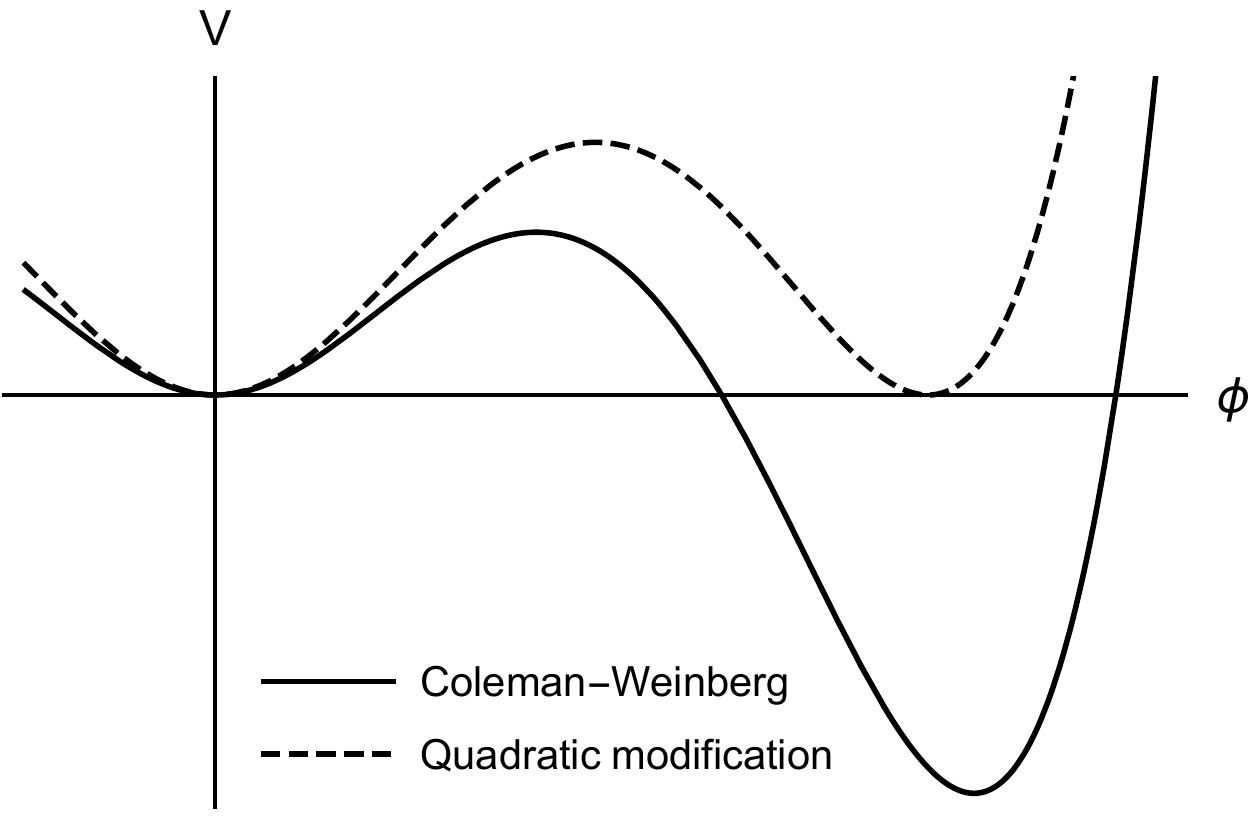}
\caption{The CW potential (\ref{potcw}) with $\Delta V/V_{\max}\simeq2.45$
and the degenerate potential $V_{0}$ for a quadratic $V_{1}$.\label{fig_potgr_cw}}
\end{figure}

For the polynomial potential (\ref{pot}), we have obtained analytical
expressions for the thin-wall approximation at different orders. 
However, one may wonder
whether this potential is too simple to represent a general situation.
Therefore, we shall now consider a Coleman-Weinberg (CW) potential
with a mass term (for a review and references, see \cite{w15}),
\begin{equation}
V(\phi)=\alpha\left[\phi^{4}\left(\ln\left(\phi^{2}/v^{2}\right)-\frac{2+\beta}{4}\right)+\frac{\beta}{2}v^{2}\phi^{2}\right],\label{potcw}
\end{equation}
where $\alpha$ and $\beta$ are positive parameters. The minima are
$\phi_{+}=0$ with $V_{+}=0$ and $\phi_{-}=v$ with $V_{-}=\frac{1}{4}\alpha\left(\beta-2\right)v^{4}$.
We have $V_{-}<V_{+}$ for $\beta<2$. Adding to this potential a
linear term $V_{1}=c\phi$, we obtain a degenerate potential only
if the difference $V_{-}-V_{+}$ is not too large. This is because
the minimum $a_{+}$ disappears if the coefficient $c$ becomes too
large. Therefore, we shall consider the quadratic modification, which
additionally simplifies significantly the expressions. We have verified
that in the cases where the linear modification is possible the results
are similar to those obtained with the quadratic modification. Adding
the term $V_{1}=\frac{1}{2}\delta m^{2}\phi^{2}$, the minimum at
$\phi=0$ remains unchanged, i.e., $a_{+}=0$. Denoting $a_{-}\equiv a$,
we have
\begin{equation}
\delta m^{2}=\alpha\left(\beta v^{2}-2a^{2}\right),\quad a^{2}=v^{2}e^{\left(\beta-2\right)/4},
\end{equation}
and
\begin{equation}
V_{0}=\alpha\phi^{2}\left[\phi^{2}\left(\ln\left(\phi^{2}/a^{2}\right)-1\right)+a^{2}\right].
\end{equation}
The potentials $V$ and $V_{0}$ are shown in Fig.~\ref{fig_potgr_cw} for $\Delta V/V_{\max}\simeq2.45$.

\begin{figure}[tb]
	\includegraphics[width=0.49\textwidth]{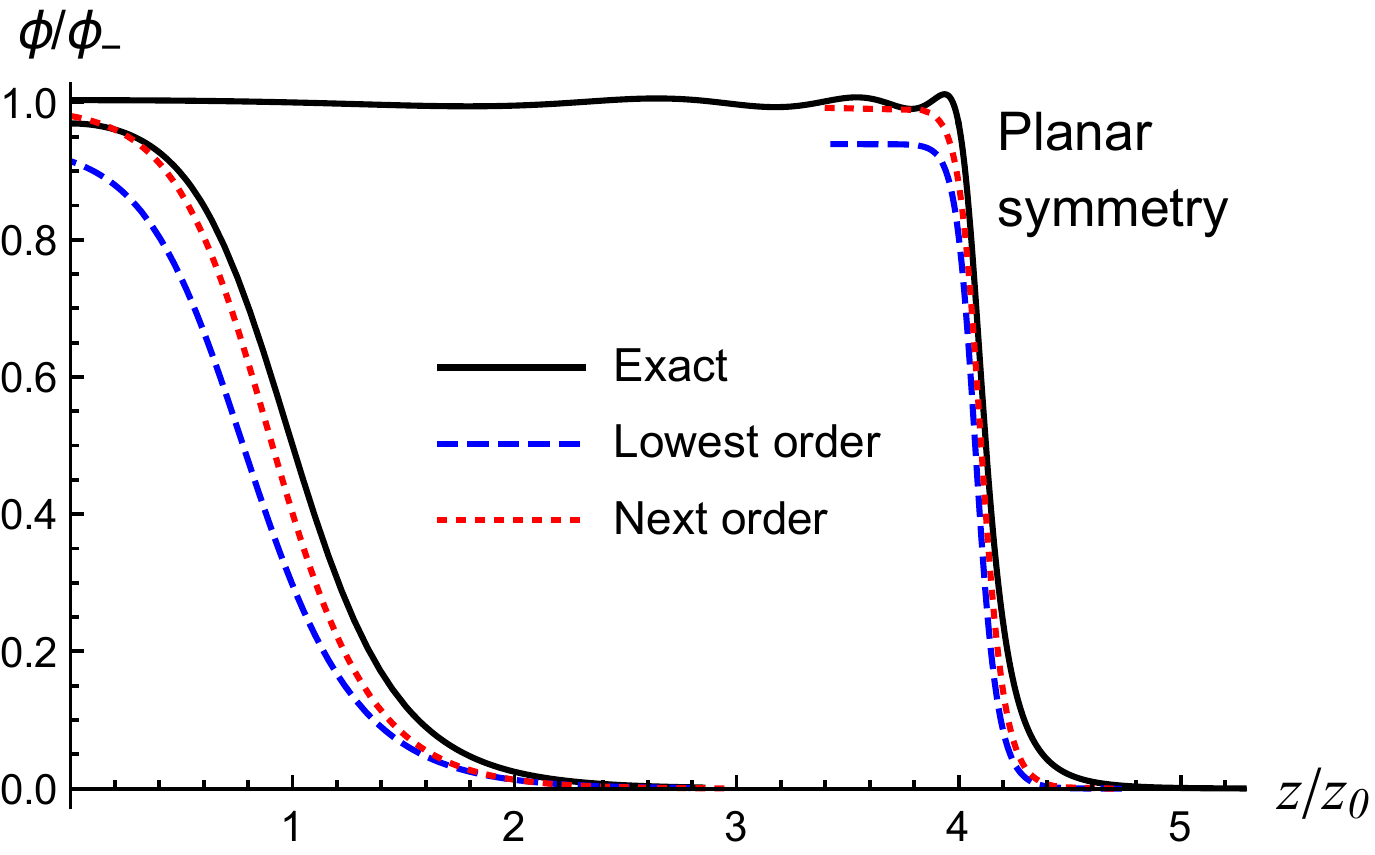}
	\hfill
	\includegraphics[width=0.49\textwidth]{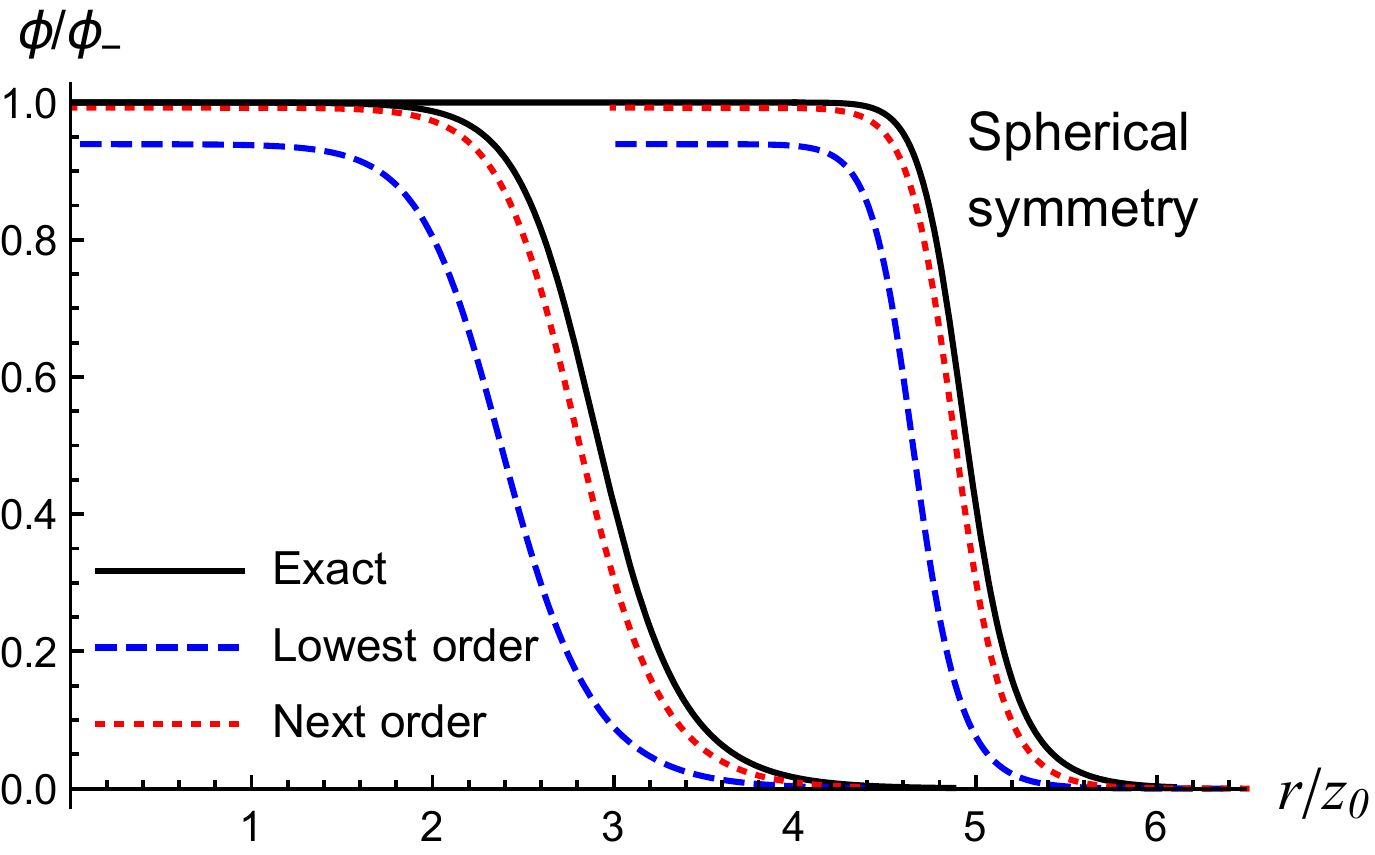}
	\caption{The solutions for the potential of Fig.~\ref{fig_potgr_cw} and their thin-wall approximations
		at $t=0$ and $t=4z_0$.\label{fig_gruesa_cw}}
\end{figure}

The zeroth-order profile and surface tension are given by Eqs.~(\ref{formulaperfil})-(\ref{formulasigma}).
Using $\phi_{*}=a/2$, we obtain
\[
n=\frac{1}{2\sqrt{2\alpha}a}\left(I_{n_{0}}-I_{n}(\phi_{0}^{2}/a^{2})\right),\quad\sigma_{0}=\sqrt{\alpha/2}a^{3}I_{\sigma},
\]
where the $I_{k}$ are integrals which do not depend on the parameters,
\begin{align}
I_{\sigma} & =\int_{0}^{1}\sqrt{x\ln x-x+1}\,dx\simeq0.4199,\\
I_{n}(x) & =\int_{\ln\left(1/4\right)}^{\ln x}\frac{dy}{\sqrt{e^{y}\left(y-1\right)+1}},\\
I_{n_{0}} & =I_{\sigma}^{-1}\int_{0}^{1}\sqrt{x\left(\ln x-1\right)+1}I_{n}(x)\,dx\simeq0.05460.
\end{align}
The leading-order corrections are given by Eqs.~(\ref{deltasigma})-(\ref{formulaC}).
We obtain
\begin{align}
\phi_{1} & =\frac{\delta m^{2}}{8\alpha a^{2}}\phi_{0}\sqrt{\left(\phi_{0}/a\right)^{2}\left[\ln\left(\phi_{0}^{2}/a^{2}\right)-1\right]+1}\,\left(I_{\phi}(\phi_{0}^{2}/a^{2})+2I_{C}\right),\\
\delta\sigma & =\delta m^{2}aI_{\delta\sigma}/4\sqrt{2\alpha},
\end{align}
where
\begin{equation}
I_{\phi}(x)=\int_{1/4}^{x}\frac{x'-G(x')}{\left(x'\ln x'-x'+1\right)^{3/2}}\frac{dx'}{x^{\prime2}},
\end{equation}
with $G(x)=I_{\sigma}^{-1}\int_{0}^{x}\sqrt{x'\ln x'-x'+1}dx'$, 
\begin{equation}
I_{C} = 
I_{\sigma}^{-1} \int_{0}^{1}I_{\phi}(x)\left[\sqrt{x\ln x-x+1}-\left(x\ln x+\frac{1-x}{2}\right)\left(I_{n_{0}}-I_{n}\right)\right]dx
\simeq -4.877,
\end{equation}
and
\begin{equation}
I_{\delta\sigma}=\int_{0}^{1}\frac{x-G(x)}{\sqrt{x\ln x-x+1}}\frac{dx}{x}\simeq -1.405.
\end{equation}
The results for the potential of Fig.~\ref{fig_potgr_cw} are shown in Fig.~\ref{fig_gruesa_cw}. 
We see that the thin-wall approximation is significantly improved by the leading-order correction.
The recursive solution is very close to the latter, so we do not show it here.
The value $\sigma_{0}$ approximates the surface tension
with an error of about 11\% for the planar case and 17\% for the spherical
case. With the correction $\delta\sigma$, the error for the planar
case remains in the same order, while for the spherical case decreases
to 4\%.

\begin{figure}[tb]
\includegraphics[width=0.49\textwidth]{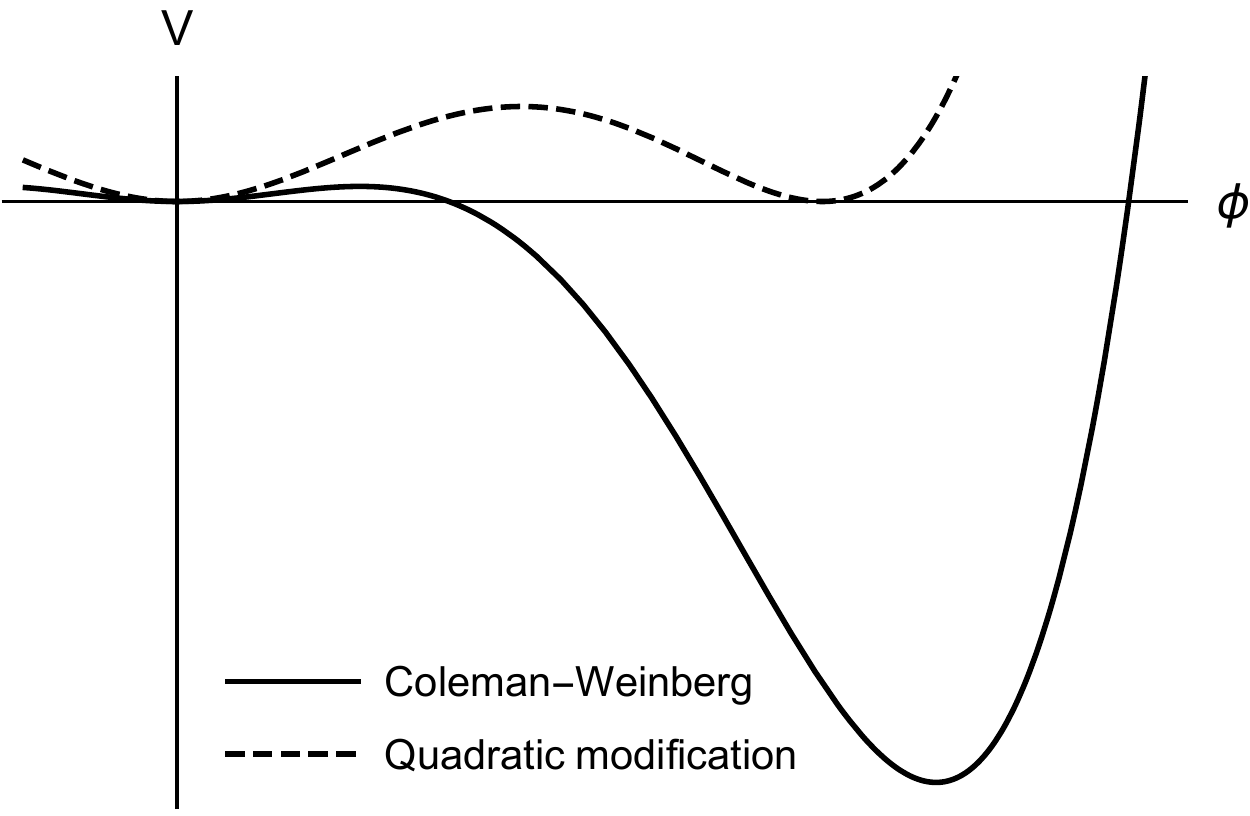}
\hfill
\includegraphics[width=0.49\textwidth]{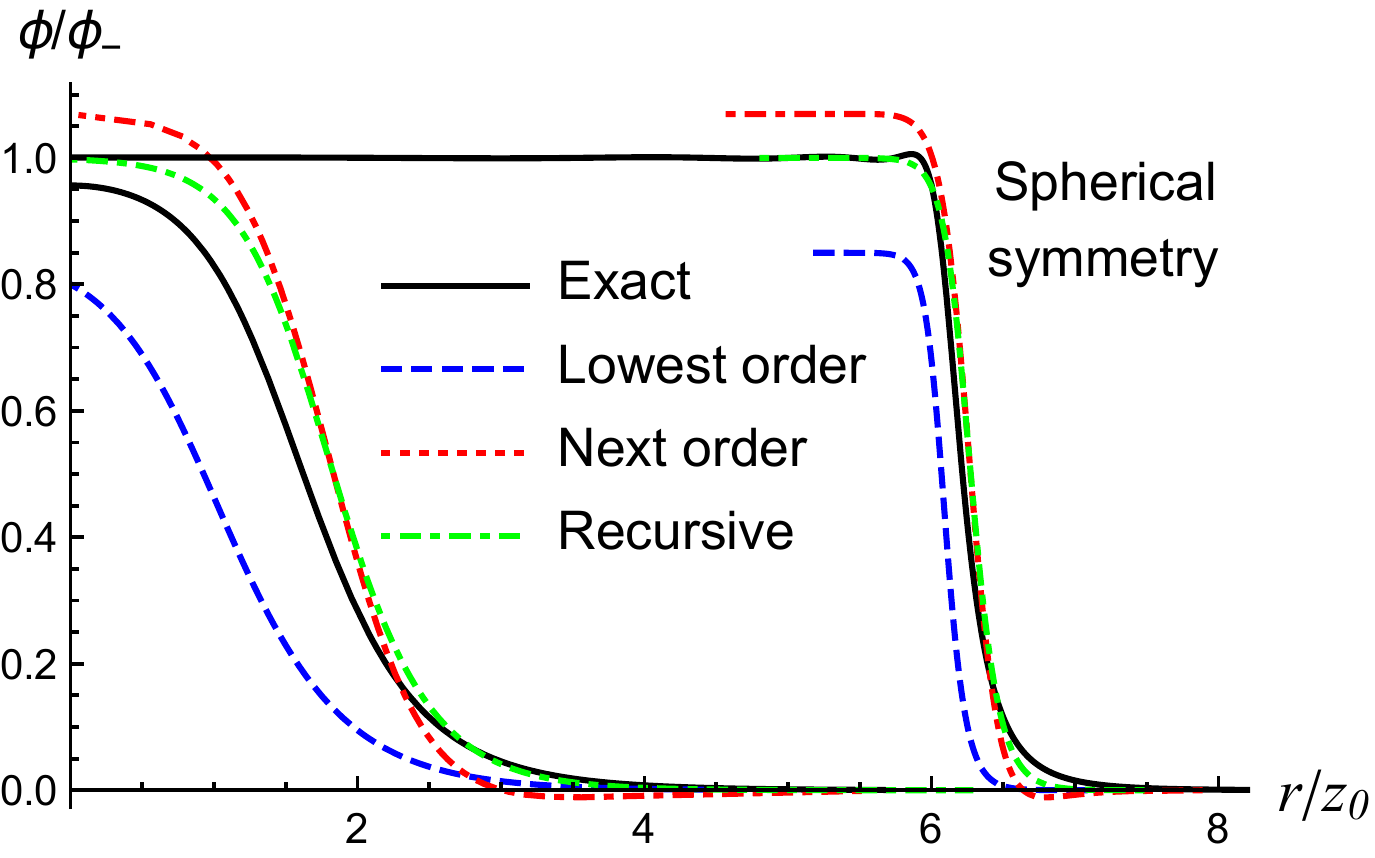}
\caption{The CW potential (\ref{potcw}) with $\Delta V/V_{\max}\simeq38$, 
	and the numerical solutions and its approximations at $t=0$ and at $t=6z_0$.\label{fig_muygruesacw}}
\end{figure}

Finally, in Figure \ref{fig_muygruesacw} we present the results
for a potential with a large value of
$\Delta V/V_{\max}\simeq38$. As expected, the approximations work
better when the field has reached the value $\phi_{-}$ inside the
bubble. The next-order approximation is quite better than the lowest-order
one, and that obtained with the recursive method is much better. 

\section{Conclusions}

\label{Conclu}

Most of the possible cosmological relics of a first-order phase transition 
depend on the propagation of bubble walls.
Two basic assumptions are often used to study those relics, namely, that the walls are infinitely thin and spherical (or even planar).
We have studied the dynamics of phase-transition bubbles beyond these
approximations. 

In the first place, we have identified and discussed a number of approximations
that are actually made when the so-called thin-wall approximation
is used. Besides neglecting the evolution of the field $\phi$ inside
the bubble, these approximations imply three main assumptions regarding
the wall profile, namely, that $\phi$ depends on a single variable
$n$ (the physical distance perpendicular to the hypersurface swept
by the wall position), that the function $\phi(n)$ varies much more 
than other quantities (which are thus assumed to be constant in the wall region), 
and that the local curvature radius
of the surface is much larger than the wall width. This latter assumption
allows to drop a term in the equation for $\phi(n)$. However, this
approximation in the field equation requires approximating the potential
$V(\phi)$ by a degenerate potential $V_{0}(\phi)$. This is because
the potential difference $\Delta V$ between the minima causes a curvature
of the hypersurface. We have discussed the appropriate construction
of the potential $V_{0}$ from the given potential $V$ by adding
a term $V_{1}$. The later is not unique and can be chosen conveniently.

We have discussed the Monge parametrization for the wall surface,
which gives an equation of motion for the wall position in a conveniently
chosen coordinate system. This equation is useful for dealing with
arbitrary deformations from a given initial shape, such as a spherical
bubble. This treatment can be applied to the surface corrugations resulting
from hydrodynamic instabilities \cite{l92,hkllm93,mm14} beyond the linear regime.
Furthermore, the expression for the stress-energy tensor in the Monge
representation can be directly applied to the method developed in
\cite{mm21a} for the calculation of gravitational waves. 

On the other hand, we have discussed two methods for improving the
basic thin-wall approximations. The first one consists in using only
the first two assumptions and calculate the field profile and the mean curvature $K$ 
(without any assumptions on the magnitude of the latter) with an iterative method. 
The second
approach consists in calculating finite-width corrections to the thin-wall
approximations. 
This method can give analytic results for the perturbative corrections to
any order. 
This approach has been used for domain walls, where the leading-order corrections
to the wall profile and equation of motion are of second order in
the wall width. For a bubble wall, in contrast, we have found that
there are non-vanishing first-order corrections. Nevertheless, to
this order the wall equation of motion keeps the same form as in the
lowest-order approximation, and only the parameter $\sigma$ (the
surface tension) changes. We shall discuss the higher order corrections
in a separate paper \cite{mm2}. 

Our methods are model-independent and provide an analytic treatment
beyond the thin-wall approximation. We have checked the approximations
to the wall profile and its equation of motion for different specific
potentials $V$ and using different modifications $V_{1}$ to obtain
the degenerate potential $V_{0}$. As specific examples we considered
spherical and planar walls, since the wall acceleration already causes
a curved hypersuface. We remark that the space-time curvature radius
$L$ depends on the potential difference $\Delta V$ and can be much
smaller than the spatial radius $R$ of the bubble. Furthermore, the
scale $L$ must be compared to the wall width $l$ in a system which
moves with the wall (while the wall width is Lorentz-contracted with
respect to the reference frame of the bubble center). 
Roughly, we have the relation $l/L\sim\Delta V/V_{\max}$,
where $V_{\max}$ is the height of the potential barrier. For the
specific examples, our method gives quite good approximations to the
numerical solution, even for $\Delta V/V_{\max}\sim1$.

In this paper we have considered the case of a vacuum phase transition.
In the presence of a hot plasma, the interaction of the scalar field
with the fluid introduces effects such as friction and bulk fluid
motions. Different levels of approximation can be used to deal
with this case, from adding an effective friction term 
to numerically solving the equations for the field-fluid
system in a lattice. In a forthcoming paper \cite{mm3} we shall extend
our analytic treatment to include the effects of the fluid.

To end, we wish to discuss the importance of the approximations investigated in this paper.
For a complete treatment of a phase transition, one has to consider the dynamics of many bubbles which nucleate, 
expand, and percolate, 
interacting with each other through collisions, 
reheating the plasma through the release of latent heat, and causing bulk fluid motions.
The consequences of the phase transition depend on the details of this global dynamics.
Ultimately, the most complete treatment will probably be a lattice simulation.
Unfortunately, lattice computations are time and resource-demanding and require very unrealistic simplifications.
The most evident limitation is perhaps the short separation of scales between the lattice spacing and the total size of the system. 
One consequence is that the bubble radius can only be a few orders of magnitude higher than the wall width. 
In a realistic cosmological phase transition, the difference is of many orders of magnitude.
Hence, for instance, the gravitational wave spectrum generated by bubble collisions (which peaks at a scale determined by the bubble size) is affected in these simulations by dynamics at the scale of the wall width 
(see \cite{mm21b} for a discussion on this issue).

Consequently, it is more usual to resort, at least partly, to analytical approximations.
For instance, the envelope approximation \cite{kt93} and similar models \cite{jt19} 
for gravitational waves from bubble collisions assume that the bubble walls and fluid shells next to them
are incomplete spherical surfaces. These assumptions are used in numerical simulations \cite{hk08,k18}
and semi-analytical calculations \cite{jt17}.
Applying our finite-width corrections to any of these calculations is straightforward. 
Moreover, in Ref.~\cite{mm21a} we generalized this kind of treatment to the case of bubble walls with deformations from the spherical shape. 
As mentioned above, the expressions derived in the present paper are relevant to that calculation.

We also emphasize that the hydrodynamic instability analysis of the wall propagation,
which results in wall corrugations \cite{l92}, 
is usually treated with the thin-wall approximation and in the linear-perturbation regime (see, e.g., \cite{hkllm93,mm14}).
Our derivation of a wall equation of motion valid for arbitrary deformations (including curvature radii which are not necessarily much larger than the wall width) is a first step towards a complete treatment of the instabilities. 
Again, the alternative of using a lattice calculation is limited by the problem of the separation of scales, as the lattice computation needs to follow the bubble growth from its nucleation, and the simulation time is not enough for the instabilities to develop.

Finally, some consequences of the phase transition depend on the wall profile.
Such is the case, e.g., with electroweak baryogenesis, 
where CP-violating currents arise from the interactions of the wall with plasma particles \cite{ckn93}.
Moreover, the wall velocity (which is a relevant parameter for most consequences of the phase transition)
also depends on the interactions with plasma particles \cite{lmt92,dlhll92}.
The calculation involves transport equations and considering the hydrodynamics around the walls (see, e.g., \cite{mp95}).
Spherical or flat walls and a $\tanh$ ansatz for the profile are usually assumed to simplify the treatment. 
Our Monge-gauge formalism aims at extending the analytic treatment beyond these symmetric walls.
As for the $\tanh$ ansatz, this specific shape corresponds to the profile obtained in the basic thin-wall approximation for a polynomial potential.
As we have seen, our perturbative approach can give an analytical profile to different orders in the wall width.

\section*{Acknowledgements}

This work was supported by Universidad Nacional de Mar del Plata,
grant EXA1091/22.

\appendix

\section{Avoiding the normal Gaussian coordinates}

\label{derivagral}

In this appendix we discuss the use of the thin-wall approximation without changing to normal Gaussian
coordinates.

\subsection{The wall equation of motion}

Let us consider the field equation (\ref{eccampo}) in an arbitrary
coordinate system and apply the first thin-wall assumption, 
namely, $\phi=\phi(n(x^{\mu}))$.
We obtain
\begin{equation}
\left(g^{\mu\nu}\partial_{\mu}n\partial_{\nu}n\right)\phi''(n)+g^{\mu\nu}\left(\partial_{\mu}\partial_{\nu}n-\Gamma_{\mu\nu}^{\rho}\partial_{\rho}n\right)\phi'(n)+V'(\phi)=0.\label{eccampo1era}
\end{equation}
We have a differential equation for $ \phi(n) $, but its coefficients in principle depend on $ x^\mu $.
Nevertheless, we can formally invert the function $n(x^{\mu})$ to change
variables from $x^{0},x^{1},x^{2},x^{3}$ to, say, $x^{0},x^{1},x^{2},n$.
Thus, for fixed $ x^{0},x^{1},x^{2} $ we have an ordinary differential equation in the variable $n$,
and we can use the usual trick of multiplying the equation by $\phi'(n)$
and then integrating with respect to $n$.
Using the second assumption, namely, that the coefficients of $\phi''$
and $\phi'$ remain approximately constant as we integrate through
the wall, we obtain 
\begin{equation}
\sigma\left(g^{\mu\nu}\nabla_{\mu}\partial_{\nu}n\right)_{n=0}=-\Delta V,\label{EOMn}
\end{equation}
with $\sigma=\int_{-\infty}^{+\infty}\phi^{\prime2}(n)dn$. 

Naively,
all we need to do to convert Eq.~(\ref{EOMn}) into an equation for the
wall position is to use the function $n(x^{\mu})$, which, according
to Eq.~(\ref{ngral}), is given by $n=F(x^{\mu})/s(x^{\mu})$ to
first order in the wall width. However, by doing so, we obtain $\nabla_{\mu}\partial_{\nu}(F/s)$
on the left-hand side of Eq.~(\ref{EOMn}), which does not give
the correct equation of motion (\ref{EOM0}), where we have $\nabla_{\mu}(\partial_{\nu}F/s)$
instead. Using the lowest-order approximation for $n(x^{\mu})$ is
too sloppy since we have second derivatives of $n$.
Indeed, if we write Eq.~(\ref{ngral}) in the form
\begin{equation}
n(x^{\mu})=F(x^{\mu})/s(x^{\mu})+c(x^{\mu})n^{2},\label{nFexac}
\end{equation}
we see that the second-order term gives a non-vanishing contribution
to $\partial_{\mu}\partial_{\nu}n|_{n=0}$. 

To obtain the coefficient $c$ to lowest order, we differentiate Eq.~(\ref{nFexac}),
\begin{equation}
\partial_{\mu}n=\partial_{\mu}F/s + F\partial_{\mu}s^{-1}+2cn\partial_{\mu}n+\mathcal{O}(n^{2}).\label{dnintermedia}
\end{equation}
Using Eq.~(\ref{nFexac}) in the second term on the right-hand side
and Eq.~(\ref{dnintermedia}) recursively in its third term, we obtain
\begin{equation}
\partial_{\mu}n=\partial_{\mu}F/s + \left(s\partial_{\mu}s^{-1} + 2c\partial_{\mu}F/s\right) n +  \mathcal{O}(n^{2}).\label{dn}
\end{equation}
According to the definitions (\ref{Nmu}), the vector $N_{\mu}=-\partial_{\mu}F/s$ satisfies
the normalization condition $N_{\mu}N^{\mu}=-1$. 
The vector\footnote{The quantity $\partial_{\mu}n$ is a vector since $n$ is defined
as the proper distance from the point $x^{\mu}$ to the hypersurface
along the geodesic.} $\partial_{\mu}n$ coincides with $N_{\mu}$ only on $\Sigma$, 
but we can readily show that it has the same normalization outside of $\Sigma$, since
in normal Gaussian coordinates we have $\partial_{\mu}n=(0,0,0,1)$, which gives
$\bar{g}^{\mu\nu}\partial_{\mu}n\partial_{\nu}n=-1$.
Therefore, taking the norm on both sides of Eq.~(\ref{dn}), we obtain
\begin{equation}
c=-\frac{1}{2}sN^{\mu}\partial_{\mu}s^{-1}+\mathcal{O}(n).
\end{equation}
Finally, from $s^{2}=-F_{,\mu}F^{,\mu}$ we obtain $\partial_{\mu}s^{-1}=-s^{-2}N^{\nu}\nabla_{\mu}\partial_{\nu}F$.
Inserting in Eqs.~(\ref{nFexac}) and (\ref{dnintermedia}), we obtain
\begin{equation}
n=\frac{F}{s}+\frac{1}{2s}N^{\rho}N^{\nu}\left(\nabla_{\rho}\partial_{\nu}F\right)n^{2}+\mathcal{O}(n^{3})
\label{n_orden2}
\end{equation}
and
\begin{equation}
\partial_{\mu}n=-N_{\mu}-\frac{1}{s}\left(\delta_{\mu}^{\rho}+N^{\rho}N_{\mu}\right)N^{\sigma}\left(\nabla_{\rho}\partial_{\sigma}F\right)n+\mathcal{O}(n^{2}).
\end{equation}
Differentiating the latter, we have
\begin{equation}
\partial_{\nu}\partial_{\mu}n|_{n=0}=\left[-\partial_{\nu}N_{\mu}+s^{-1}N_{\nu}\left(\delta_{\mu}^{\rho}+N^{\rho}N_{\mu}\right)N^{\sigma}\left(\nabla_{\rho}\partial_{\sigma}F\right)\right]_{n=0}.
\end{equation}
Contracting with $g^{\mu\nu}$, the second term vanishes. Inserting
this result in Eq.~(\ref{EOMn}), we obtain 
\begin{equation}
\sigma\left(g^{\mu\nu}\partial_{\nu}N_{\mu}-\Gamma_{\mu\nu}^{\rho}N_{\rho}\right)_{n=0}=\Delta V,
\end{equation}
which coincides with Eq.~(\ref{EOMNmu}).

To obtain the parameter $\sigma$ we go back to the field equation
(\ref{eccampo1era}). Using the normalization condition $g^{\mu\nu}\partial_{\mu}n\partial_{\nu}n=-1$
for the coefficient of $\phi''$ and just neglecting the term $\phi'$
(third thin-wall assumption), we obtain 
\begin{equation}
\phi''(n)=V'(\phi),\label{ecfisup3}
\end{equation}
which is the same equation for the kink profile previously derived
in normal Gaussian coordinates.

\subsection{The stress-energy tensor}

Let us now consider the stress-energy tensor, given by Eq.~(\ref{Tmunucpo}).
Using the assumption $\phi=\phi(n(x^\mu))$, we have 
\begin{equation}
T_{\mu\nu}=\phi^{\prime2}(n)\left[\partial_{\mu}n\partial_{\nu}n+\frac{1}{2}g_{\mu\nu}\right]+g_{\mu\nu}V(\phi),
\label{Tmunufield}
\end{equation}
where we have used the normalization condition $\partial_{\alpha}n\partial^{\alpha}n=-1$.
Inside the wall we can use Eq.~(\ref{ecfisup3}), which gives the
relation $V=\frac{1}{2}\phi^{\prime2}+V_{+}$, and we obtain the stress-energy
tensor of the wall, 
\begin{equation}
T_{\mu\nu}^{\mathrm{wall}}=\phi^{\prime2}(n)\left[\partial_{\mu}n\partial_{\nu}n+g_{\mu\nu}\right].
\end{equation}
Finally, with the approximation $\phi^{\prime2}(n)=\sigma\delta(n)$ and using $ \partial_{\mu}n|_{n=0}=-N_{\mu} $,
we obtain the result (\ref{Tmunuwall}).

In Ref.~\cite{mm21a}, we used a similar derivation for a spherical
wall with deformations to study the generation of gravitational waves.
However,  we assumed a wall profile
\begin{equation}
	\phi(\mathbf{r},t)=\phi_{0}\left(\hat{n}\cdot(\mathbf{r}-\mathbf{r}_{w})\right),\label{1erasup_espacial}
\end{equation}
where $\hat{n}$ is the normal vector to the (time dependent) three-dimensional
surface $S$ instead of the hypersurface $\Sigma$. The argument
of $\phi_{0}$ in Eq.~(\ref{1erasup_espacial}) is similar to the
variable $n$ given by Eq.~(\ref{gaussianasinvert}), but the time
dependence is only taken into account in the wall position $\mathbf{r}_{w}(t)$.
To compare this approach with our present approach, we note that
only the transverse and traceless projection of the space components $ T_{ij} $ generate
gravitational radiation, so we can drop the terms proportional to $ g_{\mu\nu} $ in Eq.~(\ref{Tmunucpo})
and consider the tensor $\tilde{T}_{ij}=\partial_{i}\phi\partial_{j}\phi$, where $i,j$ denote
the rectangular coordinates $x,y,z$. 
Parametrizing the surface as $r=r_{w}(\theta,\phi,t)$ and using the thin-wall approximation, one
obtains \cite{mm21a}
\begin{equation}
\tilde{T}^{\mathrm{wall}}_{ij}=\sigma\frac{\left(\hat{r}_{i}-\partial_{i}r_{w}\right)\left(\hat{r}_{j}-\partial_{j}r_{w}\right)}{\sqrt{1+(\nabla r_{w})^{2}}}\,\delta(r-r_{w}),\label{TijGWold}
\end{equation}
where $\hat{r}=\mathbf{r}/r$. The square root in Eq.~(\ref{TijGWold})
comes from the normalization of $\hat{n}$. 
This expression is in agreement with our result for the Monge gauge, Eq.~(\ref{Tmunu_hibrido}),
except for the denominator, which must be replaced with the quantity 
\begin{equation}
s=\sqrt{1+(\nabla r_{w})^{2}-(\partial_{t}r_{w})^{2}}.
\end{equation}
Therefore, the approximation used in Ref.~\cite{mm21a} is valid
for small velocities. We shall consider the corrections to those results
elsewhere.

\subsection{The equation of motion from the stress-energy tensor}

When the stress-energy tensor $T_{\mu\nu}$ has a delta-function singularity,
it is usual to define the \emph{surface stress-energy tensor on the
hypersurface}, $S_{\mu\nu}$, as the integral of $T_{\mu\nu}$ with
respect to the proper distance $n$ perpendicular to $\Sigma$ \cite{misner}.
In our case, according to Eqs.~(\ref{Tabwall}-\ref{Tmunuwall}),
we have $T_{\mu\nu}^{\mathrm{wall}}=S_{\mu\nu}\delta(n)$, where in
normal Gaussian coordinates we have
\begin{equation}
S_{ab}=\sigma\gamma_{ab},\quad S_{\mu n}=0,
\end{equation}
and in covariant form, 
\begin{equation}
S_{\mu\nu}=\sigma\left(g_{\mu\nu}+N_{\mu}N_{\nu}\right).\label{Smunu}
\end{equation}
The result $S_{n\mu}=0$ is general and means that the momentum flux
is on the worldvolume of the surface \cite{misner}. We can also write
the EOM $\sigma\gamma^{ab}K_{ab}=-\Delta V$ in terms of this quantity.
Using Eq.~(\ref{Tmunubulk}), we have 
\begin{equation}
S^{ab}K_{ab}=-\Delta{T^{n}}_{n}.\label{eomS}
\end{equation}
This equation also has more general validity (see, e.g., Ref.~\cite{bkt87}).
An additional equation of motion for the surface layer can be obtained
using junction conditions derived from the Einstein field equation \cite{misner}, 
\begin{equation}
{S_{a}}^{b}{}_{|b}+\Delta{T_{a}}^{n}=0,\label{eomMisner}
\end{equation}
where the vertical stroke denotes a covariant derivative with respect
to the intrinsic geometry of the hypersurface. This covariant derivative
is obtained by projecting the expression orthogonally onto the hypersurface.
Taking into account that ${S_{a}}^{n}=0$ and using the projector
${\delta_{\rho}}^{\nu}+N_{\rho}N^{\nu}$, we can write Eq.~(\ref{eomMisner})
covariantly as
\begin{equation}
\left(g_{\rho\nu}+N_{\rho}N_{\nu}\right)\nabla_{\mu}S^{\mu\nu}=\left(g_{\rho\nu}+N_{\rho}N_{\nu}\right)\Delta{T_{\nu}}^{\mu}N_{\mu}.
\end{equation}
To check that our surface stress-energy tensor fulfills this equation,
we first note that in our case the right-hand side vanishes according
to (\ref{Tab_0}). To see that the left-hand side vanishes too, we
take the covariant derivative of Eq.~(\ref{Smunu}) and then apply
the projector,
\begin{equation}
\left(g_{\rho\nu}+N_{\rho}N_{\nu}\right)\nabla_{\mu}S^{\mu\nu}=\sigma N^{\mu}\nabla_{\mu}N_{\rho}.
\end{equation}
In the normal Gaussian system this result takes the form $\partial_{n}N_{\rho}-\Gamma_{n\rho}^{\sigma}N_{\sigma}=\Gamma_{n\rho}^{n}=0$.
Therefore, in our case Eq.~(\ref{eomMisner}) is trivially satisfied,
while (\ref{eomS}) is actually equivalent to the wall equation of
motion.

\section{The bubble interior}

\label{interior}

Although in this paper we focus on the bubble wall, we shall dedicate this appendix to discuss
the behavior of the field inside the bubble. 

\subsection{Field evolution}

\begin{figure}[tb]
	\centering
	\includegraphics[height=4.6cm]{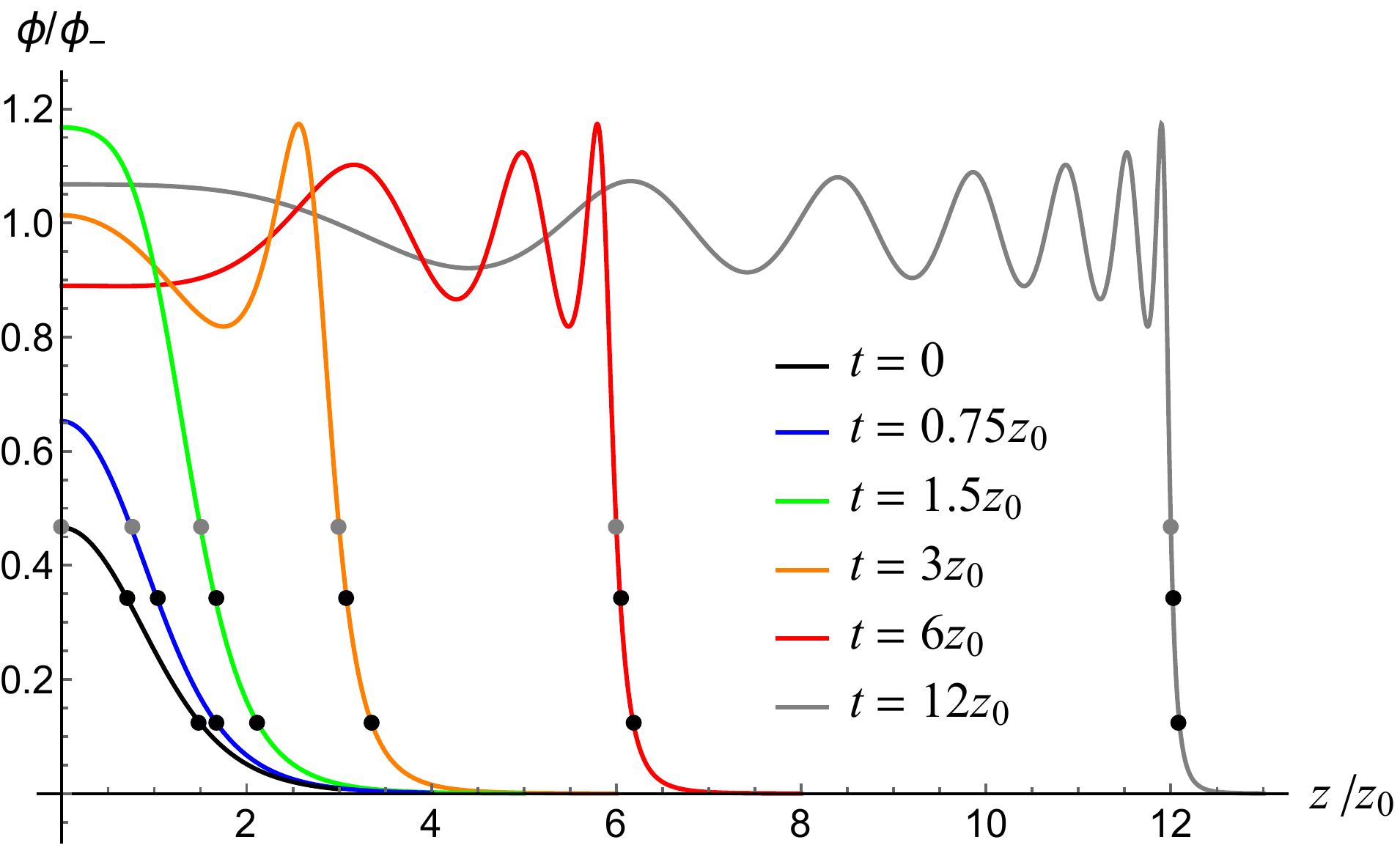}
	\hfill
	\includegraphics[height=4.6cm]{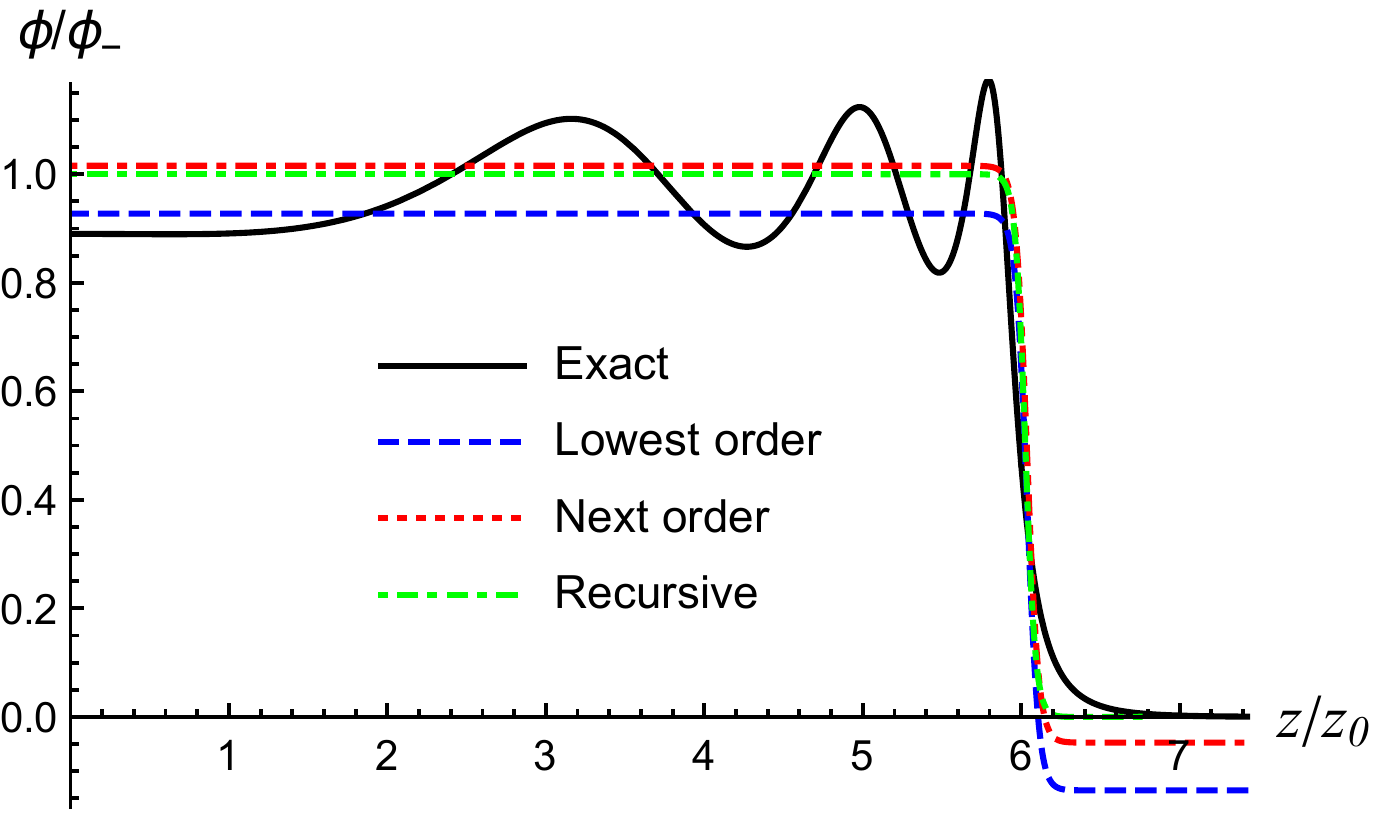}
	\caption{Left: Planar-symmetry profiles for the potential
		of Fig.~\ref{fig_potmuygruesa} at several times. 
		Right: The different approximations to the profile at a given time.\label{figperftss}}
\end{figure}

We have seen that
the exact solution of the field equation has oscillations that are generated
behind the wall and decay toward the bubble center. For the spherical
symmetry case, this behavior is appreciable only if the ratio $\Delta V/V_{\mathrm{max}}$
is high enough, as in the case of Figs.~\ref{fig_potmuygruesa}-\ref{fig_muygruesa}. 
The planar symmetry solution for the same potential 
(equivalently, the solution in 1+1 dimensions) 
is shown in Fig.~\ref{figperftss}.
It exhibits strong oscillations which are sustained in time.
This is because, in the mechanical analogy to Eqs.~(\ref{ec4dint})-(\ref{condec4dint}), the friction decreases with time $\tau$, so it can become really small if $ \phi_i $ is sufficiently far from $ \phi_- $.
We indicated the value $\phi=\phi_i$ with a gray dot in Figs.~\ref{figperfilt} and \ref{figperftss}.
This fixed point on the profile moves at the speed of light and does not represent
a physical motion. 

The lightlike hypersurface   $r=r_{i}(t)$ where the
field takes the value $\phi_{i}$ corresponds to $\rho=0$ and separates
the solution $\phi=\tilde{\phi}(\sqrt{t^{2}-r^{2}})$  for $r<t$ 
from the solution $\phi=\bar{\phi}(\sqrt{r^{2}-t^{2}})$  for $r>t$. 
Thus, inside the lightcone we have the bubble interior, where field oscillations take place, 
while outside the lightcone we have the bubble wall where the field reaches (asymptotically) the value $\phi_+$.
In Ref.~\cite{cghw20}, the wall was defined (motivated by the thin-wall approximation)
as the section of the field profile between the points $r_1(t)$ and $r_2(t)$ corresponding to the values 
$\phi_1=\phi_i (1-\tanh(-1/2))/2$ and $\phi_2=\phi_i (1-\tanh(1/2))/2$, respectively. 
These points are indicated by black dots on the profiles in Fig.~\ref{figperftss}.
As the bubble expands, their separation decreases due to Lorentz contraction.

\begin{figure}[tb]
	\centering
	\begin{minipage}[c]{8.5cm}
		\includegraphics[width=8.5cm]{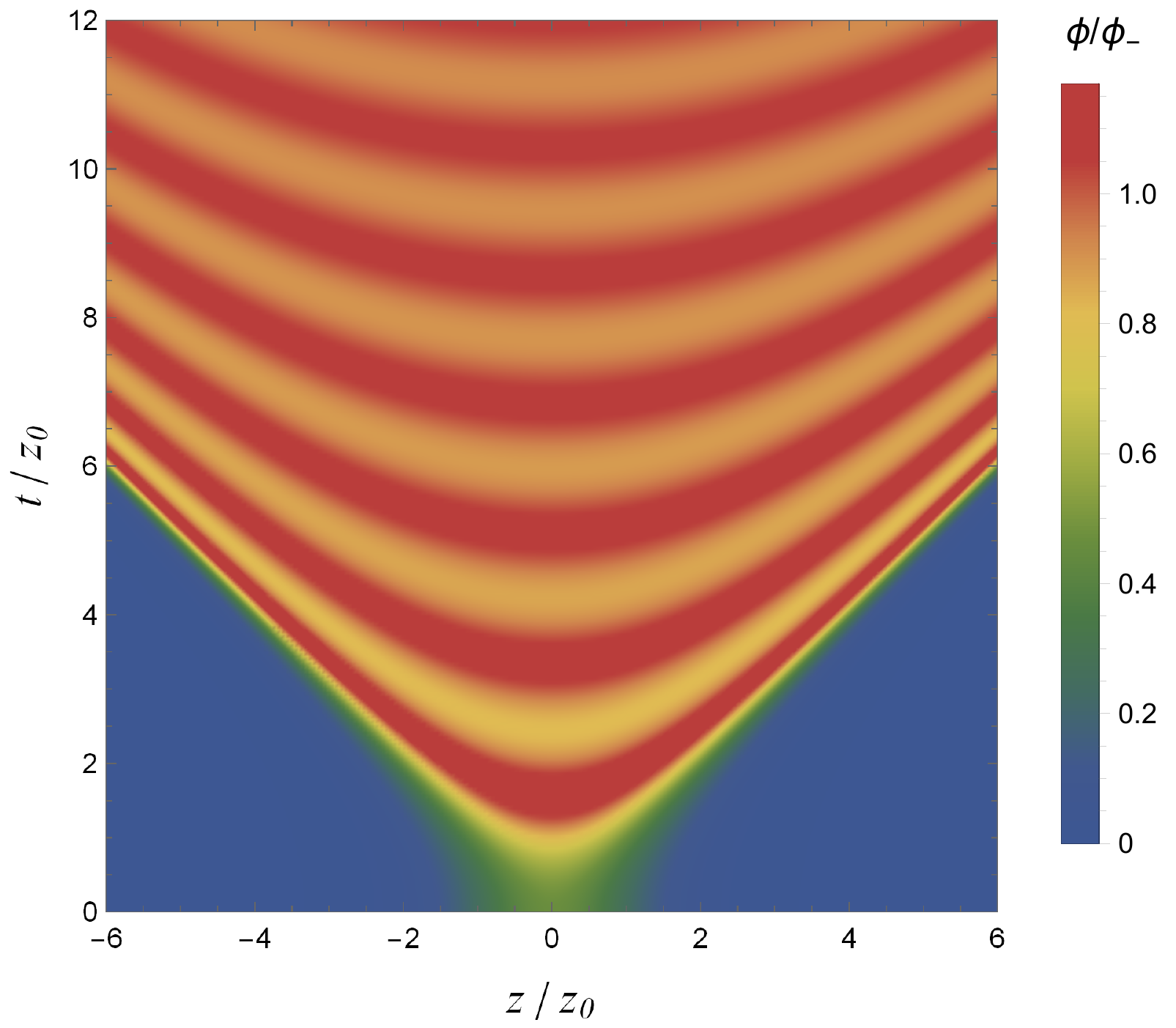}
	\end{minipage}
	\hfill
	\begin{minipage}[c]{6.8cm}
		\includegraphics[width=7cm]{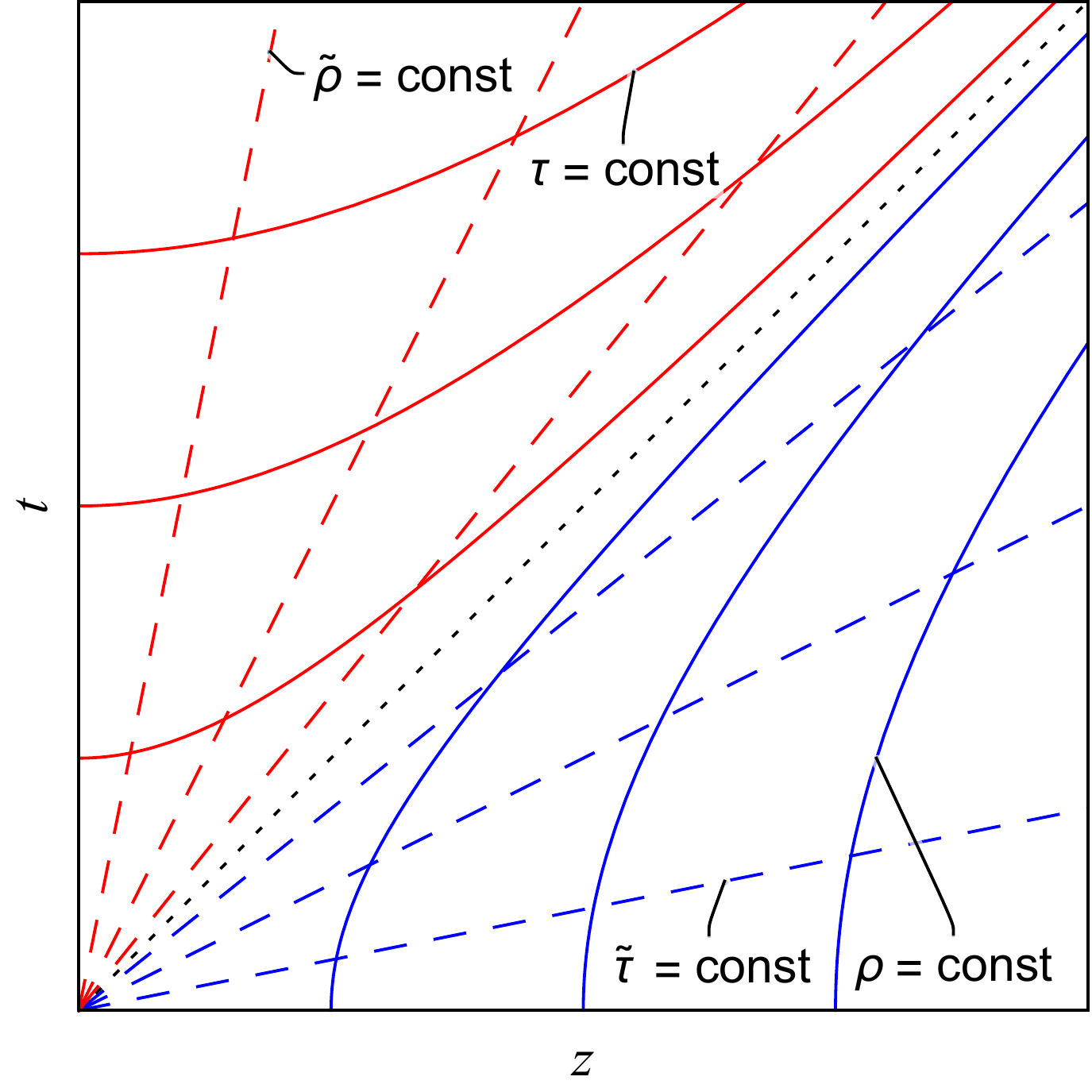}
	\end{minipage}	
	\caption{Left: Evolution of the field $\phi(t,z)$ for the potential of Fig.~\ref{fig_potmuygruesa}. 
		Right: Milne and Rindler patches of Mikowski spacetime in the quadrant $z\geq 0,t\geq0$.\label{spacetimess}}
\end{figure}

In the 1+1-dimensional case, the complete bubble profile (which includes two walls) is obtained by reflection at $z=0$.
The left panel of Fig.~\ref{spacetimess} shows the field evolution in the $zt$ plane after the nucleation of the bubble ($t\geq 0$).
The curves of constant $\phi$ are hyperbolas and are shown in the right panel for $t\geq 0$ and $x\geq 0$ (solid curves).
Outside the lightcone the hyperbolas are timelike. 
Any of these curves represents the trajectory of a fixed point in the wall profile (such as the black points of Fig.~\ref{figperftss}).
Inside the lightcone we have field oscillations. 
There, the hyperbolas are spacelike and represent the set of points in space which are in phase.
In each of these regions the field depends on a single real variable ($\rho$ or $\tau$), so the description of the dynamics is simpler in a coordinate system where this variable is a coordinate.

The coordinates adapted to the timelike hyperbolas are the well-known Rindler coordinates $(\tilde\tau,\rho)$,
which can be introduced via $t=\rho\sinh\tilde\tau$, $z=\rho\cosh\tilde\tau$ and cover the exterior of the lightcone
(see the right panel of Fig.~\ref{spacetimess}).
We have $\rho^2=z^2-t^2$ and $\tanh\tilde\tau=t/z$, so the field is given by $\phi(\rho,\tilde\tau)=\bar\phi(\rho)$
and does not depend on the time $\tilde\tau$. 
Thus, for observers at rest in this frame, the wall profile is fixed.
These observers are constantly accelerated in Minkowski space (their worldlines are the blue hyperbolas in Fig.~\ref{spacetimess}).
Inside the lightcone we introduce similar coordinates $(\tau,\tilde\rho)$ through
$t=\tau\cosh\tilde\rho$, $z=\tau\sinh\tilde\rho$,
which give the so-called Milne spacetime
(see, e.g., \cite{GronHervik}). 
We have $\tau^2=t^2-z^2$ and $\tanh\tilde\rho = z/t$,
so the field is given by $\phi(\tau,\tilde\rho)=\tilde\phi(\tau)$.
Thus, for comoving observers in the Milne Universe, the field is homogeneous and only oscillates in time.
These observers move with constant velocities in Minkowski coordinates (their worldlines are the red dashed straight lines in Fig.~\ref{spacetimess}).
The above discussion is essentially valid in the 3+1 dimensional case, with the replacement $z\to r$.

\subsection{Approximations}

In the right panel of Fig.~\ref{figperftss}, we plotted the different approximations to one of the profiles on the left.
In this case, the basic thin-wall approximation does not give a very good approximation for the wall profile, although
it gives the wall position with good precision. At higher order and
with the recursive improvement, we obtain a profile which matches
the value $\phi_{+}$ outside the bubble, while inside, it matches
the value $\phi_{-}$, which is the average of the oscillating solution. 
The thin-wall approximation (to any order)
will never reproduce the field oscillations inside the bubble.
Nevertheless, the bubble interior can also be treated analytically
with suitable approximations.

\begin{figure}[tb]
\centering
\includegraphics[width=0.45\textwidth]{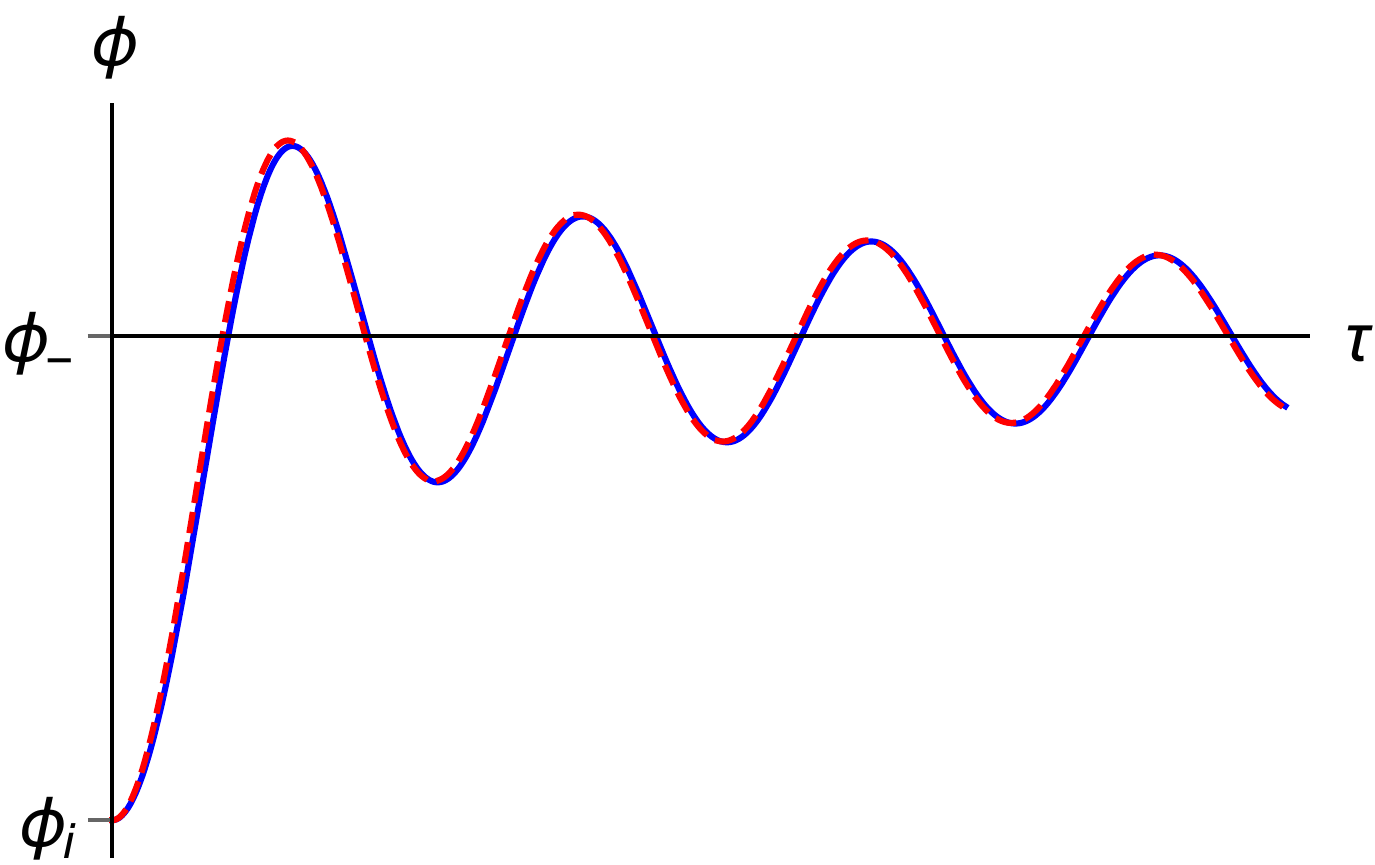} 
\hfill
\includegraphics[width=0.45\textwidth]{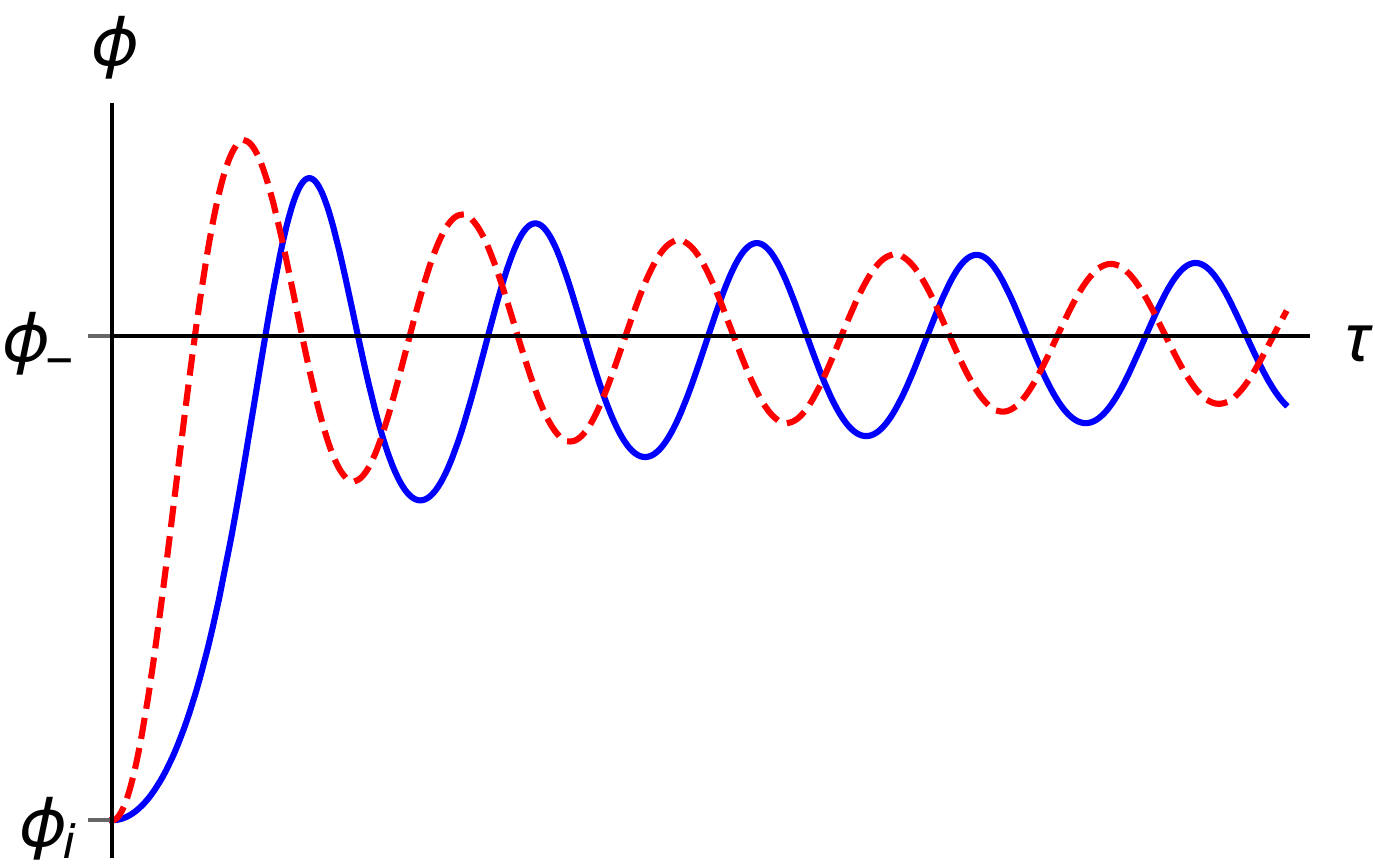}
\caption{The solution $\tilde{\phi}(\tau)$ (solid line) and the approximation
(\ref{fiintap}) (dashed line) for the planar case. Left: The solution
for the potential of Figure \ref{fig_potgruesa} ($\Delta V/V_{\max}\simeq2.6$).
Right: The solution for the potential of Figure \ref{fig_potmuygruesa}
($\Delta V/V_{\max}\simeq52$).\label{figinterior}}
\end{figure}

The solution $\tilde{\phi}$
is governed by Eq.~(\ref{ec4dint}) and satisfies the conditions
(\ref{condec4dint}). Since $\phi_{i}$ is generally close to $\phi_{-}$
(except in extreme cases like that of Fig.~\ref{figperftss}), we
can use the small-oscillations approximation. Thus, we approximate
the potential as
\begin{equation}
V(\phi)\simeq V_{-}+\frac{1}{2}M_{-}^{2}\left(\phi-\phi_{-}\right)^{2},\label{potcuad}
\end{equation}
with $M_{-}^{2}=V''(\phi_{-})$. Redefining $\tilde{\phi}=\phi-\phi_{-}$,
Eq.~(\ref{ec4dint}) becomes the equation of a damped harmonic oscillator,
\begin{equation}
\frac{d^{2}\tilde{\phi}}{d\tau^{2}}+\frac{j}{\tau}\frac{d\tilde{\phi}}{d\tau}=-M_{-}^{2}\tilde{\phi},\label{ec4dint-1}
\end{equation}
where $\tau$ is the time and the friction force decreases
in time. The general solution is
\begin{equation}
\tilde{\phi}=\left[c_{1}J_{\frac{j-1}{2}}\left(M_{-}\tau\right)+c_{2}Y_{\frac{j-1}{2}}\left(M_{-}\tau\right)\right]\tau^{-\frac{j-1}{2}},
\end{equation}
where $J_{n}(x)$ is the Bessel function of the first kind and $Y_{n}(x)$
that of the second kind. As $Y_{n}(x)$ diverges at $x=0$ while $J_{n}(x)/x^{n}=2^{-n}/n!+\mathcal{O}(x^{2})$
for small $x$, we must set $c_{2}=0$ and we obtain
\begin{equation}
\tilde{\phi}=\left(\phi_{i}-\phi_{-}\right)2^{n}n!J_{n}(M_{-}\tau)\tau^{-n},\quad n={(j-1)}/{2}.
\label{fiintap}
\end{equation}
In Fig.~\ref{figinterior}, we check this approximation for the planar
case, where the field oscillations are stronger. We see that the approximation
is very good except when $\Delta V/V_{\max}$ is really large. 
 
To plot the analytic approximation (\ref{fiintap})
(dashed lines in Fig.~\ref{figinterior}), we used the value of $\phi_{i}$ obtained numerically
from the shooting method for the solution $\phi$. 
Notice that, in the analytical approximations for the wall profile, we just replaced $\phi_{i}$ by $\phi_{-}$,
but we cannot do that in Eq.~(\ref{fiintap}). 
For a fully analytical
approximation, we could estimate the parameter $ \phi_i $ as follows. Besides
approximating $V$ by a parabola near its minimum $\phi_{-}$, we
also approximate it by an inverted parabola near its maximum
$\phi_{\max}$. We may join these two parabolas with a line of constant
potential, say, $V=V_{+}$. The parabola at the maximum can also be
joined with the other minimum, $\phi_{+}$, with such a constant potential.
Thus, we have a piecewise-defined potential, and we can solve the equation
for $\phi$, Eq.~(\ref{ec4d}), in each region analytically. We will
only need to match the analytical solutions, and the value of $\phi_{i}$
will be determined by the conditions (\ref{cc}). 

This method also gives an alternative approximation for the entire bubble profile. 
However, the process of matching the solutions for the different potential
regions is straightforward but tedious, and is certainly out of the scope of the present paper.
Similar but simpler approximations using only parabolas (which are matched directly instead of joining them with horizontal lines)  
were considered in Ref.~\cite{b16} for the bounce profile
and in Ref.~\cite{t99} for the bubble profile at any time $t$.
The latter gives the correct qualitative behavior of the field both outside and inside the bubble.
However, for the wall part of the profile, such approximations
will never be as good as those discussed in the previous sections of
this paper. 
On the other hand, this seems to be the only way to obtain analytic approximations for 
the field evolution inside the bubble.

\bibliographystyle{jhep}
\bibliography{papers}
\end{document}